\documentclass[10pt,journal,compsoc]{IEEEtran}
\ifCLASSOPTIONcompsoc
\usepackage[nocompress]{cite}
\else
\usepackage{cite}
\fi
\ifCLASSINFOpdf
\else
\fi
\usepackage{graphicx}
\usepackage{amsmath}
\usepackage{amssymb}
\usepackage{amsthm}
\usepackage{subfigure}
\usepackage{blkarray}
\usepackage{threeparttable}
\usepackage{multirow}
\usepackage{makecell}
\usepackage{colortbl}
\usepackage{array}
\usepackage{diagbox}
\usepackage{booktabs}
\usepackage{tikz}
\usepackage{fbox}
\usepackage{bm}
\usepackage{bbm}
\usepackage{url}
\usepackage[linesnumbered,ruled,boxed,norelsize,vlined,commentsnumbered]{algorithm2e}
\usepackage[colorlinks,linkcolor=red,anchorcolor=blue,citecolor=black]{hyperref}
\usepackage{comment}
\usepackage{marginnote}
\usepackage{sidecap}
\usepackage{accents}

\newcommand{\revise}[1]{{\color{black}{#1}}}

\newcommand{\revisemajor}[1]{{\color{black}{#1}}}

\makeatletter
\def\widebar{\accentset{{\cc@style\underline{\mskip10mu}}}}
\def\Widebar{\accentset{{\cc@style\underline{\mskip8mu}}}}
\makeatother

\newcommand{\tr}{{\mathop{\mathrm{tr}}}}
\newcommand{\atan}{{\mathop{\mathrm{atan}}}}

\newcommand{\RDOS}{{\mathop{\mathrm{RDOS}}}}

\newcommand{\diag}{{\mathop{\mathrm{diag}}}}
\newcommand{\argmin}{{\mathop{\mathrm{argmin}}}}
\newcommand{\ie}{{\textit{i.e.}}}
\newcommand{\etal}{{\textit{et al.}}}
\newcommand{\eg}{{\textit{e.g.}}}
\newcommand{\best}[1]{\textbf{{\color{black}{#1}}}}

\newtheorem{theorem}{Theorem}

\newtheorem{definition}{Definition}
\newtheorem{proposition}{Proposition}

\definecolor{Cauchy}{rgb}{0.904705882352941	0.191764705882353	0.198823529411765}
\definecolor{LMS}{rgb}{0.294117647058824	0.544705882352941	0.749411764705882}
\definecolor{HGMM}{rgb}{0.371764705882353	0.717647058823529	0.361176470588235}

\definecolor{SAREfit}{rgb}{0.685882352941177	0.403529411764706	0.241176470588235}
\definecolor{RANSAC}{rgb}{0.2510    0.2510    0.5020}
\definecolor{BayFit}{rgb}{1	0.548235294117647	0.100000000000000}

\definecolor{LSA}{rgb}{0.294117647058824	0.544705882352941	0.749411764705882}
\definecolor{LSO}{rgb}{0.904705882352941	0.191764705882353	0.198823529411765}
\definecolor{WLS}{rgb}{0.371764705882353	0.717647058823529	0.361176470588235}
\definecolor{Taubin}{rgb}{0.64	0.64	0.64}
\definecolor{Prasad}{rgb}{0.865	0.811	0.433}
\definecolor{Wu}{rgb}{0.971764705882353	0.555294117647059	0.774117647058824}
\definecolor{SAREfit}{rgb}{0.685882352941177	0.403529411764706	0.241176470588235}
\definecolor{BayFit}{rgb}{1	0.548235294117647	0.100000000000000}
\graphicspath{{./figure/}}
\begin{document}
	%
	\title{A Bayesian Approach Toward Robust Multidimensional Ellipsoid-Specific Fitting}
	
	%
	%
	%
	%

	\author{Mingyang~Zhao,
		Xiaohong~Jia,
		Lei~Ma,
		Yuke Shi,
		Jingen Jiang,
		Qizhai Li,\\
		Dong-Ming Yan
		and~Tiejun Huang
		\IEEEcompsocitemizethanks{
			\IEEEcompsocthanksitem M. Zhao is with the Hong Kong Institute of Science \& Innovation
			Chinese Academy of Sciences. E-mail: migyangz@gmail.com\protect
			\IEEEcompsocthanksitem X. Jia, Y. Shi, and Q. Li are with the Academy of Mathematics and Systems Science, CAS, Beijing, China. E-mail: \{xhjia, shiyuke, liqz\}@amss.ac.cn\protect
			\IEEEcompsocthanksitem L. Ma and T. Huang are with the BAAI and Peking University, Beijing, China. E-mail: \{lei.ma, tjhuang\}@pku.edu.cn\protect
			\IEEEcompsocthanksitem J. Jiang is with the School of Computer Science and Technology, Shandong University, Qingdao, China. E-mail: xiaowuga@gmail.com\protect
			\IEEEcompsocthanksitem D. Yan is with the Institute of Automation, CAS, Beijing, China. E-mail:  yandongming@ia.ac.cn
		}
		}

	%
	%

	\markboth{IEEE TRANSACTIONS ON PATTERN ANALYSIS AND MACHINE INTELLIGENCE}%
	{Shell \MakeLowercase{\textit{et al.}}: A Bayesian Approach Toward Robust Multidimensional Ellipsoid-Specific Fitting}
	%



\IEEEtitleabstractindextext{
\begin{abstract}
This work presents a novel and effective method for fitting {multidimensional} ellipsoids (\ie, ellipsoids embedded in $\mathbb{R}^n$) to scattered data in the contamination of noise and outliers. \revise{Unlike conventional algebraic or geometric fitting paradigms that assume each measurement point is a noisy version of its nearest point on the ellipsoid, we approach the problem as a \emph{Bayesian parameter estimate process} and maximize the posterior probability of a certain ellipsoidal solution given the data. We establish a more robust correlation between these points based on the \emph{predictive distribution} within the Bayesian framework, \ie, considering each model point as a potential source for generating each measurement. Concretely, we incorporate a uniform prior distribution to constrain the search for primitive parameters within an ellipsoidal domain, ensuring \emph{ellipsoid-specific} results regardless of inputs. We then establish the connection between measurement point and model data via Bayes' rule to enhance the method's robustness against noise.} Due to  independent of spatial dimensions, the proposed method not only delivers high-quality fittings to challenging elongated ellipsoids but also generalizes well to multidimensional spaces. To address outlier disturbances, often overlooked by previous approaches, we further introduce a {uniform distribution} on top of the \emph{predictive distribution} to significantly enhance the algorithm's robustness against outliers.  Thanks to the uniform prior, our maximum a posterior probability coincides with a more tractable maximum likelihood estimation problem, which is subsequently solved by a numerically stable \emph{Expectation Maximization} (EM) framework.  Moreover, we introduce an \emph{$\varepsilon$-accelerated technique} to expedite the convergence of EM considerably. We also investigate the relationship between our algorithm and conventional least-squares-based ones, during which we theoretically prove our method's superior robustness. To the best of our knowledge, this is the first \revise{comprehensive} method capable of performing multidimensional ellipsoid-specific fitting within the Bayesian optimization paradigm under diverse disturbances. We evaluate it across lower and higher dimensional spaces in the presence of heavy noise, outliers, and substantial variations in axis ratios. Also, we apply it to a wide range of practical applications such as \revise{\emph{microscopy cell counting}}, \emph{3D reconstruction}, \emph{geometric shape approximation}, and \emph{magnetometer calibration} tasks. In all these test contexts, our method consistently delivers flexible, robust, ellipsoid-specific performance, and achieves the state-of-the-art results. 
\end{abstract}
		
}

\maketitle

	\IEEEdisplaynontitleabstractindextext

	%
	\IEEEpeerreviewmaketitle
	
\begin{figure}[t]
	\centering
	\subfigure[2D approximation]{
		\includegraphics[width=0.105\textwidth]{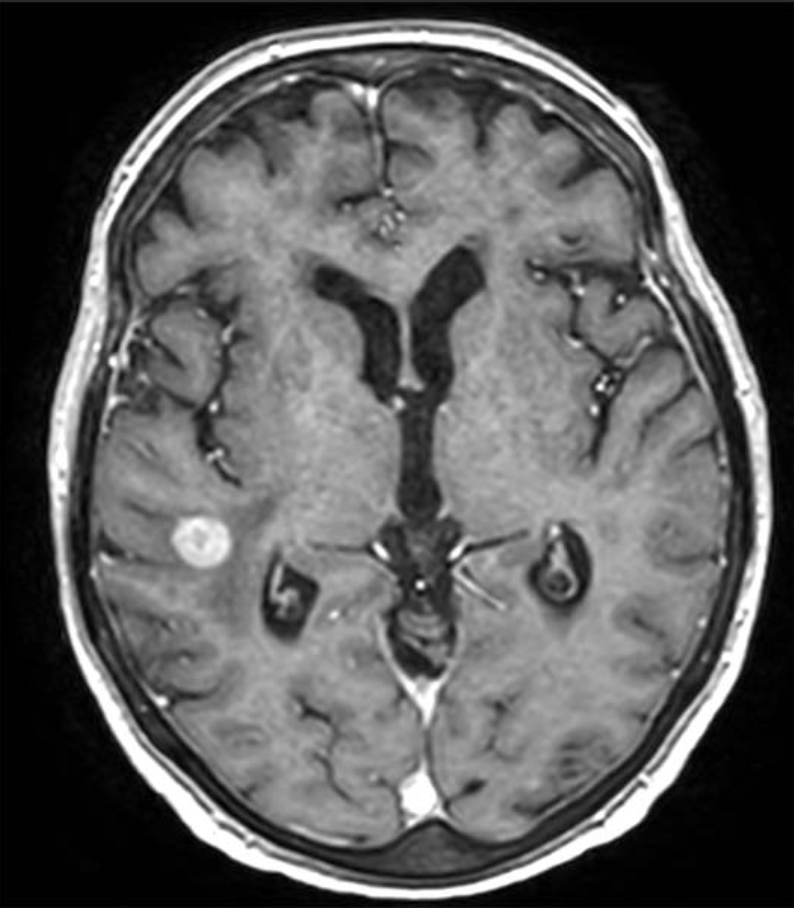}
		\includegraphics[width=0.105\textwidth]{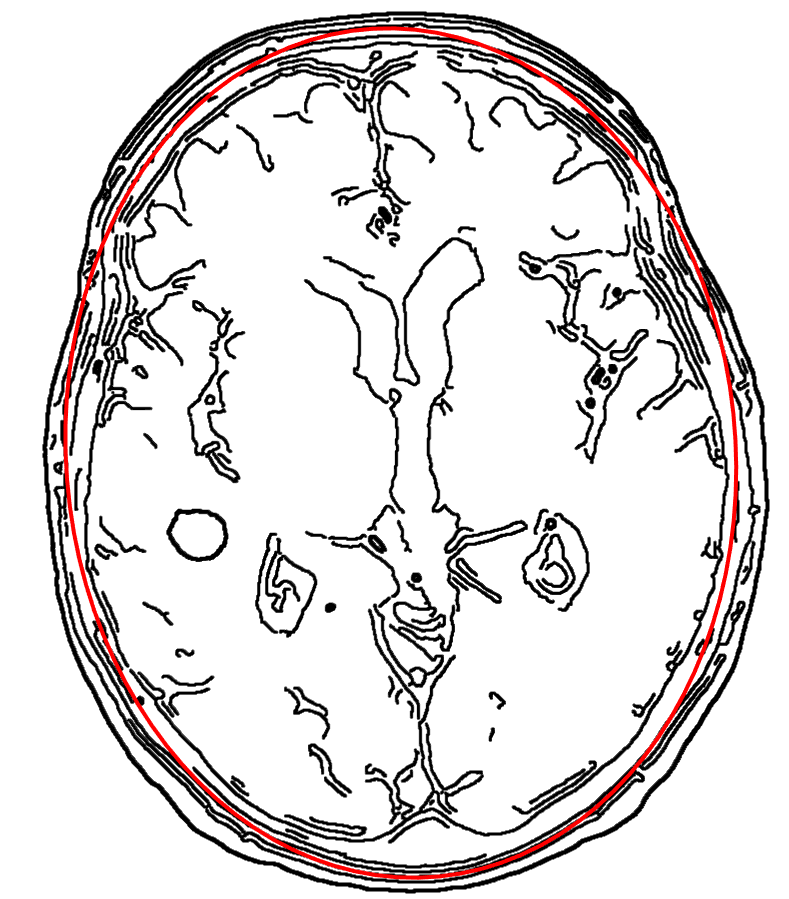}
	}
	\subfigure[3D approximation]{\includegraphics[width=0.11\textwidth]{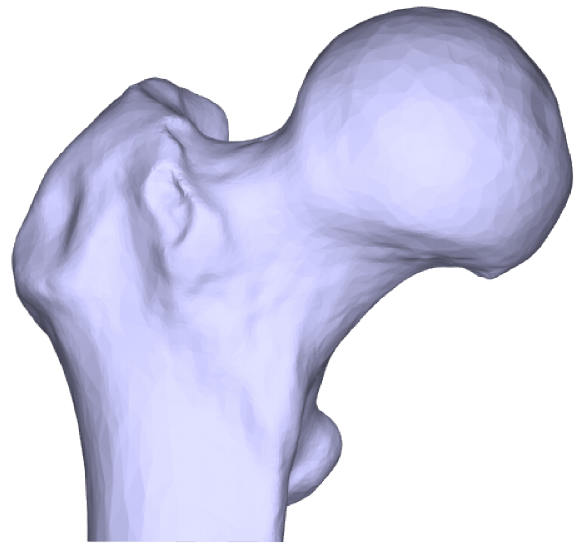}
		\includegraphics[width=0.13\textwidth]{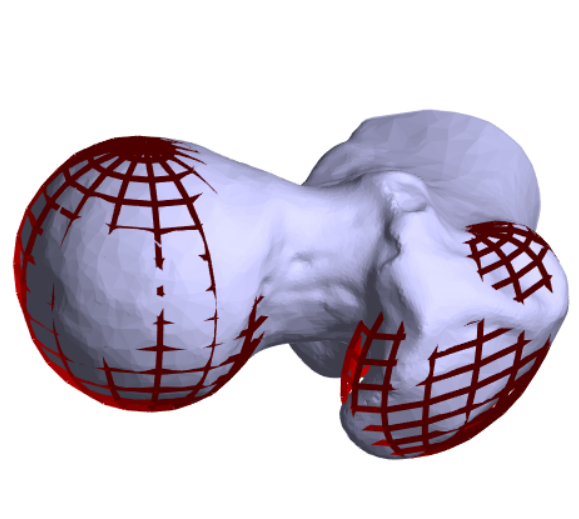}
	}
\vskip -0.3cm	
	\caption{Geometric shape approximation using elliptical curves and ellipsoidal surfaces based on the developed method. (a) Ellipse fitting to a 2D brain MRI slice in the presence of heavy outliers (\ie, other brain structures). Note that we perform ellipse fitting on the edge map directly \emph{without any preprocessing step} such as arc detection or structure removal. (b) Ellipsoid fitting to the anomalous 3D medical femur data~\cite{DataBone}.
	}
	\vskip -0.5cm
	\label{fig:ex1}
\end{figure}

	\IEEEraisesectionheading{\section{Introduction}\label{sec:introduction}}
	\label{sec:intro}
	\revise{\subsection{Background}}
    \IEEEPARstart{T}{he} fitting of quadratic curves \revise{(conics)} or surfaces \revise{(quadrics)} to unorganized point clouds is a fundamentally important problem in computer vision and pattern analysis communities. The ellipse and ellipsoid primitives hold a special interest in practice because the perspective projection of circular objects surrounded us in 2D images are ellipses, whereas the ellipsoid is the \emph{uniquely bounded and centered quadric} which enables geometric shape approximation (see Fig.~\ref{fig:ex1} as an example), object segmentation, orientation, and recognition effectively.

	\revise{Benefiting from these advantages, ellipse and ellipsoid fitting find diverse applications in various fields.} {For instance, ellipse fitting is widely employed for camera calibration~\cite{kim2005geometric}, pose estimation~\cite{liu2014relative}, object tracking~\cite{kothari2021ellseg}, and cell counting in biomedical sciences~\cite{kothari2009automated,panagiotakis2020region}.} \revise{In the context of motion capture}, Shuai~\etal~\cite{shuai2016motion} utilized the ellipsoidal skeleton to approximate 3D human bodies.  Hemerly \etal~\cite{6843947} solved the magnetometer calibration problem \revise{by using ellipsoid fitting and analyzing} the eigenvalues and eigenvectors. \revise{Moreover}, in multidimensional spaces, ellipsoidal regions can be used to facilitate pattern classification tasks~\cite{lee2005adaptive}.

\revise{Given the significant importance and wide-ranging applicability of ellipsoid-specific fitting in real-world scenarios, numerous pioneering efforts have emerged to solve this problem.} Most current algorithms \revise{center around} the \emph{least-squares} principle, aiming to minimize either the \emph{algebraic distance}~\cite{fitzgibbon1999direct,ying2012fast,lin2015fast,kesaniemi2017direct}--that is, the residual error of the implicit quadratic equation away from zero--or the \emph{geometric distance}~\cite{gander1994least,ahn2001least,ahn2002orthogonal,liu2008revisit}, which represents the shortest Euclidean (orthogonal) distance \revise{between} data points to the ideal ellipsoidal surface.

These approaches typically perform well when dealing with simple and low-degradation 2D and 3D measurements. This is achieved by imposing a variety of ellipsoid-specific constraints, such as linear, quadratic, or positive semi-definite constraints. However, their extension to higher-dimensional spaces is not always immediate. To the best of our knowledge, there is still lack of a \revise{universal} constraint that can guarantee ellipsoid-specific fitting consistently across multidimensional spaces.

\revisemajor{Furthermore}, prior approaches often exhibit a propensity to produce biased fittings in scenarios where data points are captured from ellipsoids featuring significant discrepancies between their longest and shortest major axes--referred to as \emph{large axis ratios}~\cite{kesaniemi2017direct}; besides, they tend to inadequately address the impact of heavy noise and outliers during the algorithmic design process~\cite{birdal2019generic}, which inevitably degrades the fitting solutions in terms of both accuracy and types. For example, a large percentage of outliers \revisemajor{in the data} can pull the \revisemajor{resulting} surface towards a paraboloid or hyperboloid rather than an ellipsoid.

\revise{\subsection{Motivation and The Proposed Work}}
{To tackle the aforementioned challenges, \revisemajor{including} ellipsoid-specific fitting \revisemajor{regardless of input data and spatial dimensions}, handling large axis ratios, and addressing sensitivity to outliers and noise, we present a novel and \revisemajor{systematic} paradigm in this work. Our approach offers the following favorable attributes: 1) The ability to \emph{accurately} fit ellipsoidal surfaces together with a compatible computational complexity. 2) The stability to keep the \emph{consistent ellipsoid-specific} results regardless of the characteristics of the input point clouds and \revisemajor{the embedded spaces}. This holds true even for measurements attained from ellipsoids with significant axis ratios or elongation. 3) The \emph{robustness} against external disturbances like heavy noise and outliers that are incurred during the data acquisition process. 4) The \emph{generalization capability} of the first three properties to dealing with data points embedded in the \emph{multidimensional space} $\mathbb{R}^n$, such as spanning from $\mathbb{R}^{3}$ to $\mathbb{R}^{10}$. By integrating these merits, our approach addresses the limitations of previous methods and offers a comprehensive solution for robust and accurate ellipsoid-specific fitting.}

Concretely, we approach the task of fitting ellipsoidal surfaces in $\mathbb{R}^n$ by a Bayesian manner. {Unlike predecessor methods that make implicit assumptions, such as treating each measurement point as a noisy version of the closest point on the surface, we establish a more robust connection between them by considering each model point as a potential source for generating the measurement data through predictive distribution within the Bayesian framework.} We make a prior assumption that the parameters are uniformly sampled from an ellipsoidal domain to guarantee results ellipsoid-specific. This stands in contrast to previous methods that require explicit constraints and struggle to extend to higher dimensions. Our approach solves a more tractable unconstrained optimization problem that is independent of spatial dimensions, thereby facilitating straightforward generalization to multidimensional spaces. To address the issue of outliers, we further introduce a uniform distribution related to the volume of the input data points. This addition significantly enhances the robustness of our method against outliers. Then, we solve a \emph{maximum a posterior probability} (MAP) estimate of a certain ellipsoidal surface using Bayes' theorem, given the data. Benefiting from the uniform distribution prior, MAP coincides with a more manageable maximum likelihood estimate problem, which can be solved by the numerically stable Expectation Maximization (EM) framework. Additionally, we incorporate a \emph{vector $\epsilon$ technique} into the primary EM algorithm, leading to a substantial acceleration of the convergence process. Moreover, we investigate the relationship between our proposed method and previous least-squares-based frameworks. In doing so, we emphasize the superior robustness of our approach. Furthermore, we theoretically prove the \revisemajor{necessity} of the uniform component in designing robust algorithms, using the \emph{Kullback-Leibler Information Criterion} (KLIC)~\cite{kullback1951information} as a foundation. 

We conduct extensive experiments to validate the accuracy and robustness of our proposed approach in challenging settings consisting of heavy noise, outliers, and large axis ratios, besides, we demonstrate its generalization capability to \revisemajor{high-dimensional} spaces. Additionally, we use a variety of real-world tasks, such as geometric shape approximation, 3D reconstruction, and magnetometer calibration, to evidence its broad applicability across different practical domains. In comparison to baseline competitors, \revisemajor{experiments show that our algorithm is ellipsoid-specific in multidimensional spaces and regardless of input data characteristics, consistently achieves highly accurate fittings, and exhibits enhanced robustness against various disturbances.}

This journal paper is an extended and enhanced version of our recent conference paper~\cite{zhao_bmvc21}. We delve much deeper into our analysis from both \emph{theoretical} and \emph{experimental} aspects. Moreover, we make substantial additional improvements and refinements:   
\begin{itemize}
	\item We formalize the ellipsoid fitting problem as a systematic Bayesian parameter estimate process under the perspective of maximum a posterior probability given the data, from which we derive the ellipsoid-specificity and provide a more comprehensive theoretical foundation and rigorous convergence analysis of the optimization process. 
	\item We extend the former presented 3D ellipsoid fitting method to multidimensional spaces and explore its relationship with conventional least-squares-based frameworks. Also, we undertake a theoretical analysis and prove the necessity of uniform distributions in depressing outliers from the KLIC standpoint.
	\item We significantly enrich comparative studies to better elaborate the advantages and practical utility of the developed method. In particular, the detailed 2D ellipse fitting, the high-dimensional ellipsoid fitting, and versatile real-world applicable experiments are conducted.
\end{itemize}

\section{Preliminaries and Related Work}
\begin{definition}
		A general \textbf{quadric} $\mathcal{Q}\in \mathbb{R}^n$ is mathematically defined by the set of zeros of a second-order smooth implicit polynomial:
		\begin{eqnarray}
			\begin{aligned}
				\mathcal{Q}(\mathbf{a},\mathbf{p})=\mathbf{a}^T\mathbf{p}&
				=\sum_{i=1}^np_{ii}x_i^2+\sum_{i=1}^{n-1}\sum_{j=i+1}^{n}p_{i,j}x_ix_j\\
				&+\sum_{i=1}^np_ix_i+p_0=0,
			\end{aligned}
			\label{eq:quadric}
		\end{eqnarray}
		where $\mathbf{a}=[x_1^2, \cdots, x_n^2, x_1x_2, \cdots, x_{n-1}x_n, x_1, \cdots, x_n,  1]^T$ is the design vector built from the input observation $\mathbf{x}=(x_1, \cdots, x_n)^T\in \mathbb{R}^n$, and $\mathbf{p}=[p_{1,1},\cdots,p_{n,n},p_{1,2},\cdots,p_{n-1,n}, p_1, \cdots, p_n, p_0]^T$ is the coefficient vector that characterizes the shape of a quadratic surface. 
	\end{definition}
	Eq.~\ref{eq:quadric} \revisemajor{defines} a general quadric in $\mathbb{R}^n$, which \revisemajor{encompasses} various geometric shapes such as cylinders and paraboloids. To \revisemajor{obtain} ellipsoid-specific results, previous approaches often impose additional constraints on the expression $\mathcal{Q}$ to ensure that it represents an ellipsoid. However, \revisemajor{the majority of} ellipsoidal constraints are subject to \revisemajor{the following condition}.
	
\begin{proposition}[Axis-Ratio-Dependent Ellipsoid-Specificity~\cite{kesaniemi2017direct}\label{prop:1}]
		The sufficient and necessary condition of ellipsoid-specificity in multidimensional spaces $\mathbb{R}^n$ holds only when the axis ratio $r_{ax}$ between the longest and the shortest major axes of an ellipsoidal surface satisfies
		\begin{eqnarray}
			r_{ax}<\mathcal{G}(n)=\sqrt{\frac{2n-2}{n-2}},
			\label{eq:axis_ratio}
		\end{eqnarray}
		where $n$ is the spatial dimension. 
	\end{proposition}
	Proposition~\ref{prop:1} \revisemajor{establishes a connection} between the axis ratio metric and ellipsoid-specificity. \revisemajor{Specifically}, when Eq.~\ref{eq:axis_ratio} fails, indicating that the measured axis ratio $r_{ax}$ exceeds the allowable threshold value $\mathcal{G}(n)$ (\eg, highly elongated ellipsoids), the fitted results only \revisemajor{represent} a \emph{subset} of the entire ellipsoid family~\cite{li2004least}, leading to potentially significant {bias} in the estimates~\cite{kesaniemi2017direct}. We summarize the values of $\mathcal{G}(n)$ for different dimensions in Table~\ref{tab:axis_ratio}. It is noteworthy that even in the lower 3D space, many reference algorithms tend to produce inaccurate fittings when $r_{ax}>2$. Moreover, as the dimension $n$ increases, the acceptable $\mathcal{G}(n)$ decreases significantly. \revisemajor{Importantly}, in practical scenarios, there is often a lack of \emph{a priori} knowledge regarding potential ellipsoidal data, further increasing the likelihood of unreliable estimates.

\begin{definition}[Ellipsoid in $\mathbb{R}^n$]\label{def:ellipsoid}
		An ellipsoidal surface $\mathcal{E}\in \mathbb{R}^{n}$ in center form can be parameterized as
		\begin{eqnarray}
			\mathcal{E}=\{\mathbf{x}\in \mathbb{R}^{n}|(\mathbf{x}-\mathbf{c}_{\mathcal{E}})^T\mathbf{B}^{-1}(\mathbf{x}-\mathbf{c}_{\mathcal{E}})=1\},
		\end{eqnarray}
		where $\mathbf{c}_{\mathcal{E}}\in \mathbb{R}^{n}$ is the ellipsoidal center and $\mathbf{B}\in \mathbb{S}^{n}_{++}$ is a positive definite matrix describing the ellipsoid shape. A direct eigenvalue decomposition of $\mathbf{B}=\mathbf{Q}^T\mathbf{\Lambda}\mathbf{Q}$ gives the geometric parameters of $\mathcal{E}$, namely the semi-axis lengths and the rotation angles between the ellipsoidal axes and the orthogonal coordinate axes, where $\mathbf{Q}\in SO(n)${\footnote{$SO(n):=\{\mathbf{Q}\in \mathbb{R}^{n \times n}|\mathbf{Q}\mathbf{Q}^{T}=\mathbf{Q}^{T}\mathbf{Q}=\mathbf{I}_n, \det{\mathbf{Q}}=1\}$.}} is an orthogonal matrix and $\mathbf{\Lambda}\in\mathbf{R}^{n\times n}$ is a diagonal matrix.
	\end{definition}

	\noindent{\textbf{Problem \revise{Formulation}.}} \emph{Given a finite set of observation data points  $\mathbf{X}=\{\mathbf{x}_i\in \mathbb{R}^n\}_{i=1}^{N}$ that are sampled potentially from an ellipsoid, possibly with noise and outliers, our objective is to 
	fit an ellipsoidal surface $\mathcal{E}$ to $\mathbf{X}$. This fitting process aims to recover the optimal geometric shape parameters of the ellipsoid that best represents the underlying structure of the data.}

   Most existing algorithms can be \revisemajor{categorized} into two categories composed of \emph{algebraic fitting} and \emph{geometric fitting}.  
	 
	\begin{table}[t]
		\centering
		\vskip -0.3cm
		\caption{The tolerated limit value $\mathcal{G}(n)$ with respect to the  ellipsoid-specific axis ratio $r_{ax}$ on various spatial dimensions $n\in\mathbb{Z}_{+}$.}  
		\renewcommand{\arraystretch}{1.15} 
		\begin{tabular}{c|cccccc}
			\toprule
			$n$&2 &3 &4&10&$\cdots$&$+\infty$\cr
			\midrule
			{$\mathcal{G}(n)$} &+$\infty$&2.0&$\sqrt{3}$ &1.5&$\cdots$&$\sqrt{2}$\\
			\bottomrule
		\end{tabular}
		\label{tab:axis_ratio}
	\end{table}

	\subsection{Algebraic Fitting}
	To find the optimal coefficient vector $\mathbf{p}$ in Eq.~\ref{eq:quadric}, algebraic fitting methods directly minimize the deviation of $\mathcal{Q}(\mathbf{a},\mathbf{p})$ with respect to zeros (\ie, the \emph{algebraic distance or equation error}), as defined by
	\begin{eqnarray}
		\sum_{i=1}^{N}\mathcal{Q}^2(\mathbf{a}_i,\mathbf{p})=\sum_{i=1}^{N}(\mathbf{p}^T\mathbf{a}_i)^2=\mathbf{p}^T\mathbf{Q}\mathbf{p},
		\label{eq:algebraic}
	\end{eqnarray}where $\mathbf{a}_i=\mathbf{a}(\mathbf{x}_i)$ is the design vector corresponding to the $i^{th}$ point $\mathbf{x}_i\in \mathbb{R}^n$, and $\mathbf{Q}=\sum_{i=1}^{N}\mathbf{a}_i\mathbf{a}_i^T$ is the \emph{scatter matrix}. 
	In 2D space, Fitzgibbon \etal~\cite{fitzgibbon1999direct} incorporated the ellipticity  constraint into a normalization factor, using $4p_{11}p_{22}-p_{12}^2=1$ to derive the first ellipse specificity solution through a generalized eigensystem. A straightforward extension to 3D space for ellipsoid-specific fittings was presented in~\cite{li2004least}, subject to the following constraint
	\begin{eqnarray}
		4I_1-I_2^2>0,
		\label{eq:ellipsoid_constraint}
	\end{eqnarray}where $I_1=p_{11}p_{22}+p_{11}p_{33}+p_{22}p_{33}-\frac{1}{4}(p_{12}^2+p_{13}^2+p_{23}^2)$, $I_2=p_{11}+p_{22}+p_{33}$. However, the above constraint is just a sufficient condition but not \revisemajor{a} necessary \revisemajor{one}. Eq.~\ref{eq:ellipsoid_constraint} ensures the best fit only when $r_{ax}\leq2$. Once this hypothesis fails, a bisection searching strategy must be performed to \revisemajor{find} a suitable approximation. Consequently, it may struggle to accurately estimate the parameters of flattened ellipsoids. Recently, Kes{\"a}niemi \etal~\cite{kesaniemi2017direct} proposed a hyperellipsoid fitting method that built upon existing 2D ellipse and 3D ellipsoid frameworks. Despite ellipsoid-specificity in $\mathbb{R}^n$, this method still faces challenges related to axis ratios.
    In cases where $r_{ax}$ surpasses the thresholds outlined in Table~\ref{tab:axis_ratio}, it necessitates either rejection or \emph{regularization} operations.
    In addition to challenges posed by relatively large axis ratios, previous approaches put little attention on outlier-contaminated settings which is a common occurrence in practice. Additionally, some methods attempt to reduce the algebraic ellipsoid fitting problem as a \emph{semi-definite programming} (SDP) process~\cite{calafiore2002approximation,ying2012fast,lin2015fast}, \revisemajor{where} the ellipsoid-specificity constraint is enforced through the  semi-definiteness \revisemajor{of a matrix}. However, these methods are also vulnerable to outlier disturbances.

While we can resort to \emph{iteratively re-weighted least-squares} or \emph{M-estimators} (\eg, Tukey~\cite{rousseeuw1991tutorial} and Huber~\cite{huber2011robust}) to mitigate the impact of large residuals caused by outliers, or using recently \revisemajor{developed} algorithms~\cite{liang2015robust,hu2021robust,mulleti2015ellipse,thurnhofer2020ellipse,zhao2021robust} for robust fitting in $\mathbb{R}^2$ or $\mathbb{R}^3$, \revisemajor{directly applying these methods in higher-dimensional spaces is not straightforward}. Note that adapting these approaches to $\mathbb{R}^n$ not only requires significant efforts (if feasible) but also presents highly nonlinear optimization challenges. In contrast, our proposed method is dimension-independent and allows for successful multidimensional fittings. \revisemajor{Moreover}, we experimentally demonstrate that our method \revisemajor{achieves} superior robustness and stability than its predecessors.

\subsection{Geometric Fitting}
	Alternatively, geometric fitting or \emph{orthogonal regression} minimize the sum of squares of the Euclidean distance from data points $\mathbf{X}=\{\mathbf{x}_i\in \mathbb{R}^n\}_{i=1}^{N}$ to the target ellipsoid $\mathcal{Q}$~\cite{gander1994least,ahn2002orthogonal}
	\begin{eqnarray}
		dist(\mathbf{X}, \mathcal{Q})=\sum_{i=1}^{N}\|\mathbf{x}_i-\bar{\mathbf{x}}_i\|_2^2=\sum_{i=1}^{N}d_{i}^2,
		\label{eq:gf}
	\end{eqnarray}where $\bar{\mathbf{x}}_i$ is the point on $\mathcal{Q}$ that is closest to $\mathbf{x}_i$, and $\|\mathbf{x}-\bar{\mathbf{x}}_i\|_2$ denotes the Euclidean distance from $\mathbf{x}$ to $\bar{\mathbf{x}}_i$. Compared \revisemajor{to} algebraic fitting, geometric methods offer a more physically \revisemajor{interpretable framework} and provide more accurate parameter estimates. However, directly solving $dist(\mathbf{X}, \mathcal{Q})$ in Eq.~\ref{eq:gf} is computationally impractical, even in 2D and 3D spaces. For example, computing  $dist(\mathbf{X}, \mathcal{Q})$ in $\mathbb{R}^3$ requires solving a sixth-order polynomial equation \revisemajor{(as derived in Section 3 of the {\emph{Supplementary Material}})}. To circumvent this computational complexity, various approximations or amendments have been proposed. {For instance, Turner~\cite{turner1974computer} and Sampson~\cite{sampson1982fitting}  independently introduced a weighting factor based on the gradient of the algebraic distance to approximate the distance between a point and an ellipse. This approximation is later known as the \emph{Sampson distance}. In contrast, Agin~\cite{agin1981fitting} proposed an \emph{average gradient constraint} to establish a link between the equation error and the orthogonal distance.} Taubin~\cite{taubin1991estimation} \revisemajor{extended} the concept of Sampson distance to spatial curves using a first-order Taylor approximation. Other methods within this framework encompass the non-iterative variant~\cite{prasad2013ellifit} and the polar-n-direction distance~\cite{wu2019efficient}.
\revisemajor{These methods are primarily tested in 2D or 3D spaces, and their generalization to multidimensional scenarios is unknown}. Additionally, most baseline approaches lack a comprehensive strategy for handling outlier disturbances. It is crucial to emphasize the practical significance of algorithms capable of accurately fitting ellipsoidal surfaces in the presence of noise and outliers.

\subsection{Bayesian Fitting}
There also exist a few studies that leverage \emph{Bayesian probabilistic techniques} for fitting low-dimensional algebraic curves and surfaces. For example, Gull~\cite{gull1989bayesian} presented a Bayesian analysis for the straight-line fitting problem, Subrahmonia \etal\cite{subrahmonia1996practical} employed an asymptotic Bayesian approximation to model implicit polynomial functions for recognizing 2D and 3D objects, Werman~\etal~\cite{werman2001bayesian} showcased the unbiased and excellent properties of the Bayesian paradigm for fitting general curves, while Beale \etal~\cite{beale2016fitting} enforced a matrix normal prior to regularize the fitting of quadrics. {Recently, Liu~\etal~\cite{liu2022robust} introduced a probabilistic method for accurately recovering superquadrics from point clouds. Benefiting from the developed Gaussian-uniform mixture model and the Expectation, Maximization, and Switching optimization framework, their method achieves remarkable performance in a series of tests. We present a more detailed discussion and comparison between superquadrics and ellipsoids in Section 10 of the \emph{Supplementary Material}.} While there have been pioneering and insightful explorations into Bayesian or probabilistic fitting, most of these approaches have primarily been applied to 2D or 3D cases. As far as we known, none of these methods explicitly \revisemajor{addresses} the multidimensional ellipsoid fitting problem within the framework of Bayesian parameter estimate.

\section{Methodology} \label{sec:method}
	\revisemajor{In this section, we outline specific steps to show how ellipsoid-specific fittings can be achieved in $\mathbb{R}^n$ through the optimization of the Bayesian parameter estimation process.}

\revisemajor{\subsection{Overview}}
\revisemajor{Given a set of input data, we employ Bayesian posterior inference to constrain the fitting result to be ellipsoid-specific, followed by the definition of the predictive distribution (Section~\ref{sec:inference}), from which we consider each model point as a potential source for the generation of each measurement sample.} {Then, in Section~\ref{sec:bayesian_posterior}, we approximate the predictive distribution using the Monte Carlo method and transform the Bayesian posterior optimization into a maximum likelihood estimation problem. In Section~\ref{sec:ellipsoid}, we provide a concrete parameterization of the ellipsoidal surface and discuss the EM parameter optimization in Section~\ref{sec:EM}. To accelerate the convergence of the EM algorithm, we further introduce a vector $\varepsilon$ technique in Section~\ref{sec:acceleration}. Additionally, we develop an initialization scheme in Section~\ref{sec:initialization} to alleviate the need for manual selection of hyperparameters. To reveal the robustness of our algorithm, we further perform a robust analysis between our method and previous least-squares fitting paradigms in Section~\ref{sec:robustness}.}

\subsection{Bayesian Inference of Ellipsoidal Parameters}\label{sec:inference}
Suppose the measurement points $\mathbf{X}=\{\mathbf{x}_i\in \mathbb{R}^n\}_{i=1}^{N}$ are sampled from an ellipsoidal surface  $\mathcal{E}(\bm{\theta})$ embedded in $\mathbb{R}^n$, where $\bm{\theta}$ represents a set of geometric parameters belonging to our interest domain referred to as the  \emph{ellipsoidal domain}, denoted as $\bm{\Theta}$. In our context, $\bm{\Theta}$ defines a product space that allows us to preset a relatively large scope to constrain the geometric parameters, determining the shape and orientation of the ellipsoid within the space \revisemajor{of} $\mathbb{R}^n$. Note that $\bm{\theta}$ can be readily recovered from the associated shape matrix $\mathbf{B}$. For example, in a 3D Euclidean space, $\bm{\theta}$ can be expressed as $\bm{\theta} = (x_0, y_0, z_0, a, b, c, \alpha, \beta, \gamma)$. Here, $(x_0, y_0, z_0)$ represents the center of the ellipsoid, $(a, b, c)$ denote the lengths of its three semi-axes, and $(\alpha, \beta, \gamma)$ correspond to the Euler angles, defining the orientation of the ellipsoid's major axis relative to the $x$, $y$, and $z$ axes. In the presence of noisy interference, we assume that each measurement point can be expressed as $\mathbf{x}_i=\mathbf{y}_i+\zeta_i, i=1, \cdots, N$, here $\mathbf{y}_i\in \mathcal{E}(\bm{\theta})$ represents the possible ground truth ellipsoidal point \revisemajor{and} $\xi_i\sim \revisemajor{\mathcal{N}}(\zeta|0,\mathbf{\Sigma}_i)$ accounts for the additive Gaussian noise sampled from the distribution with zero mean and covariance matrix $\mathbf{\Sigma}_i\in \mathbb{R}^{n\times n}$.

\revisemajor{If we {consider} $\bm{\theta}$ as a random variable, our goal is to maximize the probability of an ellipsoid instance with parameter $\bm{\theta}$, given the measurement data $\mathbf{X}$.}
From the \emph{Bayesian posterior inference} or Bayes' theorem, we have 
\begin{eqnarray*}
		p(\mathcal{E}(\bm{\theta})|\mathbf{X})\!=\!\frac{p(\mathbf{X}|\mathcal{E}(\bm{\theta}))p(\mathcal{E}(\bm{\theta}))}{p(\mathbf{X})}\!=\!\frac{p(\mathbf{X}|\mathcal{E}(\bm{\theta}))p(\mathcal{E}(\bm{\theta}))}{\int_{\bm{\theta\in \Theta}} p(\mathbf{X}|\mathcal{E}(\bm{\theta}))p(\mathcal{E}(\bm{\theta}))d\bm{\theta}},
\end{eqnarray*}where $p(\mathbf{X}|\mathcal{E}(\bm{\theta}))$ is the likelihood function, $p(\mathcal{E}(\bm{\theta}))$ \revisemajor{represents the prior distribution over the ellipsoidal domain $\bm{\Theta}$}, and the normalization factor $p(\mathbf{X})$ is the {marginal likelihood} that describes the distribution of  measurements $\mathbf{X}=\{\mathbf{x}_i\in \mathbb{R}^n\}_{i=1}^{N}$ \revisemajor{and} is independent of $\bm{\theta}$.

In turn, the \emph{predictive distribution} that generates different observations $\mathbf{x}_i$ based on the target ellipsoid $\mathcal{E}(\bm{\theta})$ is
\begin{equation}
			p(\mathbf{x}_i|\mathcal{E}(\bm{\theta}))\!=\!\int_{\mathbf{y}\in \mathcal{E}(\bm{\theta})}p(\mathbf{x}_i|\mathbf{y})p(\mathbf{y}|\mathcal{E}(\bm{\theta}))d\mathbf{y}\!=\!\mathbb{E}_{p(\mathbf{y}|\mathcal{E}(\bm{\theta}))}[p(\mathbf{x}_i|\mathbf{y})],
			\label{eq:integral_expec}
\end{equation}\revisemajor{where} $\mathbb{E}$ represents the \emph{expectation operator}. {It is important to emphasize that Eq.~\ref{eq:integral_expec} establishes a connection between individual data points on the ellipsoidal surface $\mathcal{E}(\bm{\theta})$ and each observation. Therefore, we consider each model point as a potential source for generating each observation. This significantly differs from predecessor algebraic and geometric approaches, which solely focus on the nearest model point or \revisemajor{the shortest Euclidean distance}, and thus are more \revisemajor{susceptible} to bias when dealing with data contaminated by heavy noise.}

\subsection{Bayesian Posterior Optimization}\label{sec:bayesian_posterior}
For simplicity, we adopt \emph{non-parametric Monte Carlo} scheme to approximate the expectation value in Eq.~\ref{eq:integral_expec}, namely, 
\begin{eqnarray}
		p(\mathbf{x}_i|\mathcal{E}(\bm{\theta}))\revisemajor{\approx}\frac{1}{M}\sum_{j=1}^Mp(\mathbf{x}_i|\mathbf{y}_j(\bm{\theta})),
		\label{eq:sum}
\end{eqnarray}where $M$ is the number of ellipsoidal points sampled from $\mathcal{E}(\bm{\theta})$. 
Crucially, to account for the impact of outliers, like~\cite{wells1997statistical}, we introduce an augmentation of a uniform distribution to Eq.~\ref{eq:sum}, then the updated equation takes the form of 
\begin{eqnarray}
p(\mathbf{x}_i|\mathcal{E}(\bm{\theta}))=(1-w)\frac{1}{M}\sum_{j=1}^Mp(\mathbf{x}_i|\mathbf{y}_j(\bm{\theta}))+w\frac{1}{V},
\label{eq:sum_uniform}
\end{eqnarray}here $V$ denotes the volume of the bounding box in terms of the input dataset $\mathbf{X}$, $w\in[0, 1]$ is a weight coefficient that balances the two types of distributions.  Assuming a uniform distribution prior over the parameter $\bm{\theta}\in\bm{\Theta}$ (with the range of parameter values previously restricted to ensure the boundedness of the probability sample space in reality), we \revisemajor{obtain} 
\begin{eqnarray}
p(\mathcal{E}(\bm{\theta})|\mathbf{X})\!=\!\frac{p(\mathbf{X}|\mathcal{E}(\bm{\theta}))p(\mathcal{E}(\bm{\theta}))}{p(\mathbf{X})}\propto p(\mathbf{X}|\mathcal{E}(\bm{\theta})).
\end{eqnarray}
Lastly, the Bayesian optimization of the posterior parameter $\bm{\theta}$, aiming to find the \emph{Maximum a Posteriori} (MAP) estimate, can be formulated as
\begin{eqnarray}
		\begin{aligned}
			\hat{\bm{{\theta}}}&=\max_{\bm{\theta}} p(\mathcal{E}(\bm{\theta})|\mathbf{X})=\max_{\bm{\theta}}
			p(\mathbf{X}|\mathcal{E}(\bm{\theta}))\\
			&=\max_{\bm{\theta}} {\Pi_{\mathbf{x}_i \in \mathbf{X}}} p(\mathbf{x}_i|\mathcal{E}(\bm{\theta}))\\
			&=\max_{\bm{\theta}} {\Pi_{\mathbf{x}_i \in \mathbf{X}}}((1-w)\frac{1}{M}\sum_{j=1}^Mp(\mathbf{x}_i|\mathbf{y}_j(\bm{\theta}))+w\frac{1}{V}),	
		\end{aligned}
\label{eq:MAP}
\end{eqnarray}here we make an independent and identically distributed (i.i.d) assumption to simplify the computational complexity. It's not hard to see that the above formulae is equivalent to a \emph{maximum likelihood estimation} problem. 
\revisemajor{This equivalence arises from the underlying assumption of a uniform distribution as the prior belief for $\bm{\theta}$}. Consequently, our objective is to optimize the subsequent \emph{negative log-likelihood function} concerning the observations
	\begin{eqnarray}
		\hat{\bm{\theta}}
		=\min_{\bm{\theta}}-\sum_{i=1}^{N}\log((1-w)\frac{1}{M}\sum_{j=1}^Mp(\mathbf{x}_i|\mathbf{y}_j(\bm{\theta}))+w\frac{1}{V}).
		\label{eq:objective}
	\end{eqnarray} 

\subsection{Parameterization of the Ellipsoidal Surface}\label{sec:ellipsoid}
	We parameterize $\mathcal{E}(\bm{\theta})$ as a \emph{unit $(n-1)$-sphere (hypersphere)} $\mathcal{S}^{(n-1)}\in\mathbb{R}^n$ to initialize the sample points  $\mathbf{Y}=\{\mathbf{y}_j(\bm{\theta})\in\mathbb{R}^n\}_{j=1}^M$ in Eq.~\ref{eq:sum}. The sample points on  $\mathcal{S}^{(n-1)}=\{\mathbf{y}_j=(\mathbf{y}_{j,1},\mathbf{y}_{j,2}, \cdots, \mathbf{y}_{j,n})\in\mathbb{R}^{n}|\|\mathbf{y}_j\|_2=1\}$ are realized by 
\begin{eqnarray}
		\left\{
		\begin{aligned}
			&\mathbf{y}_{j,1}=\cos(\psi_1),\\
			&\mathbf{y}_{j,2}=\sin(\psi_1)\cos(\psi_2),\\
			&\mathbf{y}_{j,3}=\sin(\psi_1)\sin(\psi_2)\cos(\psi_3),\\
			&\quad \vdots\\
			&\mathbf{y}_{j,n-1}=\sin(\psi_1)\cdots\sin(\psi_{n-2})\cos(\psi_{n-1}),\\
			&\mathbf{y}_{j,n}=\sin(\psi_1)\cdots\sin(\psi_{n-2})\sin(\psi_{n-1}),\\
		\end{aligned}
		\right.
		\label{eq:parameterization}
\end{eqnarray}where $\psi_1, \psi_2, \cdots, \psi_{n-2}\in[0, \pi]$, $\psi_{n-1}\in[0,2\pi)$ are the angular coordinates. {According to definition \ref{def:ellipsoid}}, $\mathcal{S}^{(n-1)}$ with a center position $\mathbf{c}_\mathcal{S}$ can be described as $(\mathbf{y}-\mathbf{c}_\mathcal{S})^T(\mathbf{y}-\mathbf{c}_\mathcal{S})=1$. Therefore, the relationship between the observations and $\mathcal{S}^{(n-1)}$ is 
\begin{equation}
		[\mathbf{x}-({\mathbf{t}}+{\mathbf{A}}\mathbf{c}_{S})]^T({\mathbf{A}}{\mathbf{A}}^T)^{-1}[\mathbf{x}-({\mathbf{t}}+{\mathbf{A}}\mathbf{c}_{S})]=1,
	\end{equation}
where $\mathbf{A}\in \mathbb{R}^{n\times n}$ and $\mathbf{t}\in\mathbb{R}^n$ are the transformation parameters that relate the unit $(n-1)$-sphere with a general ellipsoidal surface. This connection leads to the derivation of the shape matrix $\mathbf{B}={\mathbf{A}}{\mathbf{A}}^T$ and the center position ${\mathbf{c}}_\mathcal{E}={\mathbf{t}}+{\mathbf{A}}\mathbf{c}_{\mathcal{S}}$. To facilitate a more convenient reconfiguration of geometric parameters, we re-parameterize $\mathcal{E}$ by $\bm{\theta}=(\mathbf{A},\mathbf{t})$. This choice allows for a straightforward conversion of geometric parameters from the $(\mathbf{A},\mathbf{t})$ representation. For example, in 3D Euclidean space, the nine geometric parameters of an ellipsoidal surface are expressed as
	\begin{equation*}
		\centering
		\begin{aligned}
			&{\mathbf{c}}_\mathcal{E}={\mathbf{t}}+{\mathbf{A}}\mathbf{c}_{\mathcal{S}},\ {a}=\sqrt{\lambda_1},\ {b}=\sqrt{\lambda_2},\ 
			{c}=\sqrt{\lambda_3},\\
			&{\alpha}=\atan2{\frac{-\mathbf{Q}_{31}}{\sqrt{(\mathbf{Q}_{11}+\mathbf{Q}_{21})^2}}},\ {\beta}=\atan2{\frac{\mathbf{Q}_{21}}{\mathbf{Q}_{11}}}, \ {\gamma}=\atan2{\frac{\mathbf{Q}_{32}}{\mathbf{Q}_{33}}},
		\end{aligned}
	\end{equation*}
where $\lambda_1$, $\lambda_2$, $\lambda_3\in \mathbb{R}$, and $\mathbf{Q}_{3 \times 3}\in SO(3)$ are the eigenvalues and the orthogonal matrix \revisemajor{obtained} via the eigen-decomposition of ${\mathbf{B}}={\mathbf{A}}{\mathbf{A}}^T$ in \revisemajor{definition}~\ref{def:ellipsoid}. Moreover, if we take an isotropic noise intensity denoted as $\mathbf{\Sigma}_i=\sigma^2\mathbf{I}$, where $\mathbf{I}\in \mathbb{R}^{n\times n}$ is an identity matrix, and update the trade-off weight $w$ automatically, then the expression in Eq.~\ref{eq:objective} transforms to
\begin{small}
	\begin{equation}
		\begin{aligned}
			{\bm{\Omega}}^*&=\min_{\bm{\Omega}}\ell(\bm{\Omega})=-\sum_{i=1}^{N}\log((1-w)\frac{1}{M}\sum_{j=1}^Mp(\mathbf{x}_i|\mathbf{y}_j(\bm{\theta}))+w\frac{1}{V})\\
			&=-\sum_{i=1}^{N}\log (\frac{1-w}{M}\sum_{j=1}^{M}\frac{1}{(2\pi\sigma^2)^{n/2}}\exp({\frac{\|\mathbf{x}_i-\mathbf{y}_m(\bm{\theta})\|^2}{-2\sigma^2}})+\frac{w}{V}),
		\end{aligned}
		\label{eq:final}
	\end{equation}
\end{small}where ${\bm{\Omega}}=(\bm{\theta},\sigma^2,w)$ are unknown variables.
	
\subsection{Optimization by the EM Framework}\label{sec:EM} 
\revisemajor{To solve for the target parameter ${\bm{\Omega}}^{*}$ in Eq.~\ref{eq:final}}, we employ the EM framework\cite{moon1996expectation}. First, we compute \revisemajor{the} posterior probability using Bayes' theorem, given the parameter $\bm{\Omega}^{t-1}$ from the previous iteration, which is termed as the expectation step (\emph{E-step}). Then, we update the new parameter ${\bm{\Omega}}^{t}$ by minimizing the expectation of the complete data negative log-likelihood function, \ie, the maximization step (\emph{M-step}). These two steps are iterated until EM converges.

\subsubsection{E-Step}
According to Jensen's inequality, we have
\begin{equation}
		\begin{aligned}
			\ell(\bm{\Omega})&\leq \sum_{i=1}^{N}\mathbb{E}_{p(\mathbf{y}_j(\bm{\theta}))}[-\log p(\mathbf{y}_j(\bm{\theta}), \mathbf{x}_i|\bm{\Omega})|\mathbf{x}_i,\bm{\Omega}^{t-1}]\\
			&=-\sum_{i=1}^{N}\sum_{j=1}^{M+1}p(\mathbf{y}_j(\bm{\theta})|\mathbf{x}_i,\bm{\Omega}^{t-1})\log p(\mathbf{y}_j(\bm{\theta}), \mathbf{x}_i|\bm{\Omega})\\
			&=\frac{\exp(-s)}{2}\sum_{i=1}^{N}\sum_{j=1}^{M}p(\mathbf{y}_j(\bm{\theta})|\mathbf{x}_i,\bm{\Omega}^{t-1})\|\mathbf{x}_i-\mathbf{y}_j(\bm{\theta})\|^2\\
			&+\frac{N_pn}{2}s-\log(w)N_o-\log (1-w)N_p\triangleq Q(\bm{\Omega},\bm{\Omega}^{t-1}),
		\end{aligned}
		\label{eq:jenson}
	\end{equation}where 
	\begin{eqnarray*}
		N_o\!=\!\sum_{i=1}^{N}p(\mathbf{y}_{M+1}|\mathbf{x}_i,\bm{\Omega}^{t-1}),
		N_p\!=\!\sum_{i=1}^{N}\sum_{j=1}^{M}p(\mathbf{y}_{j}(\bm{\theta})|\mathbf{x}_i,\bm{\Omega}^{t-1}).
	\end{eqnarray*}$p(\mathbf{y}_{M+1}|\mathbf{x}_i,\bm{\Omega}^{t-1})$ indicates the posterior probability that $\mathbf{x}_i$ is sampled from the uniform distribution component $\frac{1}{V}$. We have ignored terms that are independent of $\bm{\Omega}$ in Eq.~\ref{eq:jenson}. Like~\cite{kendall2017uncertainties}, in practice, we use $s:=\log\sigma^2$ for a  numerically stable computation, since it avoids the potential division by zero. As a result, the posterior probability via Bayes' theorem is formulated as 
	\begin{eqnarray*}
		\begin{aligned}
			&p({\mathbf{y}_{M+1}}|\mathbf{x}_i, \bm{\Omega}^{t-1})
			\!=\!\frac{1}{1+\frac{V}{M}\frac{1-w}{w}\sum_{k=1}^{M}f(\mathbf{x}_i, \mathbf{y}_k(\bm{\theta}))},\\
			&p(\mathbf{y}_m|\mathbf{x}_i, \bm{\Omega}^{t-1})
			\!=\!\frac{f(\mathbf{x}_i, \mathbf{y}_k(\bm{\theta}))}{\sum_{k\!=\!1}^{M}f(\mathbf{x}_i, \mathbf{y}_k(\bm{\theta}))\!+\!(2\pi\exp(s))^{n/2}\frac{w}{1\!-\!w}\frac{M}{V}},
		\end{aligned}
	\end{eqnarray*}
	where 
	\begin{eqnarray*}
		f(\mathbf{x}_i, \mathbf{y}_k(\bm{\theta}))=\exp({\frac{-\|\mathbf{x}_i-\mathbf{y}_k(\bm{\theta})\|^2}{2}\exp(-s)})
	\end{eqnarray*} represents the exponential of the square of the isotropic \emph{Mahalanobious distance} between $\mathbf{x}_i$ and $\mathbf{y}_k(\bm{\theta})$.
	
\subsubsection{M-Step}
In M-step, instead of directly minimizing the loss function $\ell(\bm{\Omega})$, we take its upper bound $Q(\bm{\Omega},\bm{\Omega}^{t-1})$ as the objective function. By solving $\nabla Q(\bm{\Omega},\bm{\Omega}^{t-1})=0$ with respect to each parameter, their closed-form solutions are summarized as  $\mathbf{t}=\frac{1}{N_p}(\mathbf{X}^{T}\mathbf{P}^T\mathbf{1}-\mathbf{A}\mathbf{Y}^{T}\mathbf{P}\mathbf{1})$, $w=\frac{N_o}{N_p+N_o},$ $s=\log(\frac{1}{N_pn}\tr(\mathbf{\widebar{X}^T}(\diag(\mathbf{P}^T\mathbf{1})\mathbf{\widebar{X}})-\mathbf{P}^T\widebar{\mathbf{Y}}\mathbf{A}^T))$, $\mathbf{A}=(\mathbf{\widebar{X}}^T\mathbf{P}^T\widebar{\mathbf{Y}})(\widebar{\mathbf{Y}}^T\diag(\mathbf{P}\mathbf{1})\widebar{\mathbf{Y}})^{-1},$ here $\mathbf{X}=[\mathbf{x}_1, \cdots, \mathbf{x}_N]^T$ and $\mathbf{Y}=[\mathbf{y}_1, \cdots, \mathbf{y}_M]^T$,  $\mathbf{P}\in\mathbb{R}^{M\times N}$ with the element  $p_{mn}=p(\mathbf{y}_m|\mathbf{x}_i, \bm{\Omega}^{t-1})$, $\widebar{\mathbf{X}}=\mathbf{X}-\frac{1}{N_p}\mathbf{O}\mathbf{P}\mathbf{X}$, $\widebar{\mathbf{Y}}=\mathbf{Y}-\frac{1}{N_p}\mathbf{O}\mathbf{P}^T\mathbf{Y}$, $\mathbf{O}=\mathbf{1}\mathbf{1}^T$ is a matrix with all ones, $\diag(\mathbf{a})$ is a diagonal matrix formed by the vector $\mathbf{a}$, and $\tr(\cdot)$ represents the trace operator of an matrix.

\subsubsection{Convergence Analysis}
\begin{theorem}[Convergence Assurance]\label{theorem:convergence}
The proposed objective function $\ell(\bm{\Omega})$ in Eq.~\ref{eq:final} converges to $\ell(\bm{\Omega}^{*})$, where $\bm{\Omega}^{*}$ is the desirable solution.
\end{theorem}
\revisemajor{Please refer to Section 2 of the \emph{Supplementary Material} for proof.}

\subsection{Acceleration of the EM Algorithm}\label{sec:acceleration}
Next, we introduce a \emph{vector $\varepsilon$ technique} \cite{wynn1962acceleration,kuroda2006accelerating} \revisemajor{into} our framework to accelerate the convergence behavior of the primary EM algorithm. To this end, we expand the unknown variables in $\bm{\Omega}$ as a vector \revisemajor{in $\mathbb{R}^n$} with \revisemajor{a} length equivalent to $n^2+n+2$. \revisemajor{We still denote this vector as $\bm{\Omega}$} for simplicity. Then, we update the $\varepsilon$-accelerated vector sequence $\{\dot{\boldmath{\bm{\Omega}}}^{(n)}\}_{n\geq 0}$ as follows.
	
\begin{definition}
		The $\varepsilon$-accelerated EM sequence $\{\dot{\boldmath{\bm{\Omega}}}^{(n)}\}_{n\geq 0}$ is  
		\begin{small}
			\begin{eqnarray*}
				\dot{\boldmath{\bm{\Omega}}}^{(n)}=\boldmath{\bm{\Omega}}^{(n+1)}\!+\!((\boldmath{\bm{\Omega}}^{(n+2)}\!-\!\boldmath{\bm{\Omega}}^{(n+1)})^{-1}\!-\!(\boldmath{\bm{\Omega}}^{(n+1)}\!-\!\boldmath{\bm{\Omega}}^{(n)})^{-1})^{-1},
			\end{eqnarray*}
		\end{small}where $[\mathbf{v}]^{-1}$ is the inverse of a vector $\mathbf{v}$ defined as $[\mathbf{v}]^{-1}=\mathbf{v}/\|\mathbf{v}\|^2$. The above steps proceed by iterations until
		\begin{eqnarray}
			{||\dot{\boldmath{\Omega}}^{(n+1)}-\dot{\boldmath{\Omega}}^{(n)}||}^{2}\leq\tau,
			\label{eq:convergence}
		\end{eqnarray}
		where $\tau=10^{-8}$ is the default convergence accuracy.
\end{definition}
	In particular, the $\varepsilon$-accelerated EM algorithm \revisemajor{maintains convergence consensus, meaning that it converges to the same optimal value as the primary EM algorithm.} 
\begin{theorem}[Convergence Consensus of the $\varepsilon$-Accelerated EM~\cite{wang2008acceleration}]
		Let the sequences $\{{\boldmath{\bm{\Omega}}}^{(n)}\}_{n\geq 0}$ generated by the primary EM converge to a stationary point $\bm{\Omega}^{*}$, then the sequences $\{\dot{\boldmath{\bm{\Omega}}}^{(n)}\}_{n\geq 0}$ generated by the $\varepsilon$-accelerated EM manner also converges to $\bm{\Omega}^{*}$.
\end{theorem}
\revisemajor{Please refer to \cite{wang2008acceleration} for proof.} 

\subsection{Initialization of $M$ and $w$}\label{sec:initialization}
To alleviate the necessity for arbitrary or manual adjustments of parameters such as the sample size $M$ from ellipsoidal surfaces and the weight $w$ in Eq.~\ref{eq:final}, we \revisemajor{propose a fully} automated initialization approach. This approach utilizes the concept of the \emph{relative density-based outlier score} (RDOS)~\cite{tang2017local} to \revisemajor{assess} the \revisemajor{outsiderness degree} for each measurement in $\mathbf{X}=\{\mathbf{x}_i\in \mathbb{R}^n\}_{i=1}^{N}$. \revisemajor{By analyzing the RDOS, we can estimate the size of outliers and obtain a meaningful initialization for both $M$ and $w$ in a single endeavor.} 

\begin{figure}[t]
		\centering
		\subfigure[$M=85, w=0.585$]{
			\begin{minipage}[b]{0.43\linewidth}
				\centering
				\includegraphics[width=\linewidth]{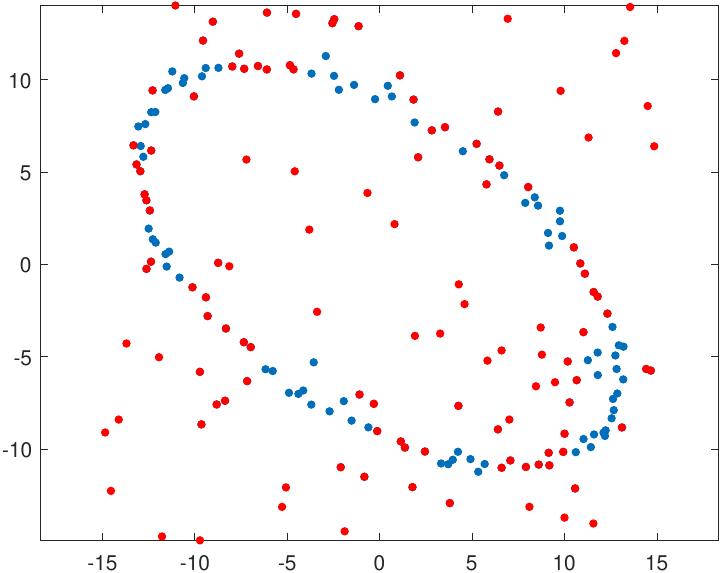}
			\end{minipage}
		}
		\subfigure[$M=918, w=0.282$]{
			\begin{minipage}[b]{0.34\linewidth}
				\centering
				\includegraphics[width=\linewidth]{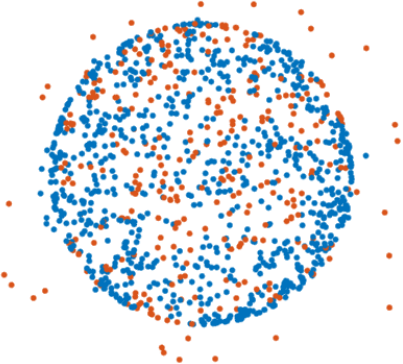}
			\end{minipage}
		}
		\caption{\revise{Visualization of using RDOS to initialize $M$ and $w$ simultaneously, where red points indicate the inferred outliers.}}
		\label{fig:rdos_vis}
		\vskip -0.3cm
	\end{figure}
	
	The asymptotic consistency of RDOS~\cite{tang2017local} gives a \emph{lower bound} for the inference of outlier points. We initialize $M=\sum_{\mathbf{x}_i\in\mathbf{X}}\mathbbm{1}(\RDOS(\mathbf{x}_i)\leq 1)$, where $\mathbbm{1}$ is an indicator function. Meanwhile, $w$ is initialized as $w={\sum_{\mathbf{x}_i\in\mathbf{X}}\mathbbm{1}(\RDOS(\mathbf{x}_i)> 2)}/{N}$. \revise{Fig.~\ref{fig:rdos_vis} presents two examples to visualize the initialization effect via RDOS. In these two panels, the ground truth values for inlier size and outlier ratio are set to} {$100$ (inliers), $0.5$, and $1,000$ (inliers), $0.2$, respectively. The estimated values for $M$ and $w$ are displayed beneath each panel, utilizing the same neighbor size $k=11$ (in default). As observed, the results suggest reasonable initialization for both parameters.} Additionally, since the target ellipsoidal surface is initially parameterized as $\mathcal{S}^n$, we utilize  $\mathbf{A}=\mathbf{I}\in\mathbb{R}^{n\times n}, \mathbf{t}=\mathbf{0}\in\mathbb{R}^{n}$, and
	$s=\log(\frac{1}{nNM}\sum_{i, m}^{N, M}\|\mathbf{x}_i-\mathbf{y}_m(\bm{\theta})\|^2)$
	to initialize $\mathbf{\Omega}$ within the EM algorithm. \revisemajor{We present a pseudo-code of the proposed method in Section 4 of the  \emph{Supplementary Material} for better understanding}. \revise{Further elaboration on the initialization of $M$ and $w$, along with additional details about RDOS, can be also found in Section 8 of the \emph{Supplementary Material}.}

\subsection{Robustness Analysis}\label{sec:robustness}	
In this section, we investigate the correlation between our method and previous least-squares fitting paradigms. We also emphasize the significance and rationale behind the \revisemajor{inclusion} of the uniform distribution within our loss function, which ensures robust fittings.

\subsubsection{Relationship to Least-Squares Fitting}	
Assuming $w=0$ and disregarding variables irrelevant to $\bm{\theta}$, similar to Eq.~\ref{eq:jenson}, the upper bound of Eq.~\ref{eq:final} reduces to 
\begin{equation}
-\sum_{i=1,j=1}^{N,M}\exp({\frac{-\|\mathbf{x}_i-\mathbf{y}_j(\bm{\theta})\|^2}{2}}\exp(-s)),
\label{eq:ls},
\end{equation} which is equivalent to optimizing a weighted least-squares problem with the weight coefficient $\exp(-s)$. If we proceed to treat $s$ as a constant, Eq.~\ref{eq:ls} degenerates to an ordinary least-squares problem. Therefore, the previous \revisemajor{established} least-squares fitting framework can be viewed as a special instance of our newly developed algorithm. 
	

\begin{figure}[t]
	\centering
	\includegraphics[width=\linewidth]{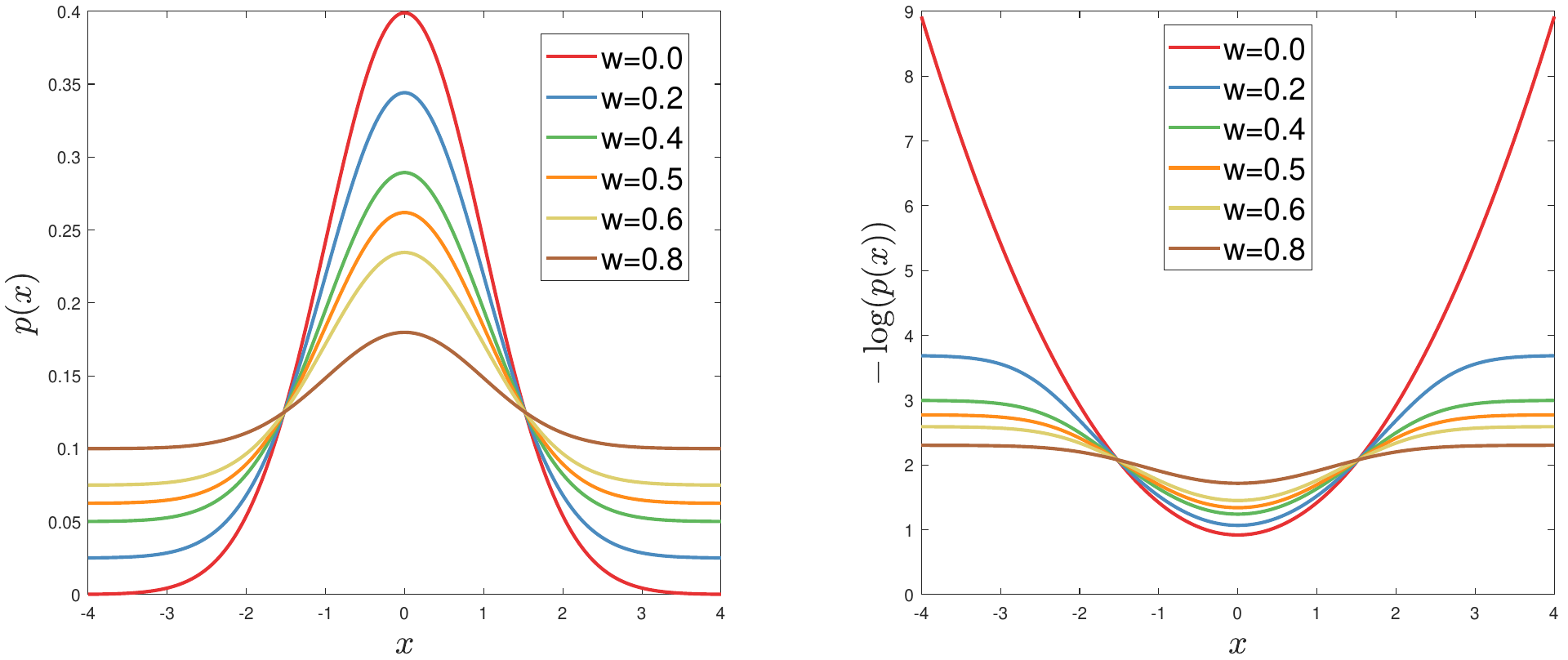}
	\vskip -0.3cm
	\caption{Left: Comparison of the probability $p(x)$ with ($w\neq0.0$) or without ($w=0.0$) the uniform distribution in 1D space. Right: The corresponding negative log-likelihood function of $p(x)$, where bounded influence functions are ensured by our method when $w\neq0.0$. }
	\label{fig:uniform}
	\vskip -0.3cm
\end{figure}

\begin{figure*}[t]
	\centering
\includegraphics[width=\textwidth]{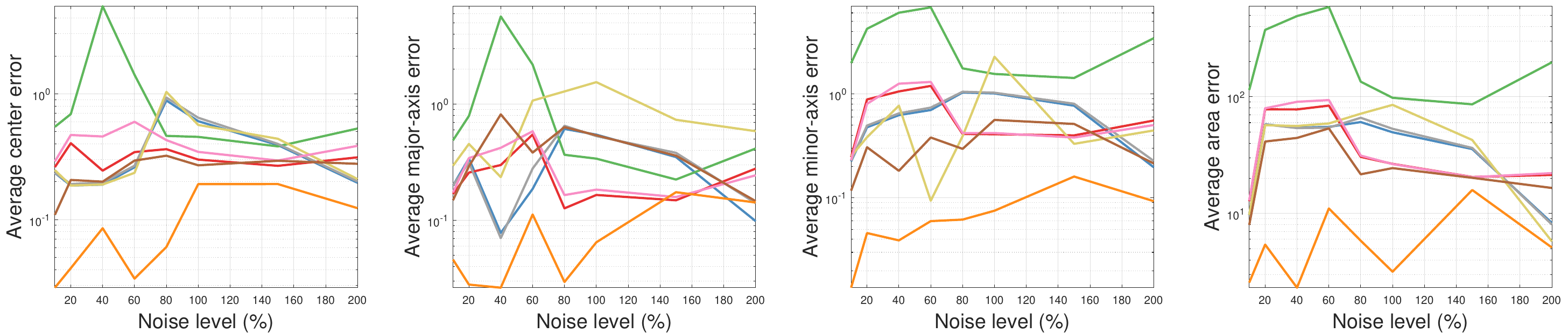}
	\centering
		\begin{tikzpicture}[align=center]
			\draw[line width=1pt,color=LSA] (-0.3,0)--(0.4,0) node[right](line){\color{black}{LSA}};
			\draw[line width=1pt,color=LSO] (1.4,0)--(2.1,0) node[right](line){\color{black}{LSO}};
			\draw[line width=1pt,color=WLS] (3.1,0)--(3.8,0) node[right](line){\color{black}{WLS}};
			\draw[line width=1pt,color=Taubin] (4.9,0)--(5.6,0) node[right](line){\color{black}{Taubin}};
			\draw[line width=1pt,color=Prasad] (6.9,0)--(7.6,0) node[right](line){\color{black}{Prasad}};
			\draw[line width=1pt,color=Wu] (8.9,0)--(9.6,0) node[right](line){\color{black}{Wu}};
			\draw[line width=1pt,color=SAREfit] (10.4,0)--(11.1,0) node[right](line){\color{black}{SAREfit}};
			\draw[line width=1pt,color=BayFit] (12.6,0)--(13.3,0) node[right](line){\color{black}{BayFit (Ours)}};
		\end{tikzpicture}
		\vskip -0.2cm
	\caption{Quantitative evaluations on noisy measurements with respect to 2D ellipse fitting,  \revise{utilizing the log-scale $y$ axis to enhance the readability of the statistical results.} We gradually increase the noise level from $10\%$ to $200\%$ and report the MSE of each compared method based on 100 tests. Remarkably, BayFit demonstrates superior fitting precision across all noise levels.   	
	}
	\label{fig:ellipe_noie}
	\vskip -0.3cm
\end{figure*}
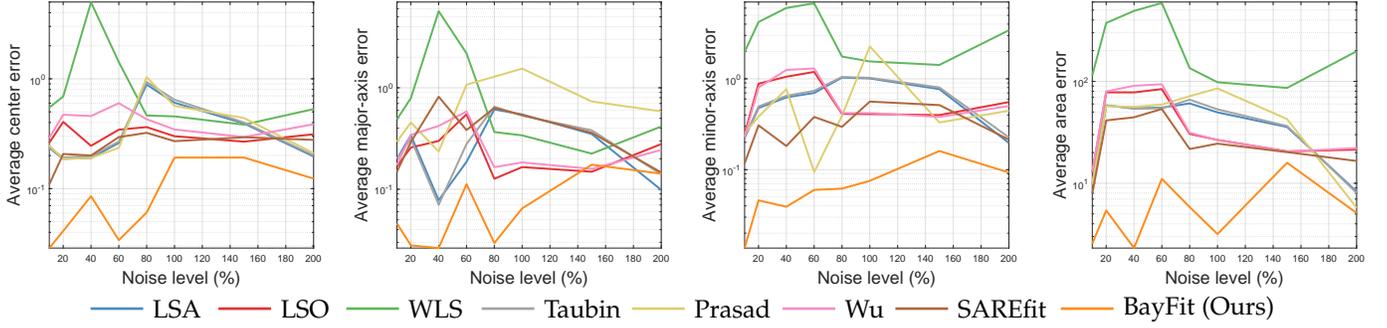

\begin{figure*}[t]
\centering
\begin{minipage}{0.49\linewidth}
\centering
\includegraphics[width=0.49\textwidth]{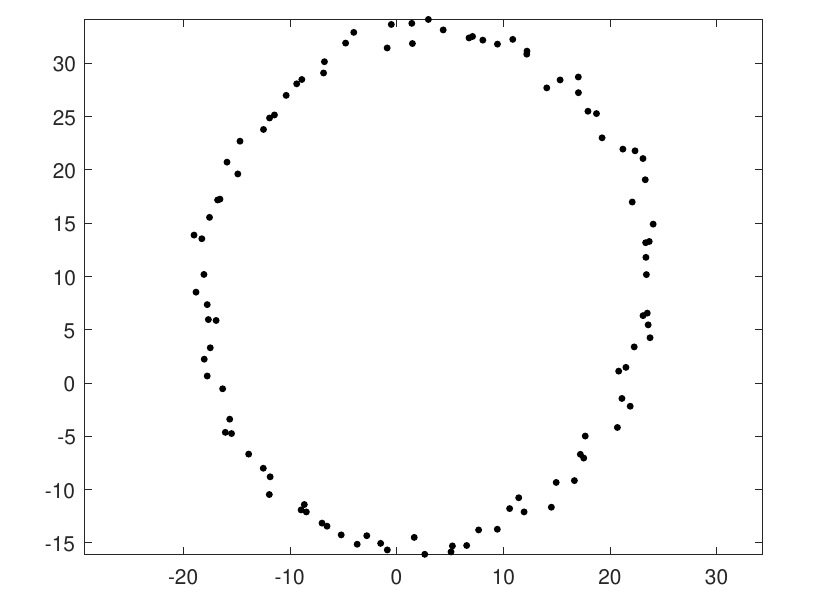}
\includegraphics[width=0.49\textwidth]{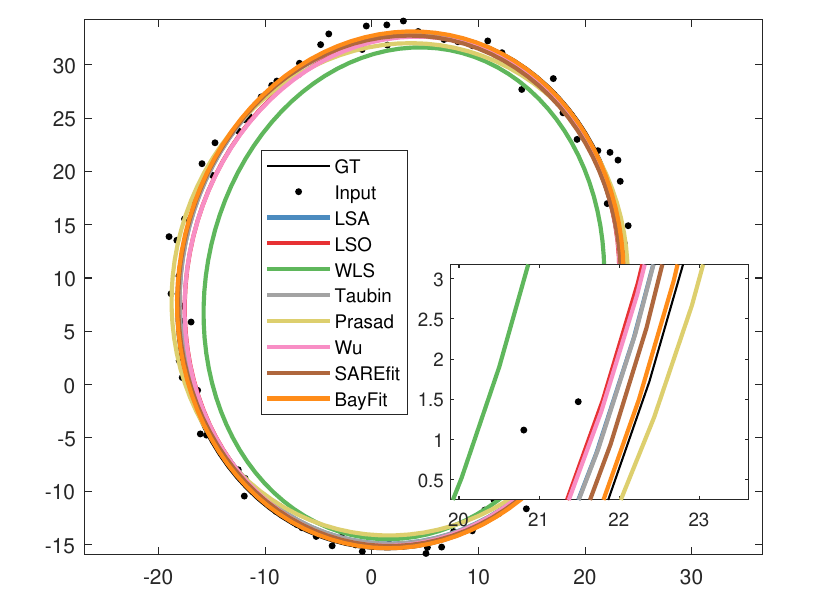}
\end{minipage}
\begin{minipage}{0.49\linewidth}
\centering
\includegraphics[width=0.49\textwidth]{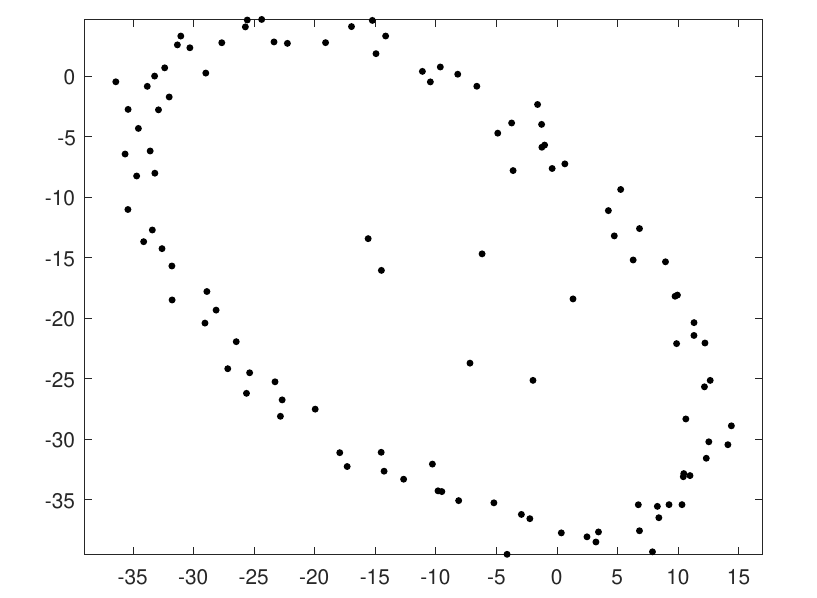}
\includegraphics[width=0.49\textwidth]{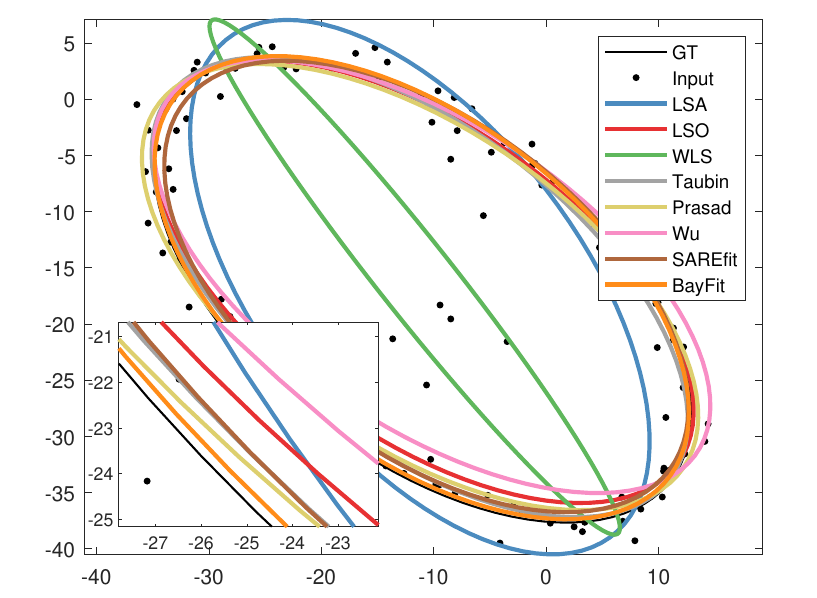}
\end{minipage}

\begin{minipage}{0.49\linewidth}
	\centering
	{{(a) $\sigma^2=50\%$}}
\end{minipage}
\begin{minipage}{0.49\linewidth}
	\centering
	{{(b) $\sigma^2=100\%$}}
\end{minipage}

\begin{minipage}{0.49\linewidth}
\centering
\includegraphics[width=0.49\textwidth]{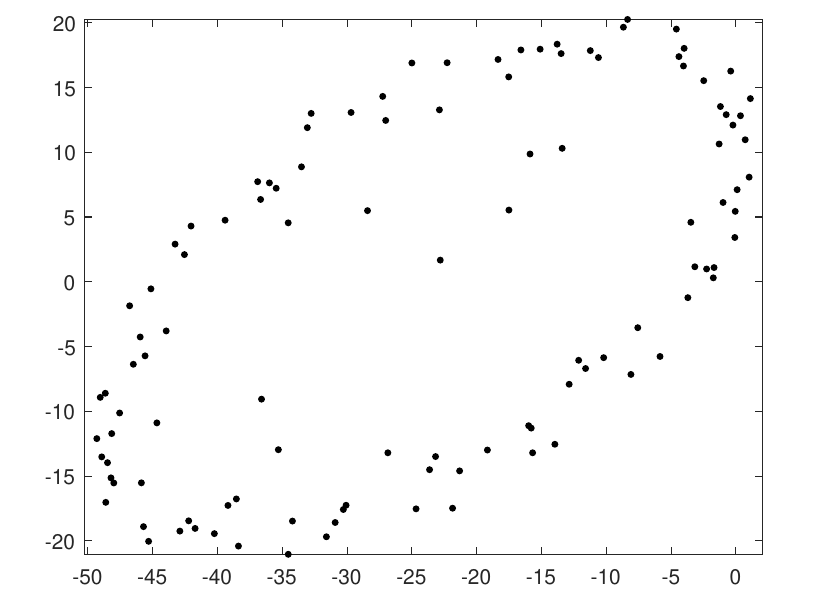}
\includegraphics[width=0.49\textwidth]{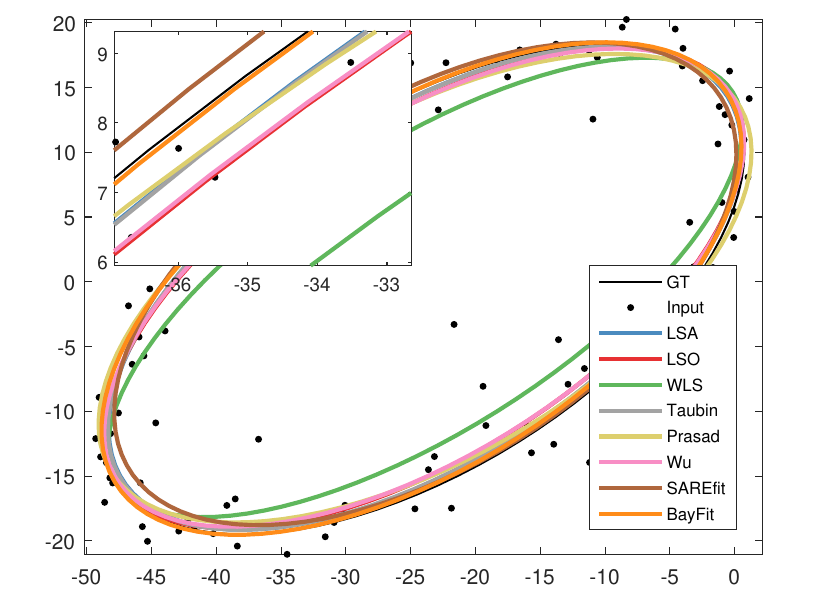}
\end{minipage}
\begin{minipage}{0.49\linewidth}
\centering
\includegraphics[width=0.49\textwidth]{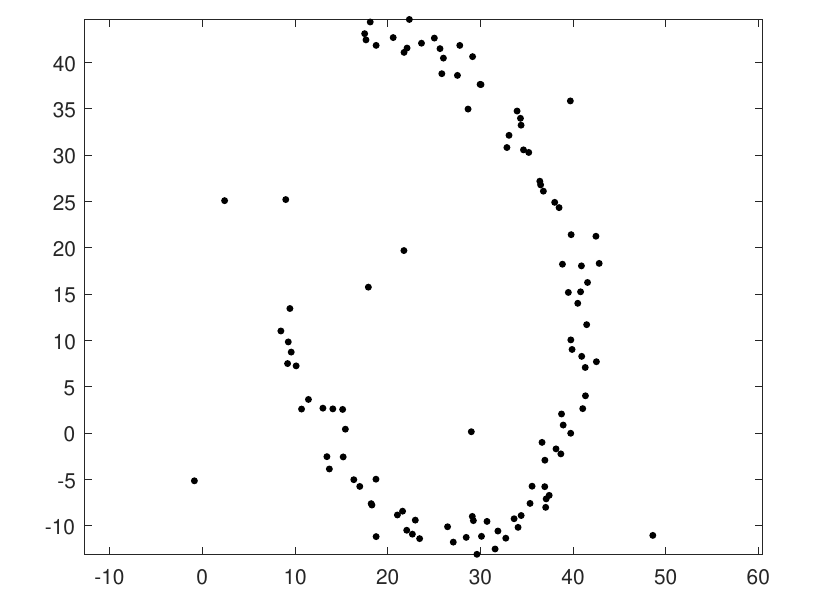}
\includegraphics[width=0.49\textwidth]{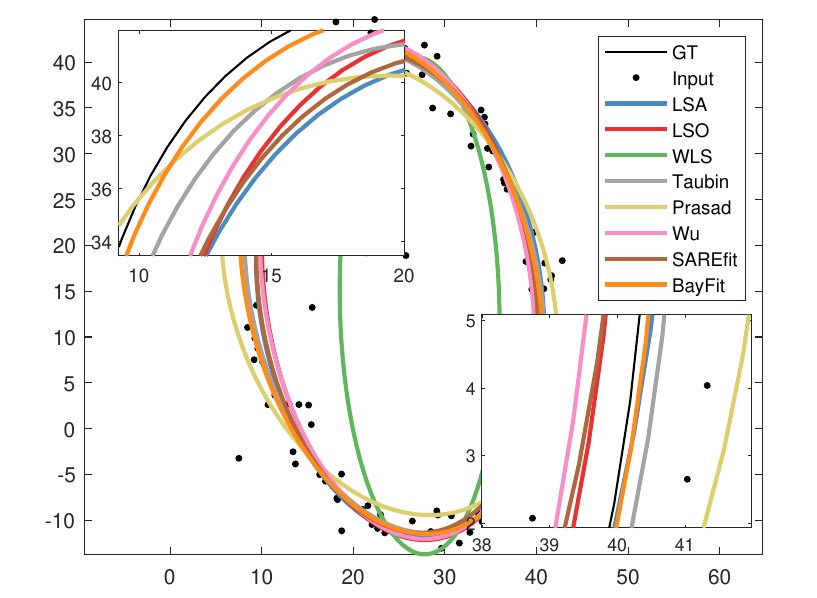}
\end{minipage}
\begin{minipage}{0.49\linewidth}
	\centering
	{{(c) $\sigma^2=150\%$}}
\end{minipage}
\begin{minipage}{0.49\linewidth}
	\centering
	{{(d) $\sigma^2=150\%$, $1/4$ occlusion}}
\end{minipage}
	\vskip -0.2cm
	\caption{Qualitative results of 2D ellipse fitting on different noise levels. The zoomed-in sub-figures suggest that BayFit has superior fitting precision than competitors, especially under the existence of heavy noise.
	}\label{fig:ellipse_noise_example}
	\vskip -0.3cm
\end{figure*}

\subsubsection{Necessity of the Uniform Distribution}
In Section~\ref{sec:bayesian_posterior}, we establish that the input measurement points $\mathbf{X}=\{\mathbf{x}_i\in \mathbb{R}^n\}_{i=1}^N$ are sampled from
\begin{equation}
	p(\mathbf{x}_i|\mathcal{E}(\bm{\theta}))=(1-w)\frac{1}{M}\sum_{j=1}^Mp(\mathbf{x}_i|\mathbf{y}_j(\bm{\theta}))+w\frac{1}{V},
	\label{eq:GMM_uniform}
\end{equation} where $w\neq0$. If we exclude the uniform distribution and \revisemajor{instead consider a family of measurable probabilistic densities given by}
\begin{equation}
	f(\mathbf{u},\mathbf{\xi})=\frac{1}{L}\sum_{i=1}^{L}\frac{1}{(2\pi \sigma^2)^{n/2}}\exp\big(-\frac{1}{2\sigma^2}||\mathbf{u}-\mathbf{x}_i||^2\big)
	\label{eq:GMM}
\end{equation} to approximate $p(\mathbf{x}_i|\mathcal{E}(\bm{\theta}))$, where $\mathbf{\xi}=(\mathbf{x}_1,\cdots, \mathbf{x}_L)$ is the unknown parameter vector, a natural question arises: {Does Eq.~\ref{eq:GMM}
possess the same \revisemajor{expressive} capability as Eq.~\ref{eq:GMM_uniform} for robust ellipsoid fitting?}

\revisemajor{In White's work~\cite{white1982maximum}, the quasi-maximum likelihood estimator $\hat{\mathbf{\xi}}_N$ is defined as the solution to the following problem} 
\begin{align*}
	\max_{\mathbf{\xi}} \ell_L(\mathbf{u},\mathbf{\xi})=\max_{\mathbf{\xi}} \frac{1}{L}\sum_{i=1}^{L}\log f(\mathbf{x}_i,\mathbf{\xi}).
\end{align*}Note that $\hat{\mathbf{\xi}}_L$ is a natural estimator for $\mathbf{\xi}_*$ which minimizes the Kullback-Leibler Information Criterion (KLIC) between $p(\mathbf{x}_i|\mathcal{E}(\bm{\theta}))$ and $f(\mathbf{u},\mathbf{\xi})$, namely
\begin{equation}
	\mathbf{\xi}_*=\mathop{\argmin}\limits_{\mathbf{\xi}} I(g:f,\mathbf{\xi})=\mathbb{E}(\log(p(\mathbf{x}_i|\mathcal{E}(\bm{\theta}))/f(\mathbf{u},\mathbf{\xi}))).
\end{equation}Based on \revisemajor{the} KLIC, we derive the following conclusion. 
\begin{theorem}[Nonequivalence between $p(\mathbf{x}_i|\mathcal{E}(\bm{\theta}))$ and $f(\mathbf{u},\mathbf{\xi})$]\label{theorem:KLIC} 
	\revisemajor{Assuming} that $\mathbb{E}(p(\mathbf{x}_i|\mathcal{E}(\bm{\theta})))$ exists and $I(g:f,\mathbf{\xi})$ has a unique minimum at $\mathbf{\xi}_*$, \revisemajor{it follows that} $\hat{\mathbf{\xi}}_L\stackrel{a.s.}\longrightarrow\mathbf{\xi}_*$ (\ie, almost surely converge) as $L\rightarrow\infty$, where $\mathbf{\xi}_*$  minimizes  $I(g:f,\mathbf{\xi})=\mathbb{E}(\log(p(\mathbf{x}_i|\mathcal{E}(\bm{\theta}))/f(\mathbf{u},\mathbf{\xi})))$ and the corresponding KLIC $I(g:f,\mathbf{\xi}_*)>0$.
\end{theorem}
\revisemajor{Theorem~\ref{theorem:KLIC} establishes the disparity between the probabilistic distributions $p(\mathbf{x}_i|\mathcal{E}(\bm{\theta}))$ and $f(\mathbf{u},\mathbf{\xi})$ by  $I(g:f,\mathbf{\xi}_*)\neq0$.} This disparity highlights that the expressive capabilities of Eq.~\ref{eq:GMM} and Eq.~\ref{eq:GMM_uniform} to represent the same observed data are not equivalent. Consequently, including the uniform distribution component in our loss function is crucial for effectively accounting for outliers. Otherwise, Eq.~\ref{eq:GMM} reduces to a least-squares optimization problem, which tends to be susceptible to  disturbances caused by outliers.

\noindent{\textbf{Example Analysis}.} We construct a standard Gaussian distribution with $M=1$ in 1D space and vary the weight $w$ in Eq.~\ref{eq:GMM_uniform} to provide insight into the robustness of the developed algorithm. As observed in the left panel of Fig.~\ref{fig:uniform}, the probability $p(x)$ with $w\neq0.0$ becomes a \emph{heavy-tailed distribution}, and thus exhibits more tolerance \revisemajor{towards} data points that are located far from the mean (\ie, outliers). Meanwhile, in contrast to $-\log(p(x))$ that has a \emph{unbounded influence function} and thereby suffers from outliers when $w=0.0$, our objective function augmented by the uniform distribution ($w\neq0.0$) is \emph{bounded}. This bounded characteristic effectively \revisemajor{mitigates the impact} of significant residuals caused by outliers.

\begin{figure*}[t]
	\centering
	\includegraphics[width=0.243\textwidth]{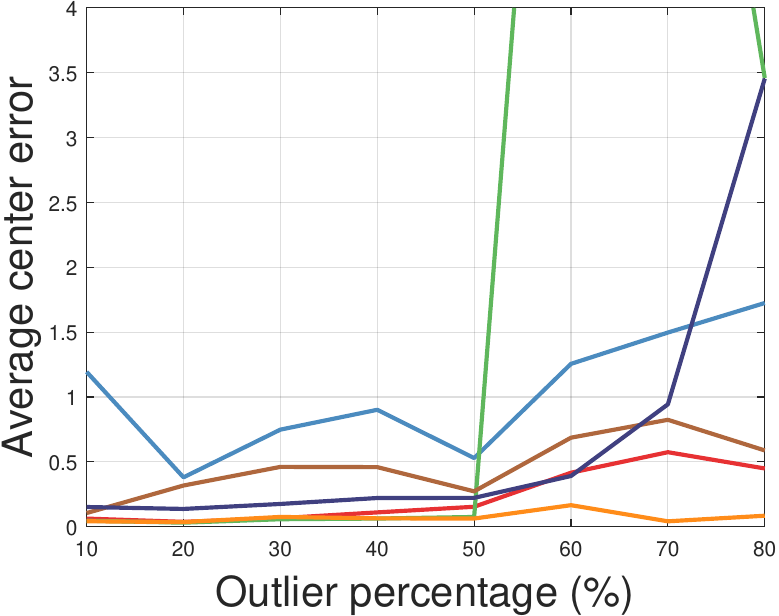}
	\includegraphics[width=0.2385\textwidth]{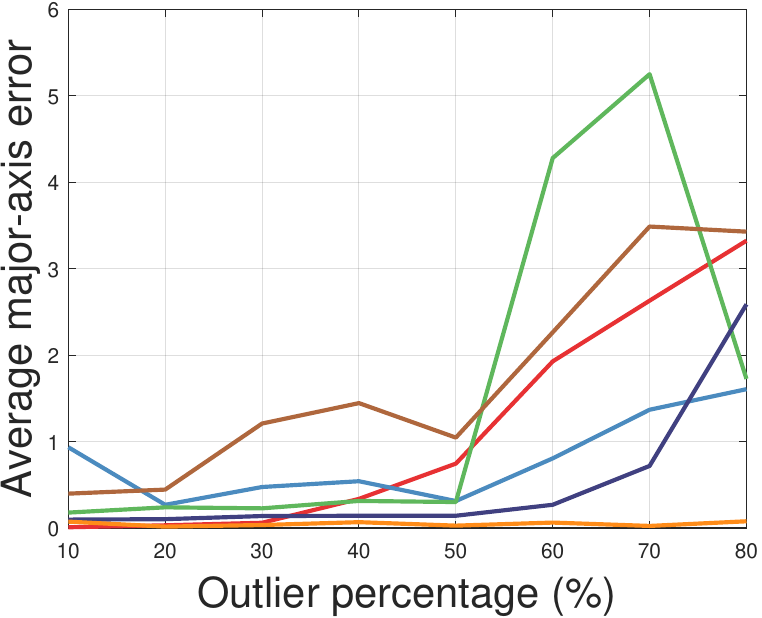}	
	\includegraphics[width=0.243\textwidth]{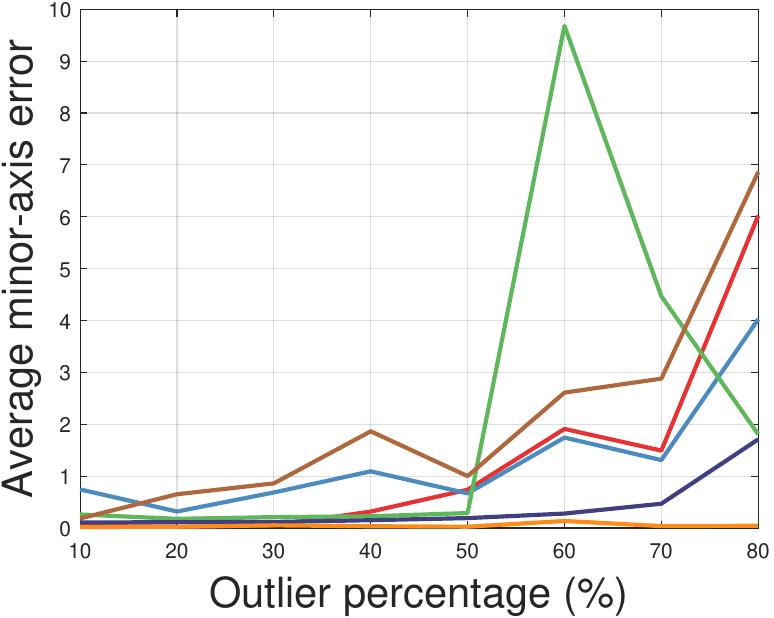}
	\includegraphics[width=0.251\textwidth]{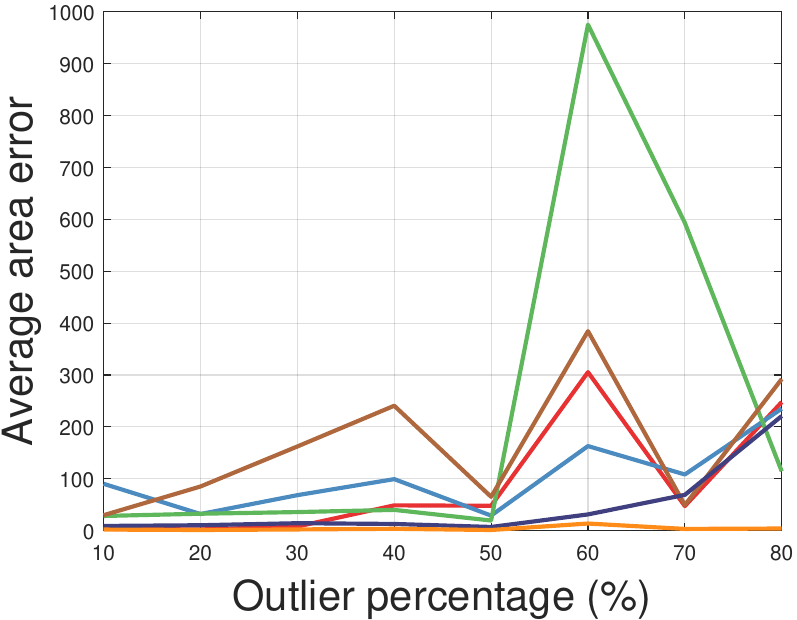}
	\centering
		\begin{tikzpicture}[align=center]
			\draw[line width=1pt,color=Cauchy] (0.3,0)--(1,0) node[right](line){\color{black}{Cauchy}};
			\draw[line width=1pt,color=LMS] (2.4,0)--(3.1,0) node[right](line){\color{black}{LMS}};
			\draw[line width=1pt,color=HGMM] (4.4,0)--(5.1,0) node[right](line){\color{black}{HGMM}};
			\draw[line width=1pt,color=SAREfit] (6.6,0)--(7.3,0) node[right](line){\color{black}{SAREfit}};
			\draw[line width=1pt,color=RANSAC] (8.9,0)--(9.6,0) node[right](line){\color{black}{RANSAC}};
			\draw[line width=1pt,color=BayFit] (11.3,0)--(12,0) node[right](line){\color{black}{BayFit (Ours)}};
		\end{tikzpicture}
	\vskip -0.3cm
	\caption{Quantitative evaluations on outlier-contaminated settings with respect to 2D ellipse fitting. The outlier percentage $\eta$ is increased from $10\%$ to $80\%$. We report MSE of all compared robust methods under 100 random tests. BayFit achieves higher breakdown points than competitors and is substantially more robust and stable against outlier disturbances.  
	}\label{fig:outlier_result}
\end{figure*}
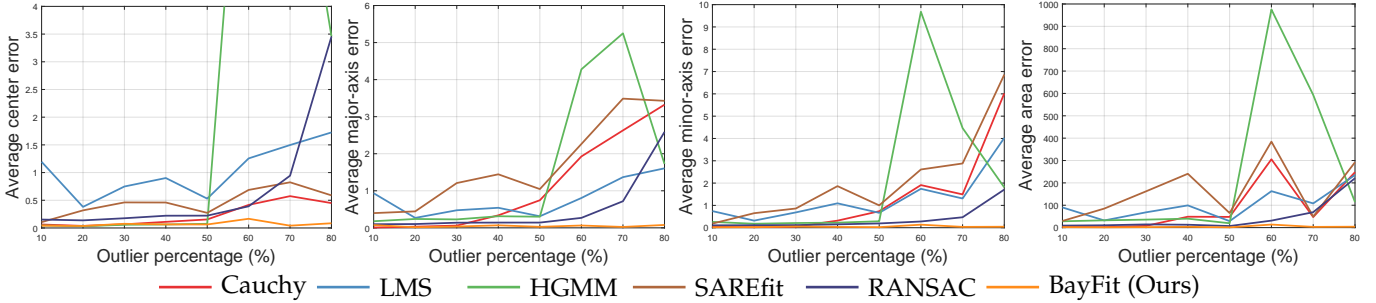

\begin{figure}[t]
	\centering
	\begin{minipage}{0.49\linewidth}
		\includegraphics[width=0.49\textwidth]{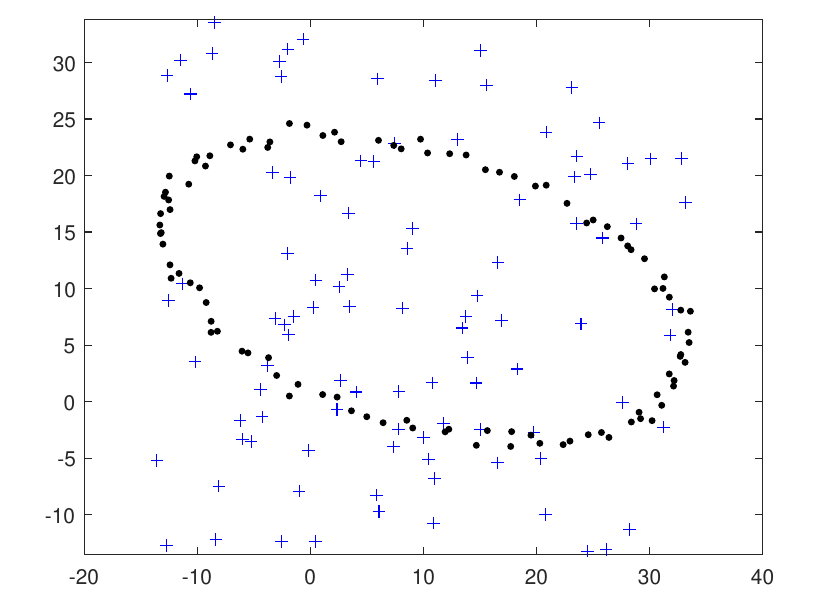}
		\includegraphics[width=0.49\textwidth]{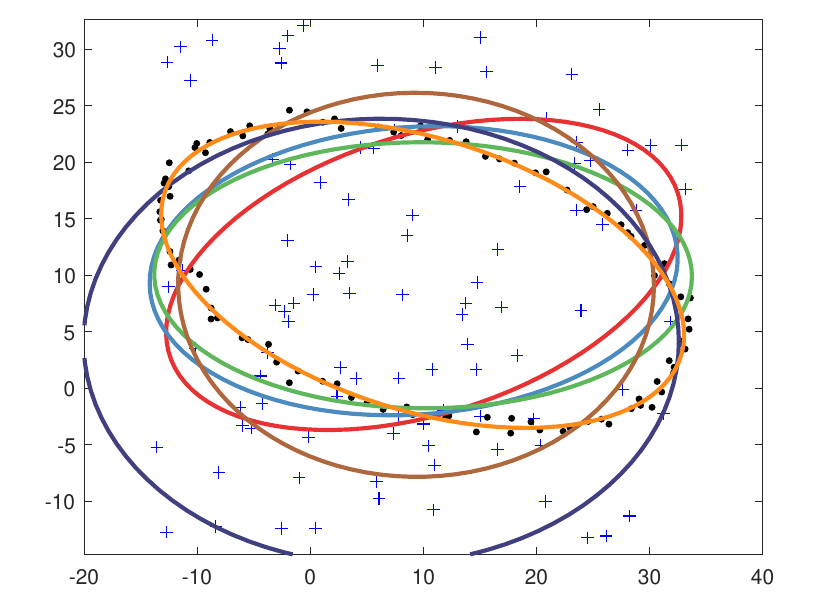}
	\end{minipage}
	\begin{minipage}{0.49\linewidth}
		\includegraphics[width=0.49\textwidth]{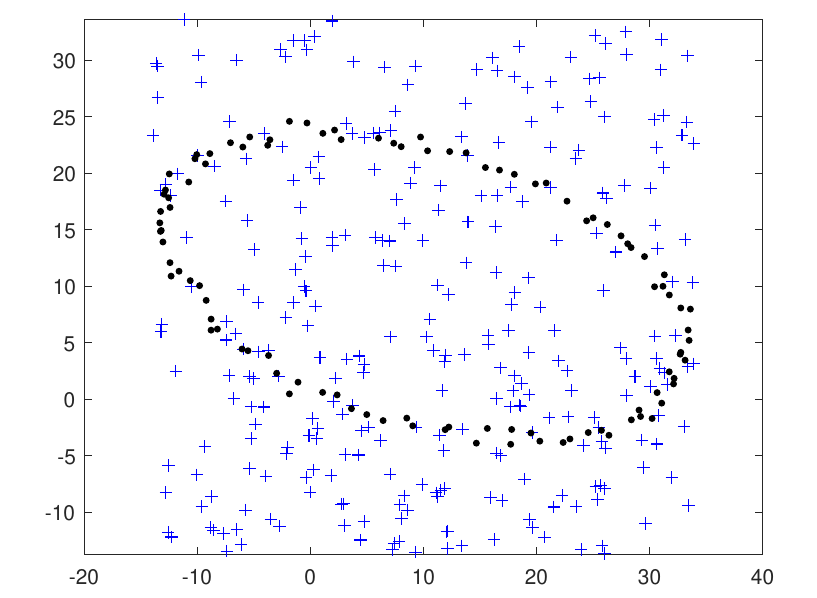}
		\includegraphics[width=0.49\textwidth]{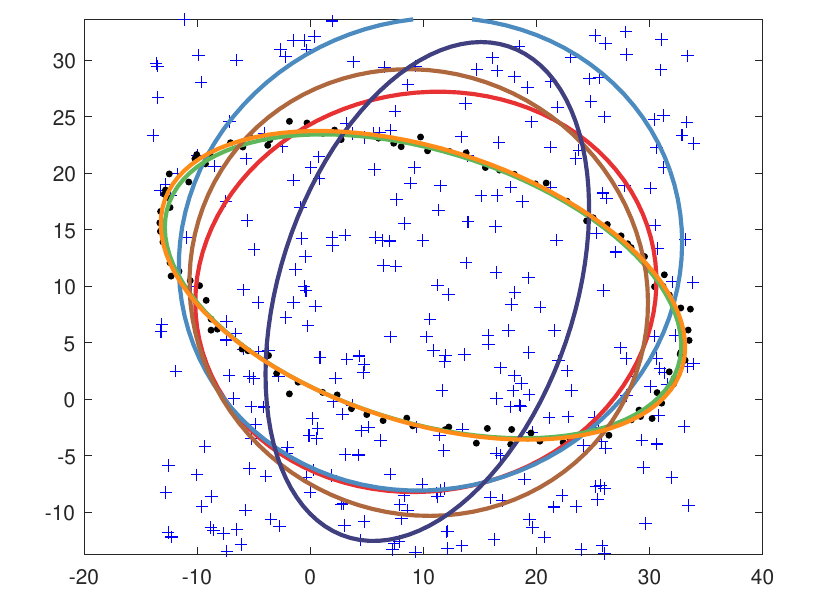}
	\end{minipage}
	\begin{minipage}{0.49\linewidth}
		\centering
		{\scriptsize{$\eta=50\%,\sigma^2=25\%$}}
	\end{minipage}
	\begin{minipage}{0.49\linewidth}
		\centering
		{\scriptsize{$\eta=75\%,\sigma^2=25\%$}}
	\end{minipage}
	
	\begin{minipage}{0.49\linewidth}
		\includegraphics[width=0.49\textwidth]{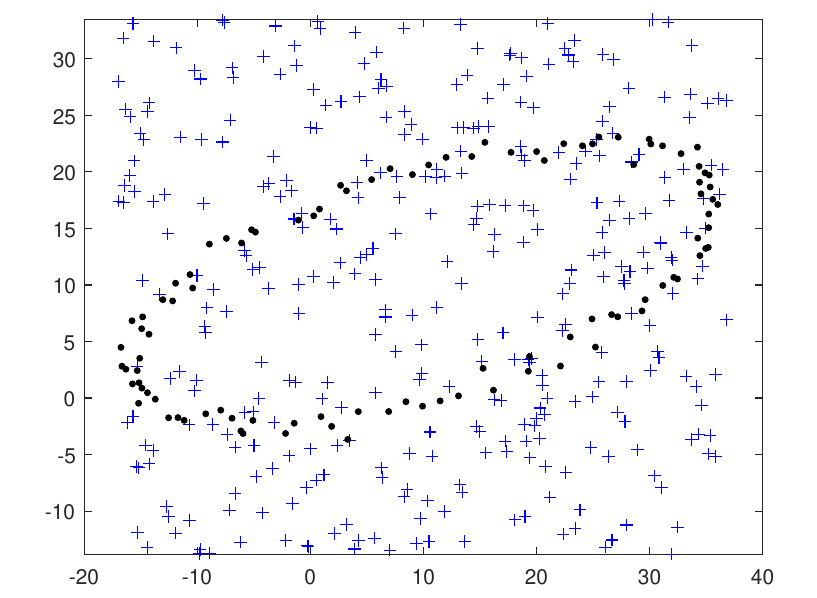}
		\includegraphics[width=0.49\textwidth]{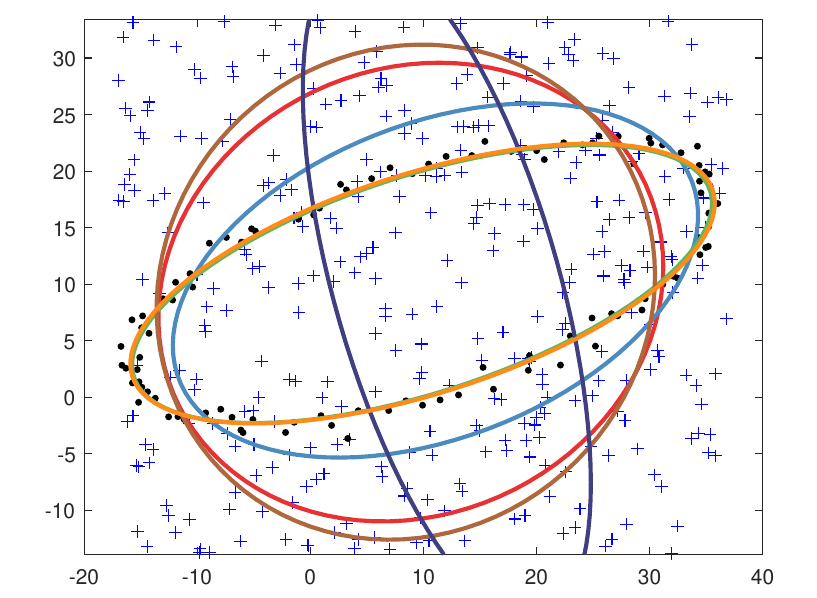}
	\end{minipage}
	\begin{minipage}{0.49\linewidth}
		\includegraphics[width=0.49\textwidth]{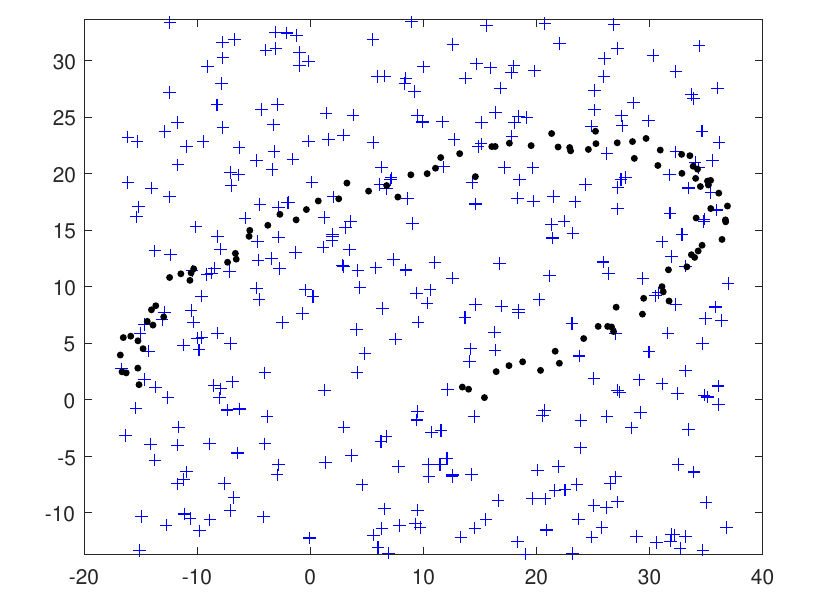}
		\includegraphics[width=0.49\textwidth]{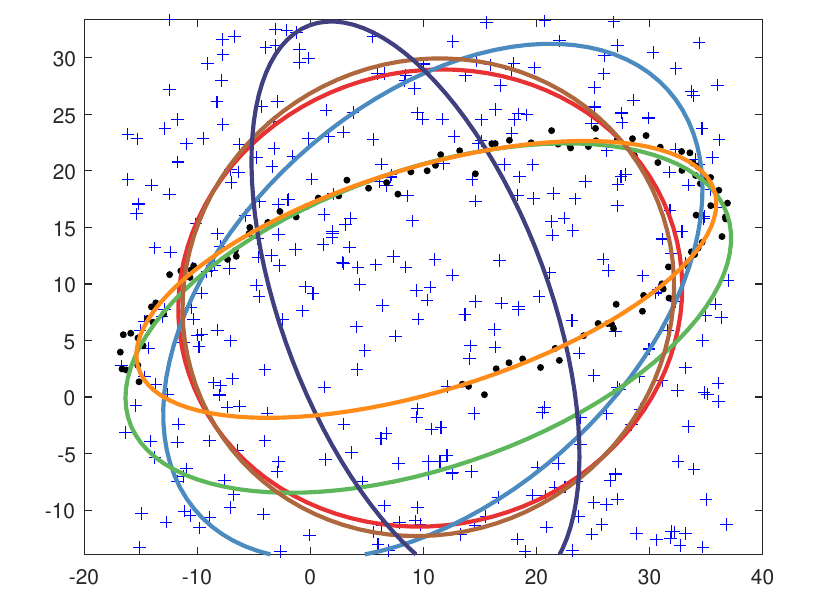}
	\end{minipage}
	\begin{minipage}{0.49\linewidth}
		\centering
		{\scriptsize{$\eta=80\%, \sigma^2=40\%$}}
	\end{minipage}
	\begin{minipage}{0.49\linewidth}
		\centering
		{\scriptsize{$\eta=80\%, \sigma^2=40\%$, $1/4$ occlusion}}
	\end{minipage}
\vskip -0.2cm
	\caption{\revisemajor{(Best viewed in color)} Qualitative comparisons of robust ellipse fitting methods on various outlier percentages and noise levels. It is observed that BayFit consistently \revisemajor{obtains} better fitting results than the compared methods.}\label{fig:outlier_exam_ellipse}
	\vskip -0.2cm
\end{figure}

\section{Experimental evaluations}\label{sec:experiment}
In this section, we conduct extensive experiments to evaluate the performance of the \revisemajor{proposed} algorithm. \revisemajor{Specifically}, the tests of fitting 2D ellipses, 3D ellipsoids, and higher-dimensional ellipsoidal surfaces are investigated separately. Additionally, we compare our method termed \emph{BayFit} with representative state-of-the-art or commonly used ones, consisting of algebraic, geometric, and robust approaches, in terms of precision, robustness, and efficiency. Concrete parameter settings and implementation details of baseline approaches are reported in Section 1 of the \emph{Supplementary Material}. We implement BayFit in Matlab and execute all tests on a desktop with AMD Core Ryzen 5 3600XT and 32 GB RAM. Our project code and datasets are made publicly available at \url{https://zikai1.github.io/}.

\subsection{Assessments of 2D Ellipse Fitting}
We first assess BayFit in 2D ellipse fitting settings, in which the geometric parameters $\bm{\theta}=[x_c,y_c,a,b,\alpha]$ of an ellipse, namely, the center position $(x_c, y_c)$, the lengths of \revisemajor{the} semi-axes $(a, b)$, and the rotation angle $\alpha$ between the major axis and the horizontal orientation, are sampled from a continuous uniform probability distribution separately. We use the built-in function $\texttt{unifrnd(lb, rb)}$ in Matlab to generate uniform samples among the interval $[lb, rb]$ with the lower bound $lb$ and the upper bound $ub$. The fitting accuracy of compared methods is quantitatively measured by the \emph{mean-squared error} (MSE) between the estimate parameters and the ground truth parameters. Moreover, we record the area difference between the ground truth and fitted ellipses. We conduct 100 trials in each experimental setup to assure statistically representative results. 
\begin{figure*}[t]
\centering
\includegraphics[width=\textwidth]{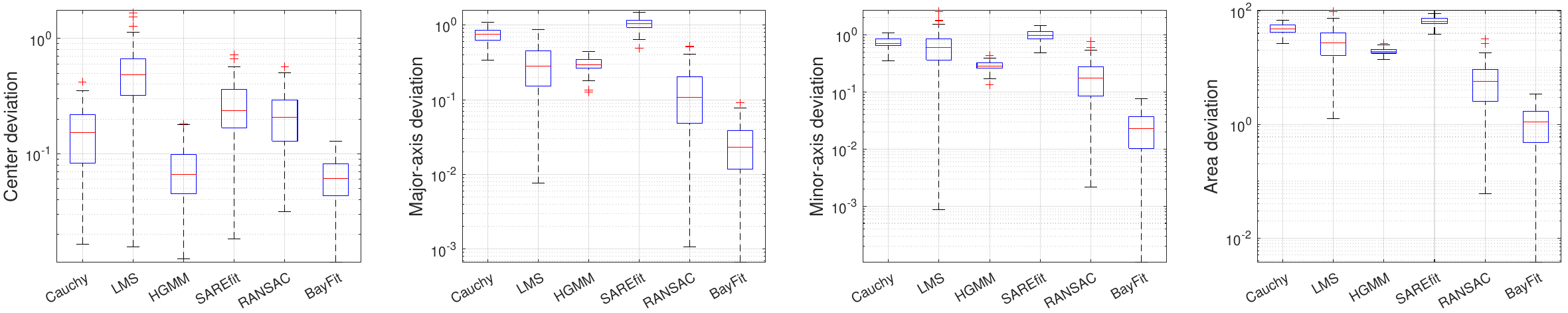}\\
\includegraphics[width=\textwidth]{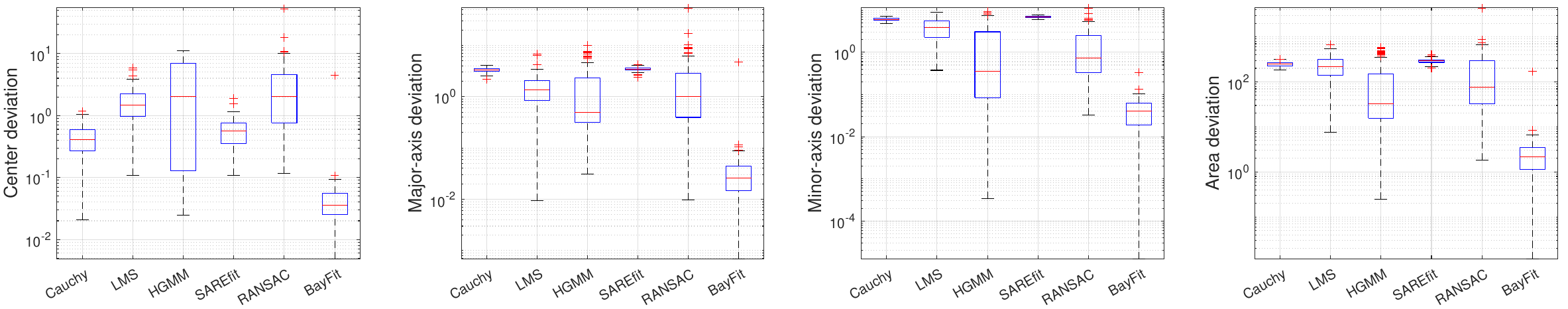}\\
\vskip -0.2cm
\caption{Boxplot results of outlier tests regarding 2D ellipse fitting, \revise{utilizing the log-scale $y$ axis to enhance the readability of the statistical results}. The outlier percentage $\eta$ from the first row to the second is equal to  $50\%$ and $80\%$, separately. Compared with the other robust methods, BayFit \revise{consistently exhibits} the lowest median value together with the smallest top and bottom quartiles, \revise{especially in cases with heavy outlier contamination}, indicating its high degree of robustness and stable fitting ability.}
\label{fig:boxplt}
\end{figure*}

\begin{figure}[!htbp]
\centering
\subfigure[Speed of noisy test]{
			\includegraphics[width=0.475\linewidth]{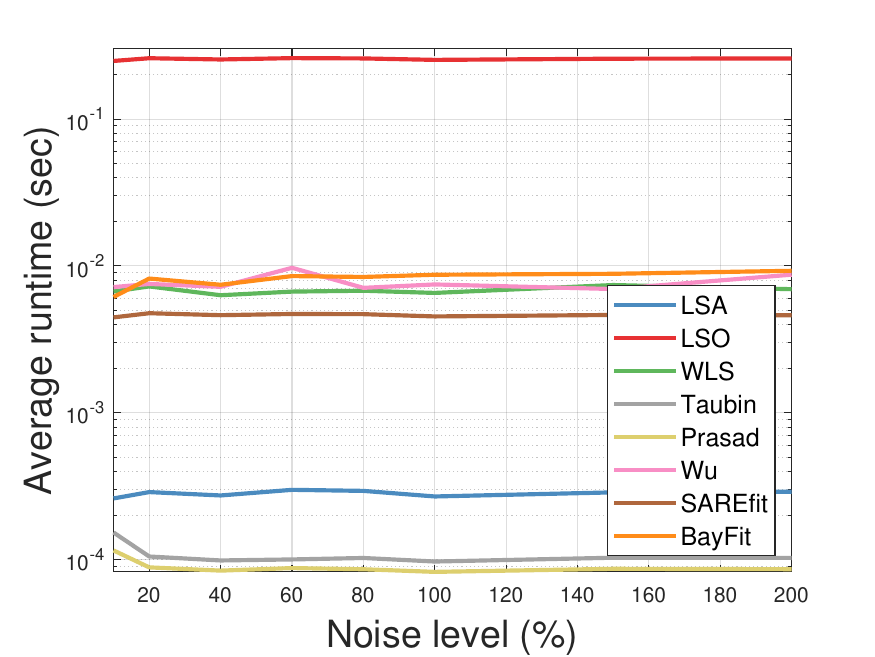}
		}
\subfigure[Speed of outlier test]{
			\includegraphics[width=0.47\linewidth]{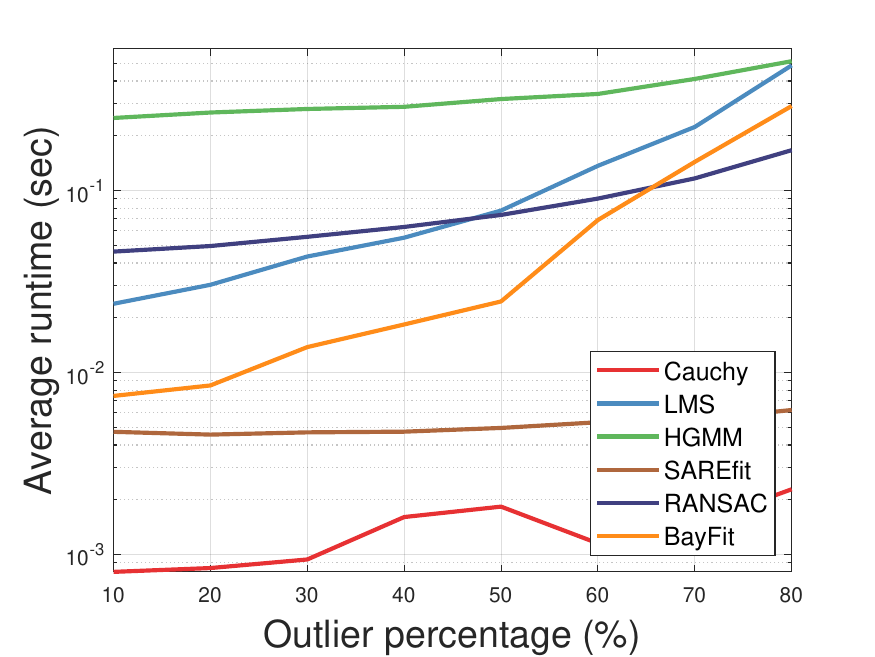}
		}
		\vskip -0.2cm
\caption{\revisemajor{Statistics of the runtime among noise and outlier experiments, where the log-scale $y$ axis is used to enhance the readability.}}
		\label{fig:noise_runtime}
		\vskip -0.3cm
\end{figure}
	
\begin{figure*}[t]
		\centering
\includegraphics[width=\textwidth]{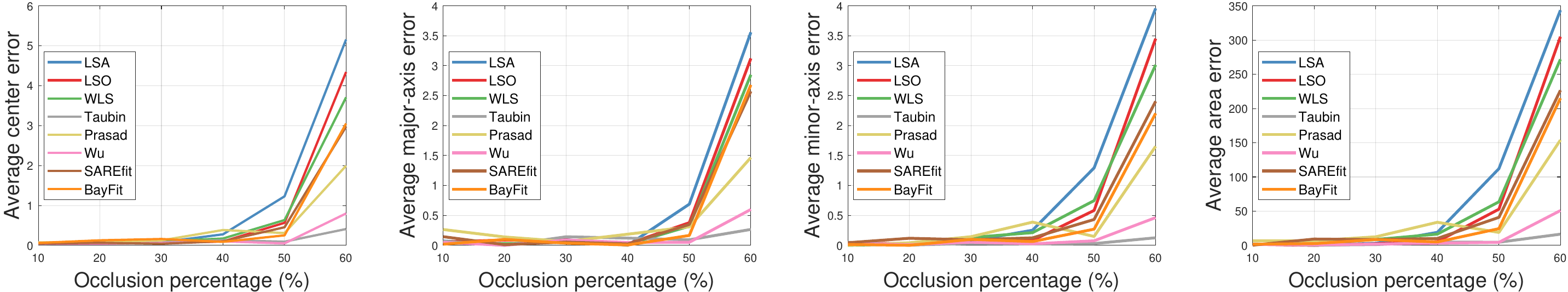}
\vskip -0.2cm
\caption{\revise{Quantitative evaluations on occlusion measurements as the occlusion level increases from $10\%$ to $60\%$. } 	}
		\label{fig:ellipe_occ}
		\vskip -0.3cm
\end{figure*}
\begin{figure}[t]
\centering
\includegraphics[width=0.235\textwidth]{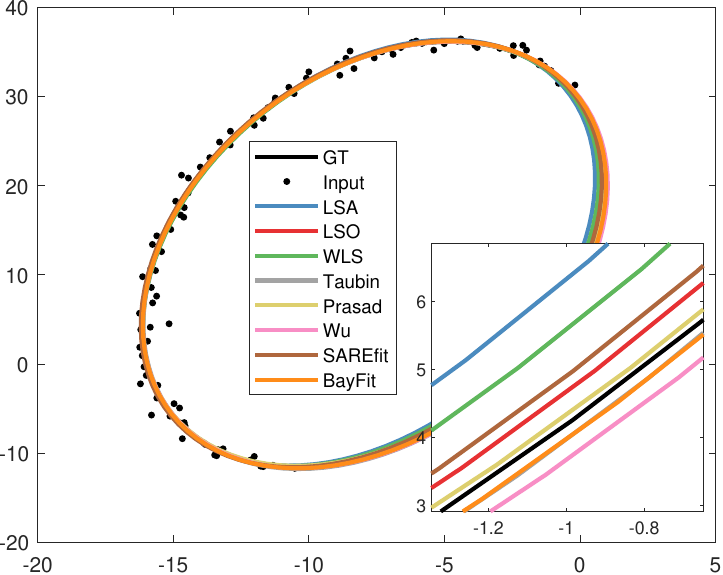}
\includegraphics[width=0.235\textwidth]{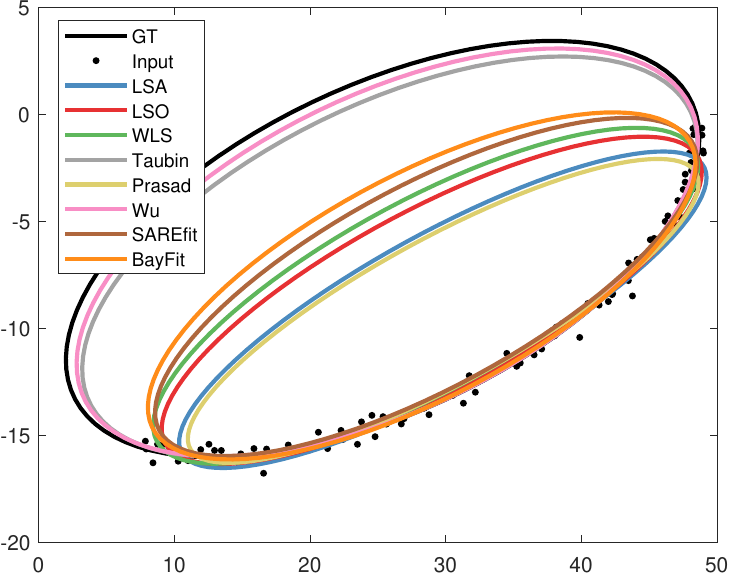}
\vskip -0.3cm
\caption{\revise{Qualitative comparisons are conducted on two occlusion scenarios, where the occlusion percentage is gradually increased from $40\%$ (left) to $60\%$ (right).}}\label{fig:ellipse_occ_example}
\vskip -0.3cm
\end{figure}
\subsubsection{Quantitative Evaluations on Noisy Settings} 
To \revisemajor{evaluate} the fitting accuracy of the proposed algorithm, we generate 100 ellipses randomly with $x_c, y_c\in\texttt{unifrnd(-30, 30)}$, $a, b\in \texttt{unifrnd(10, 30)}$, $\alpha \in \texttt{unifrnd($0^\circ, 180^\circ$)}$, and the sample magnitude equates 100. We add a series of Gaussian noise with zero mean and varying variance $\sigma^2\in[10\%, 200\%]$ {(the percentage is used to make the decimals integer for conciseness, \eg, $\sigma^2=0.1=10\%$)} as well as several random outliers to the data samples. Fig.~\ref{fig:ellipse_noise_example} presents a couple of test samples illustrating this \revisemajor{experimental} setup. The proposed method BayFit is compared with popular or representative state-of-the-art approaches, including the least-squares ellipse fitting based on the algebraic distance (LSA)~\cite{fitzgibbon1999direct} and the orthogonal distance (LSO)~\cite{ahn2002orthogonal}, the gradient-weighted least-squares (WLS)~\cite{zhang1997parameter}, Taubin~\cite{taubin1991estimation}, Prasad~\cite{prasad2013ellifit}, Wu~\cite{wu2019efficient}, and SAREfit~\cite{thurnhofer2020ellipse}. The statistical results are reported in Fig.~\ref{fig:ellipe_noie}, where we adopt the log-scale $y$ axis to enhance the result readability. As observed, all approaches exhibit fluctuations as the noise intensity increases. WLS shows significant deviations and more fluctuations compared to the other competitors, \revisemajor{particularly in} axis and area estimates. {The non-monotonic behavior may be attributed to the presence of heavy noise. When $\sigma^2$ is extremely high, the sample points exhibit significantly scattered patterns, which make them resemble to outliers.} Although BayFit is also affected by noise, it consistently achieves the highest accuracy in parameter estimate \revisemajor{across} various noise interference. For instance, our method maintains substantially lower center deviations compared to the other approaches, even when $\sigma^2$ reaches up to $200\%$. We present several qualitative results in Fig.~\ref{fig:ellipse_noise_example}.

\subsubsection{Quantitative Evaluations on Outlier Settings} 
Subsequently, we conduct a comprehensive investigation into the robustness of BayFit against outliers. The elliptical parameters are still randomly sampled from continuous uniform distributions. Since direct least-squares-based fitting paradigms are sensitive to outlier disturbances, we adopt robust ellipse fitting approaches for comparison. Specifically, we \revisemajor{include} the M-estimator using the Cauchy loss function (Cauchy)~\cite{zhang1997parameter}, the least median of squares (LMS)~\cite{rousseeuw1984least}, RANSAC~\cite{bolles1981ransac}, SAREfit~\cite{thurnhofer2020ellipse}, and HGMM~\cite{zhao2021robust}. We contaminate the elliptical points with uniformly distributed outliers, in which their spatial scope is relative to the semi-major axis $a$ of the target ellipse, \ie, $(x_{out}, y_{out})\in [x_c-a, x_c+a]\times [y_c-a, y_c+a]$. We systematically increase the outlier ratio $\eta$ from $10\%$ to $80\%$ with the interval equivalent to $10\%$.  Several test samples are presented in the first and third columns of Fig.~\ref{fig:outlier_exam_ellipse}. Fig.~\ref{fig:outlier_result} summarizes the quantitative results, from which we conclude that our method consistently outperforms competitors by a significant margin across various outlier percentages. Even in the presence of $80\%$ outliers, our method is capable of providing correct and stable estimates. In contrast, \revisemajor{the} competitors exhibit considerably lower breakdown points. For instance, HGMM maintains a $50\%$ breakdown point in all metrics, while SAREfit achieves even lower breakdown points in the semi-axis estimate.

Furthermore, Fig.~\ref{fig:boxplt} presents the results of 100 estimate deviations under three outlier ratios: $\eta=\{10\%, 50\%, 80\%\}$, \revisemajor{aiming to} provide a detailed insight into the fitting accuracy. \revise{It can be seen that BayFit achieves the smallest median value, as well as the $25^{th}$ $(Q_1)$ and $75^{th}$ $(Q_3)$ percentiles across almost all experimental tests. Additionally, BayFit demonstrates a relatively low box height and few extreme outliers marked as "+" beyond the box. These findings collectively suggest an exceptionally stable performance, even in the presence of significant outlier interference. We present several robust ellipse fitting results in Fig.~\ref{fig:outlier_exam_ellipse}.}

\subsubsection{Efficiency of the Designed Algorithm} 
\revisemajor{In addition to evaluating the accuracy and robustness}, we also assess the efficiency of our proposed method. Fig.~\ref{fig:noise_runtime} provides a summary of the average runtime for all tested algorithms under previous noisy and outlier-contaminated conditions. Fig.~\ref{fig:noise_runtime}(a) verifies that our method achieves comparable computational speed with most of the competitors, whereas LSO that is based on the orthogonal distance principle consumes the most time. \revisemajor{Among the robust approaches, BayFit is faster than HGMM and LMS. Even when the outlier percentage $\eta$ reaches up to 60\%, BayFit still maintains computation times of less than 0.1s.}

\revise{\subsubsection{Quantitative Evaluations on Occlusion Settings}} 
{Moreover, we investigate the effectiveness of our designed method in scenarios involving occlusion. Fig.~\ref{fig:ellipe_occ} presents the quantitative evaluation results of all compared approaches \revisemajor{as} the occlusion percentage $o$ \revisemajor{is increased} from $10\%$ to $60\%$. We observe that BayFit achieves comparable high fitting precision to the other approaches when  $o\leq50\%$. However, noticeable fitting deviations become evident for all compared algorithms when $o$ exceeds $50\%$. Taubin and Wu \revisemajor{obtain} relatively higher fitting stability due to their integration of geometric information, such as the Sampson distance. Our method secures a position among the top four best fitting approaches. \revisemajor{Moreover, it exhibits the best robustness} or stability against noise, outliers, large axis ratios, as well as ellipsoid-specificity settings. We present two qualitative comparison results in Fig.~\ref{fig:ellipse_occ_example}. }

\begin{figure}[t]
	\centering
	\includegraphics[width=0.234\textwidth]{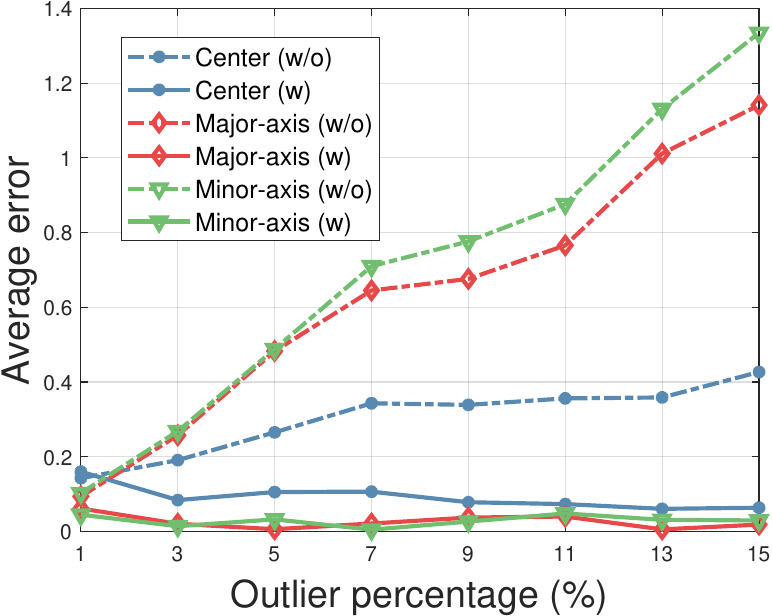}
	\includegraphics[width=0.232\textwidth]{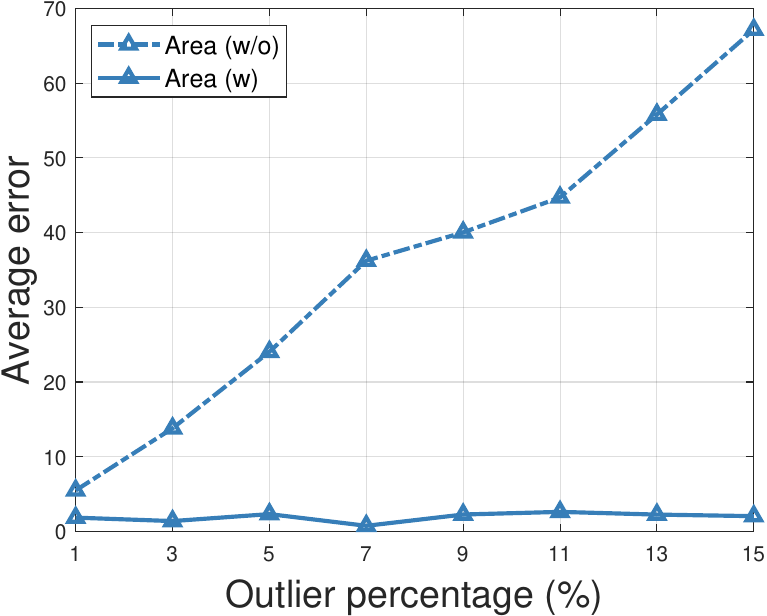}
	\vskip -0.3cm	
	\caption{\revise{Ablation study of the uniform distribution. The outlier percentage $\eta$ is gradually increased from $1\%$ to $15\%$.}  
	}\label{fig:ablation_result}
\vskip -0.3cm
\end{figure}

\revise{\subsubsection{Ablation Study of the Uniform Distribution}}
{Apart from the theoretical analysis of the uniform distribution in Section \ref{sec:robustness}, we further carry out an ablation study to demonstrate the necessity of the uniform distribution for robust fitting. We set $w=0$ in Eq.~\ref{eq:GMM_uniform} and gradually increase the outlier ratio $\eta$ from $1\%$ to $15\%$. The statistical results are reported in Fig.~\ref{fig:ablation_result}, comparing cases \emph{with (w)} or \emph{without (w/o)} the uniform distribution component. It is evident that BayFit, when equipped with the uniform distribution, exhibits significantly greater robustness against outlier disturbances across all metrics. \revisemajor{On the other hand}, BayFit without the uniform distribution \revisemajor{component} is more susceptible to the impact of outliers. We present a set of qualitative comparisons in Section 13 of the \emph{Supplementary Material} for reference.

} 
	
\begin{figure*}[t]
	\centering
\includegraphics[width=\textwidth]{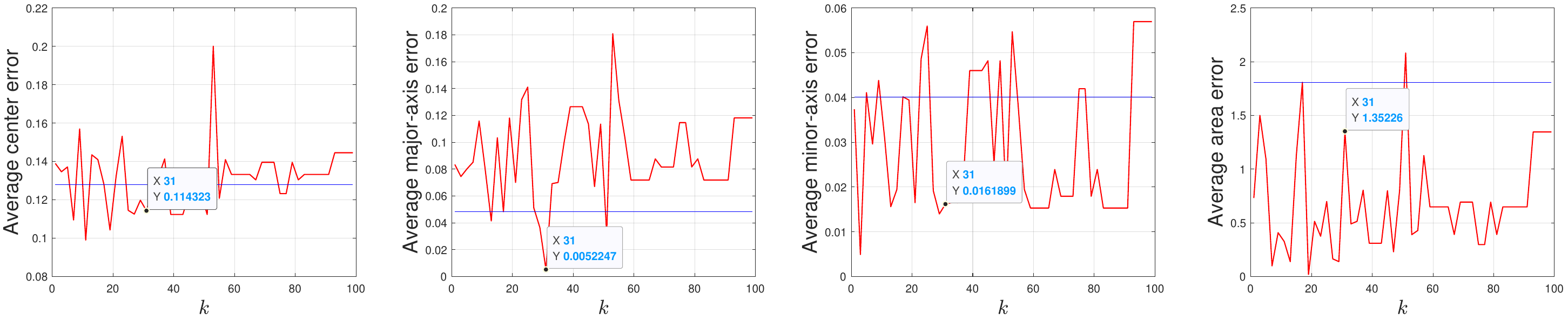}
	\vskip -0.2cm
	\caption{\revise{We vary $k$ within the range of $[1, 100]$ with $\Delta k=2$ to accomplish a simultaneous initialization of $M$ and $w$, where $k=31$ provides a higher-quality initialization. The horizontal blue line indicates the results attained through manual initialization, where $w$ is directly set to $0.3$.}}\label{fig:ablation_k}
	\vskip -0.3cm
\end{figure*}

\revise{\subsubsection{Ablation Study of $k$}}
{We also conduct an ablation study to assess the effectiveness of the proposed initialization strategy involving $M$ and $w$ by varying the parameter $k$ in $\RDOS$. To this end, we generate a random ellipse with an outlier ratio equivalent to $40\%$ and vary $k\in[1, 100]$ with \revisemajor{a step size of} $\Delta k=2$. Fig.~\ref{fig:ablation_k} presents a summary of the statistical fitting results,} {where the horizontal blue line indicates the results \revisemajor{obtained} by manually setting $w=0.3$ directly. As observed, the proposed initialization strategy jointly accomplishes the dual task of determining $M$ and $w$ in a single endeavor and offers a reasonable estimation for both parameters. For instance, selecting $k=31$ yields a relatively improved fitting outcome, eliminating the need for blind and labor-intensive manual adjustments of $w$ or $M$. More experimental analysis and discussion are presented in Section 8 of the \emph{Supplementary Material}.
} 

\begin{figure}[t]
	
	\begin{minipage}{.48\textwidth}
	\centering
		\includegraphics[width=0.24\linewidth]{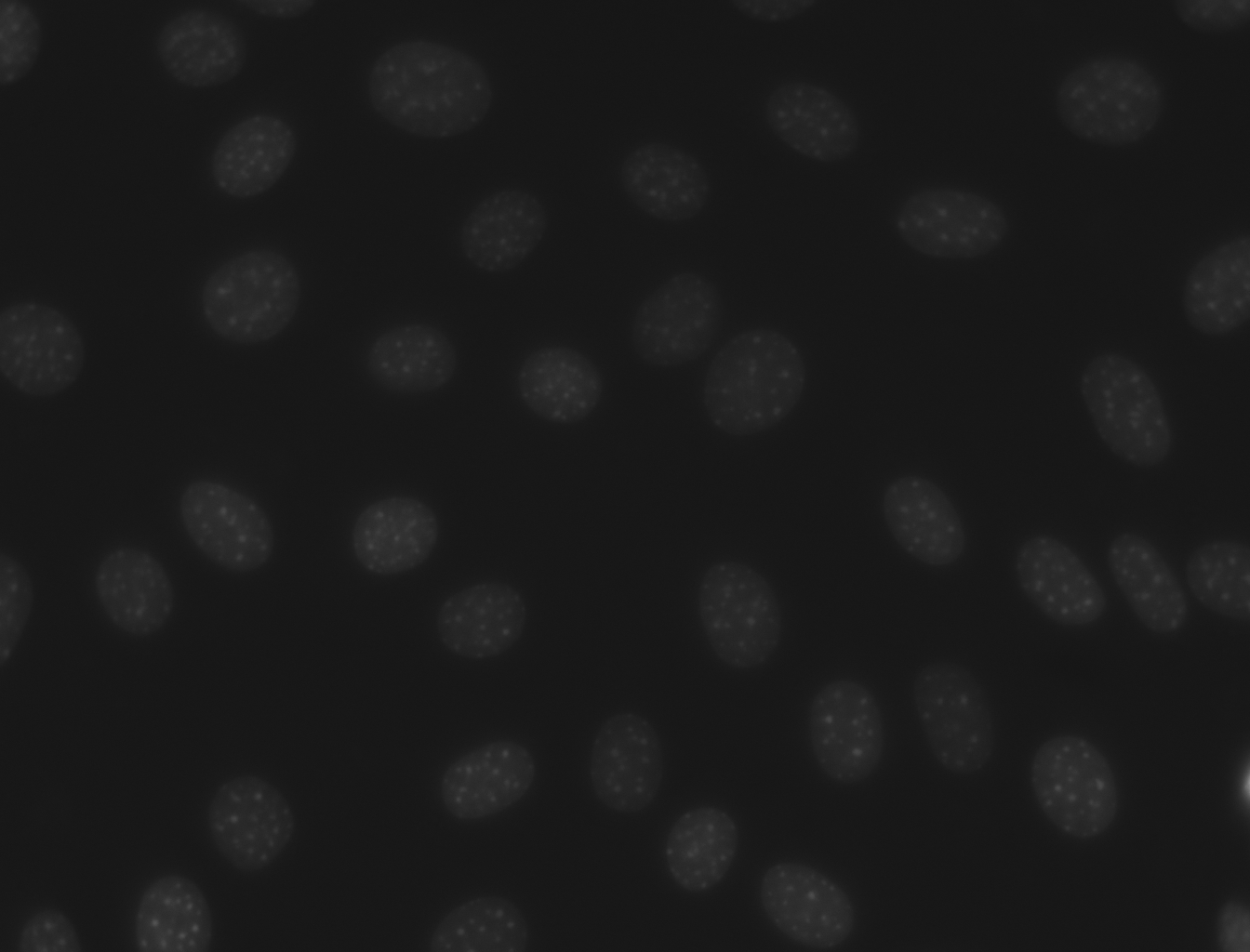}
		\includegraphics[width=0.24\linewidth]{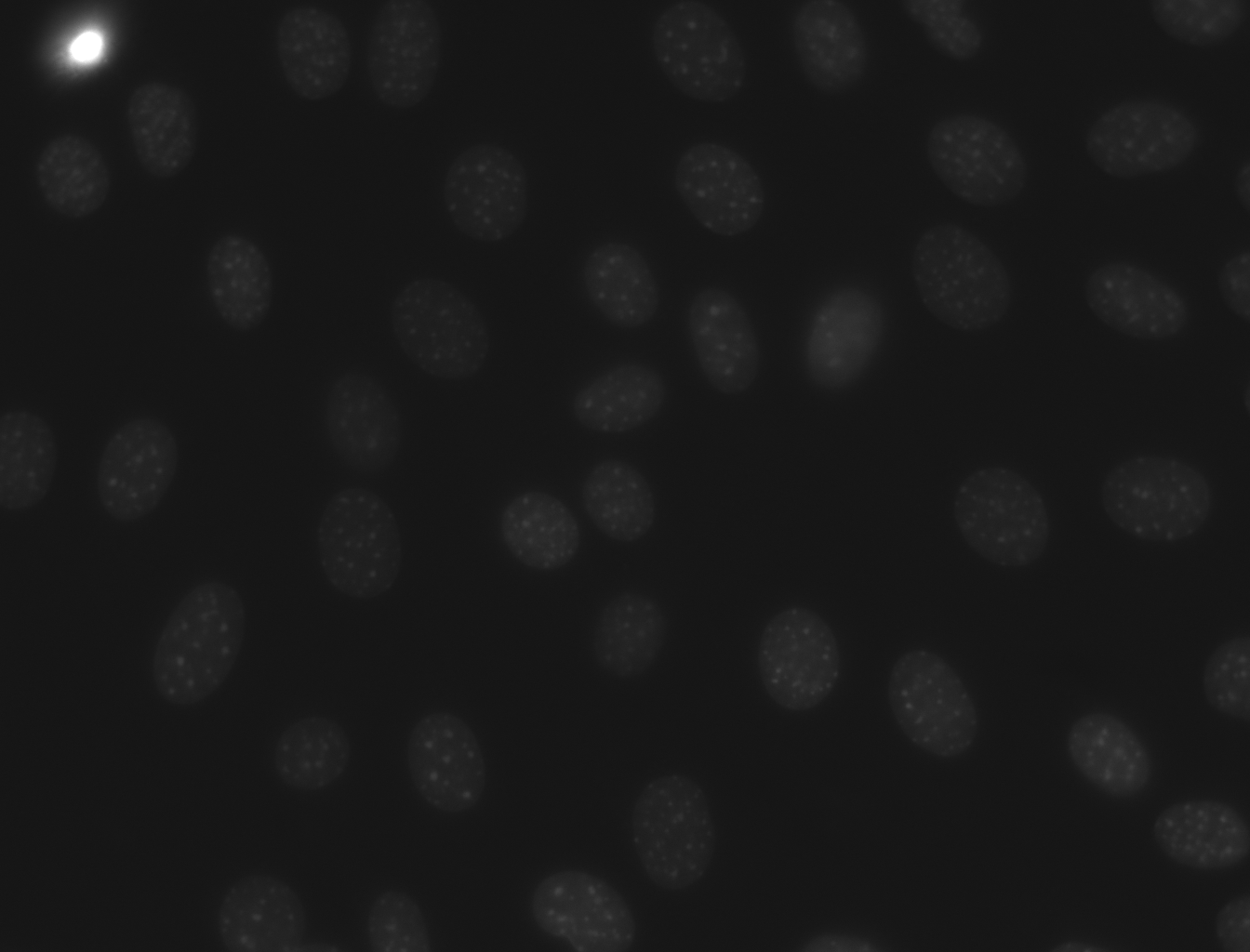}
		\includegraphics[width=0.24\linewidth]{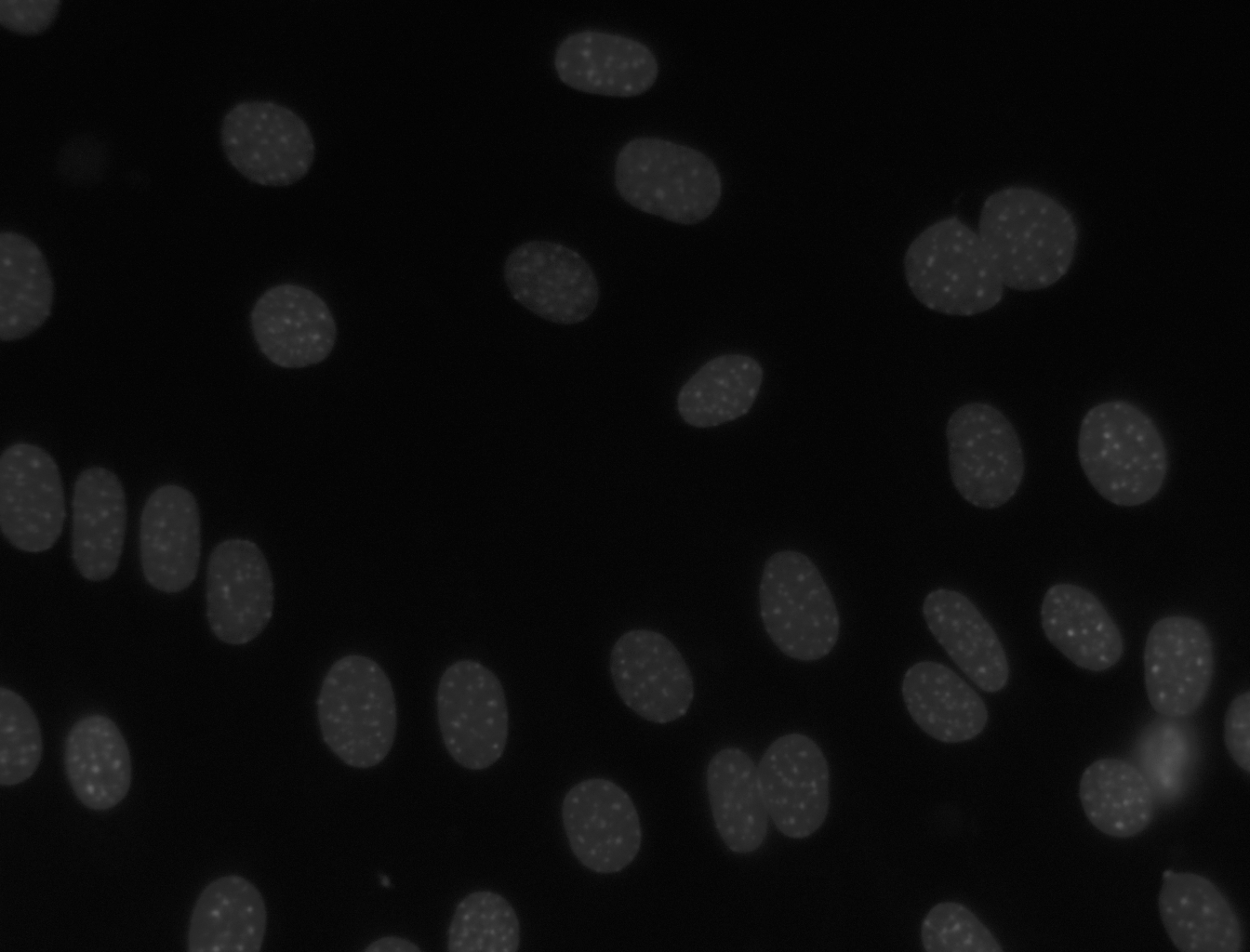}
		\includegraphics[width=0.24\linewidth]{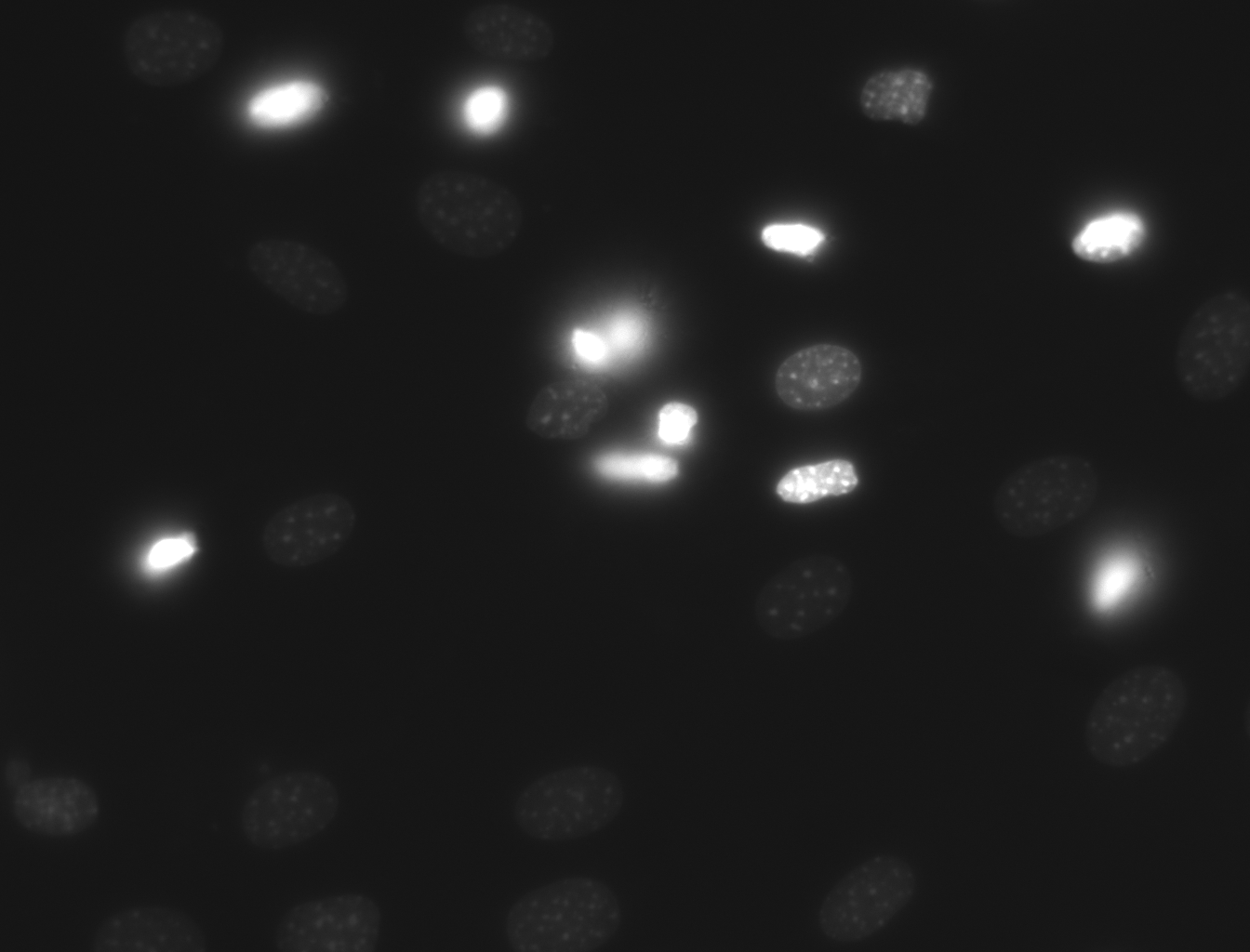}
	\end{minipage} 

\vskip 0.1cm

	\begin{minipage}{.48\textwidth}
		\centering
		\includegraphics[width=0.24\linewidth]{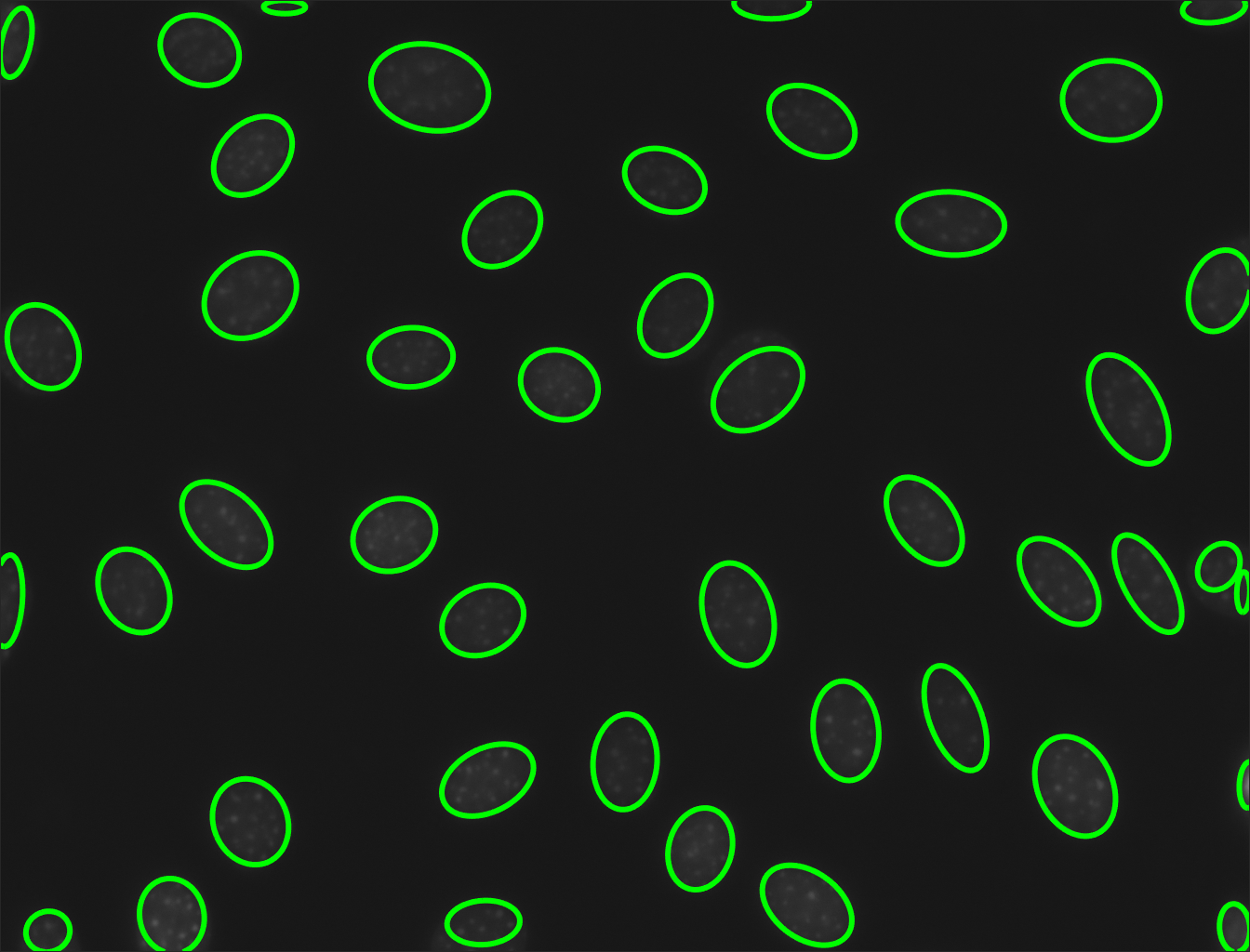}
		\includegraphics[width=0.24\linewidth]{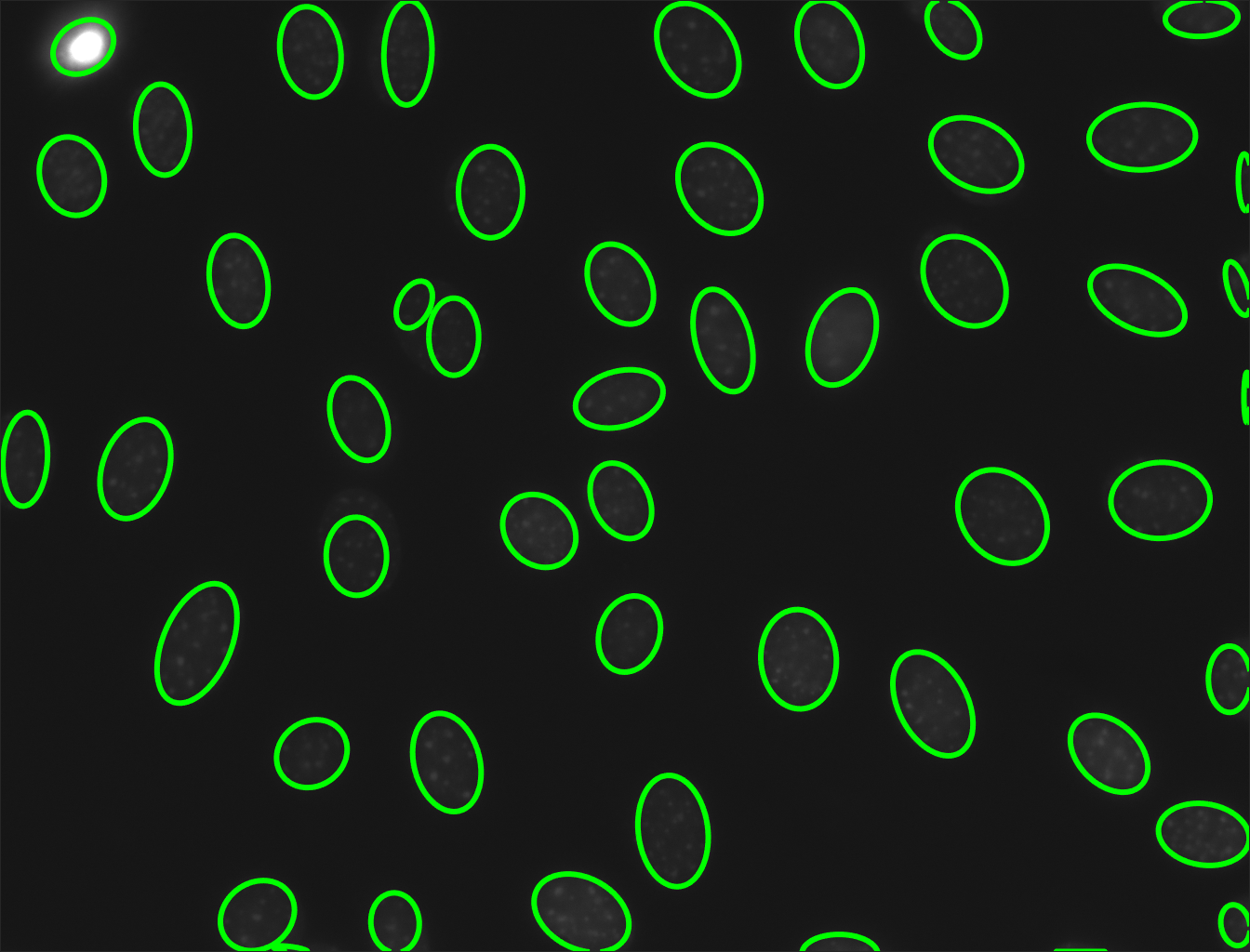}
		\includegraphics[width=0.24\linewidth]{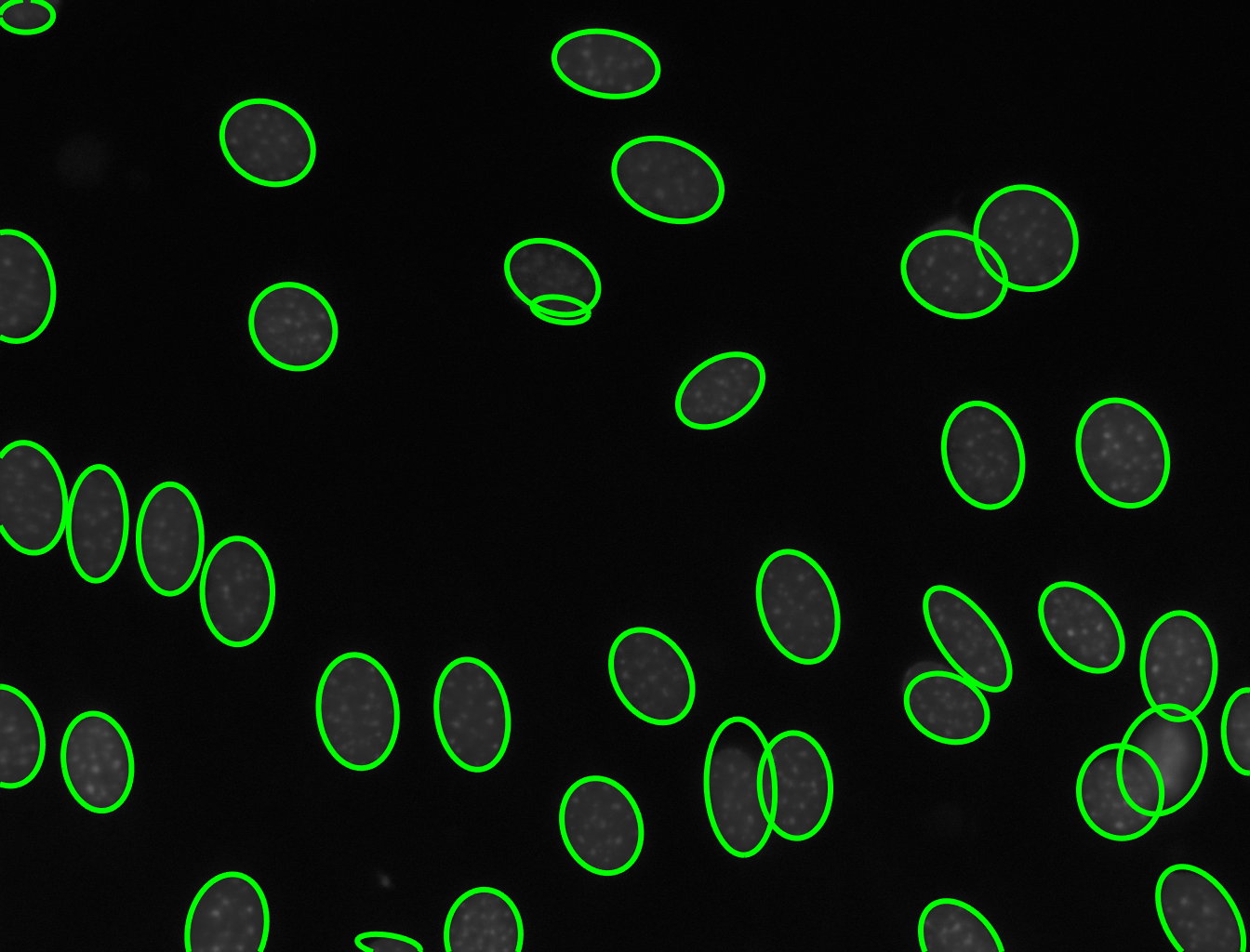}
		\includegraphics[width=0.24\linewidth]{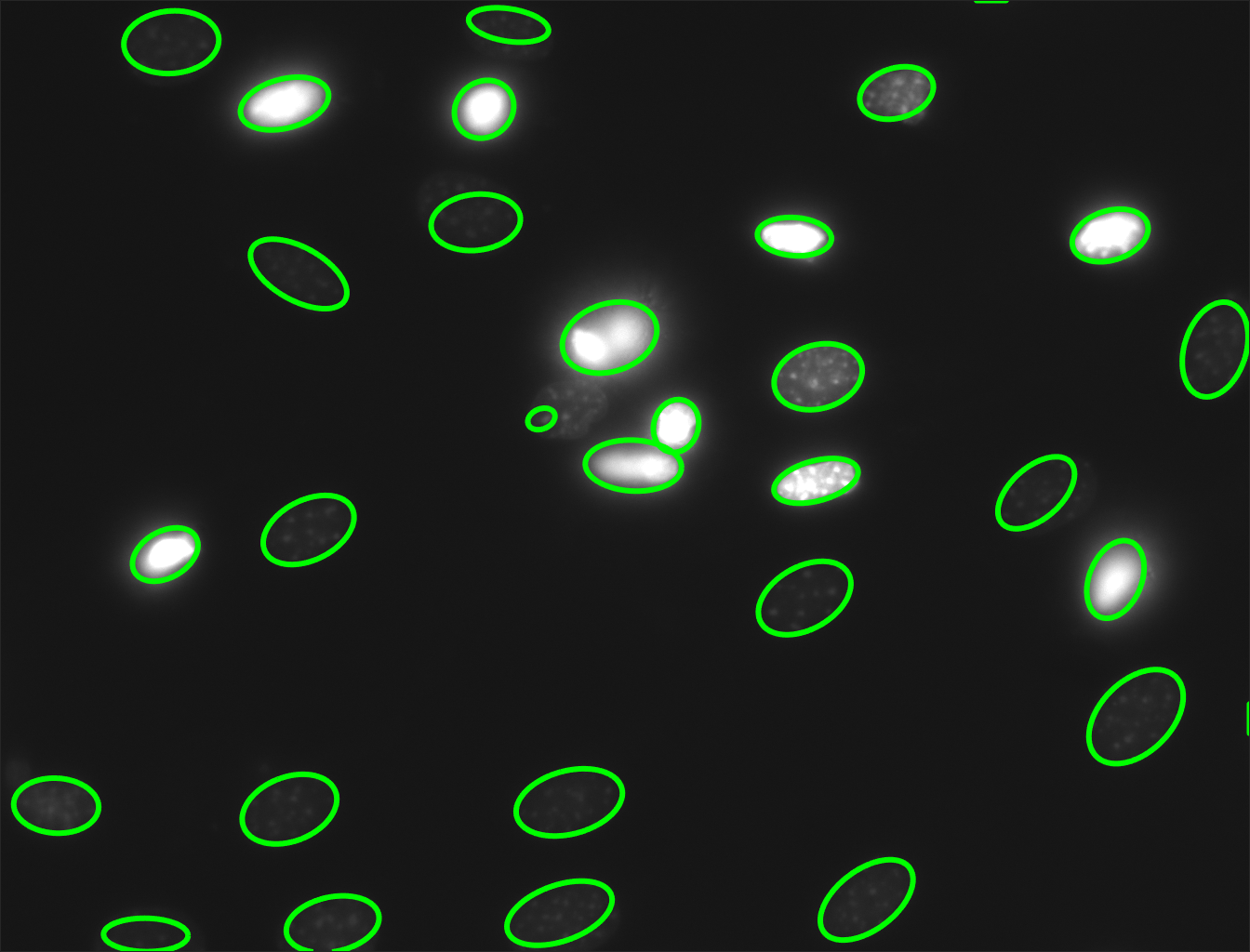}
	\end{minipage}
	
	\caption{\revisemajor{(Best viewed in color)} Cell counting in microscopy fluorescence images via BayFit.}
	\label{fig:fluorescence}
	\vskip -0.3cm
\end{figure}

\begin{figure}[t]	
\begin{minipage}{.48\textwidth}
  	\centering
\includegraphics[width=0.223\linewidth]{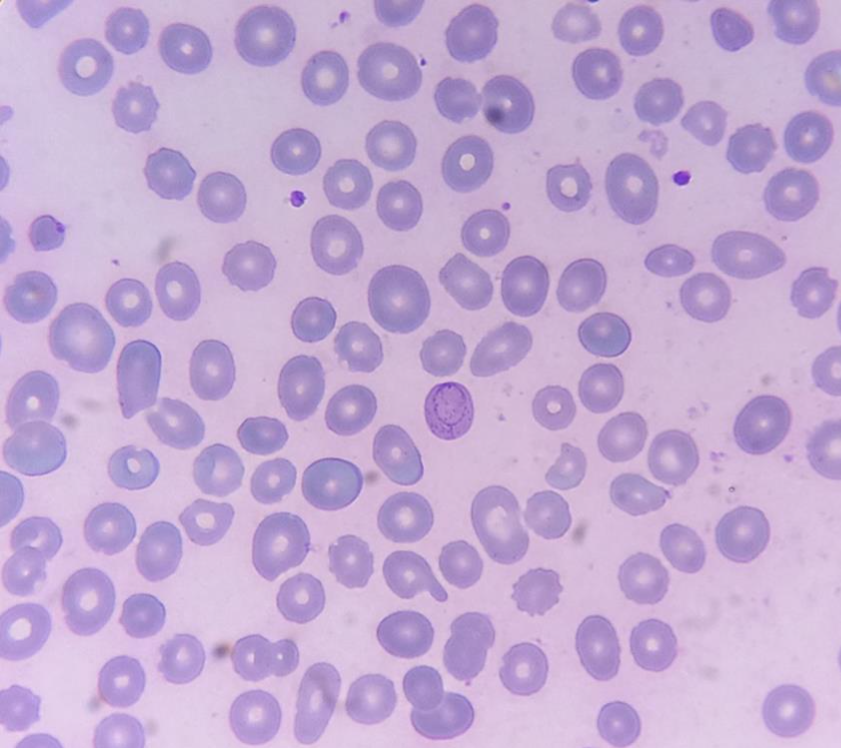}
\includegraphics[width=0.225\linewidth]{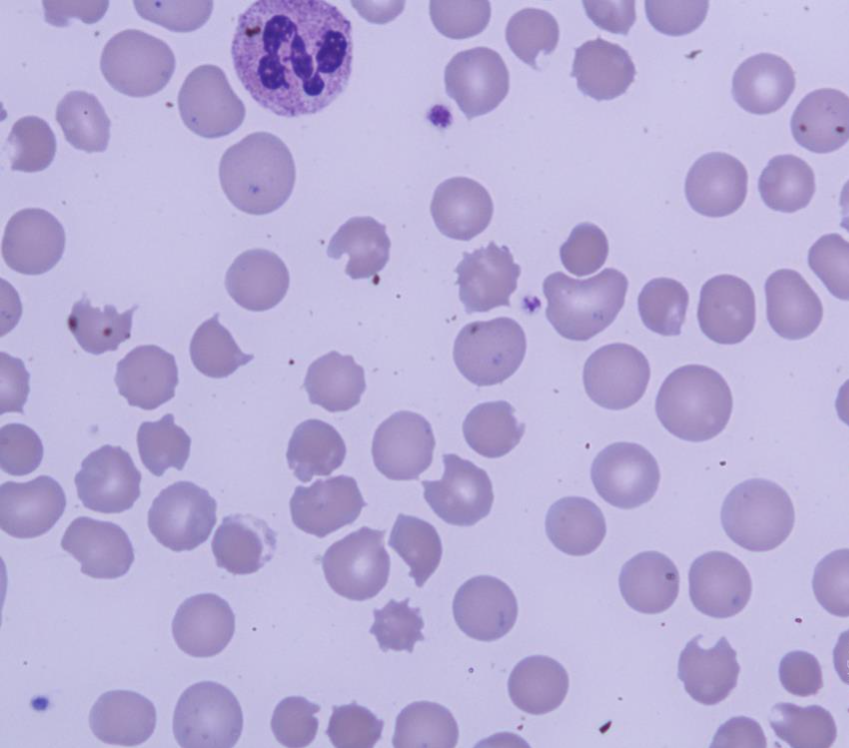}
\includegraphics[width=0.225\linewidth]{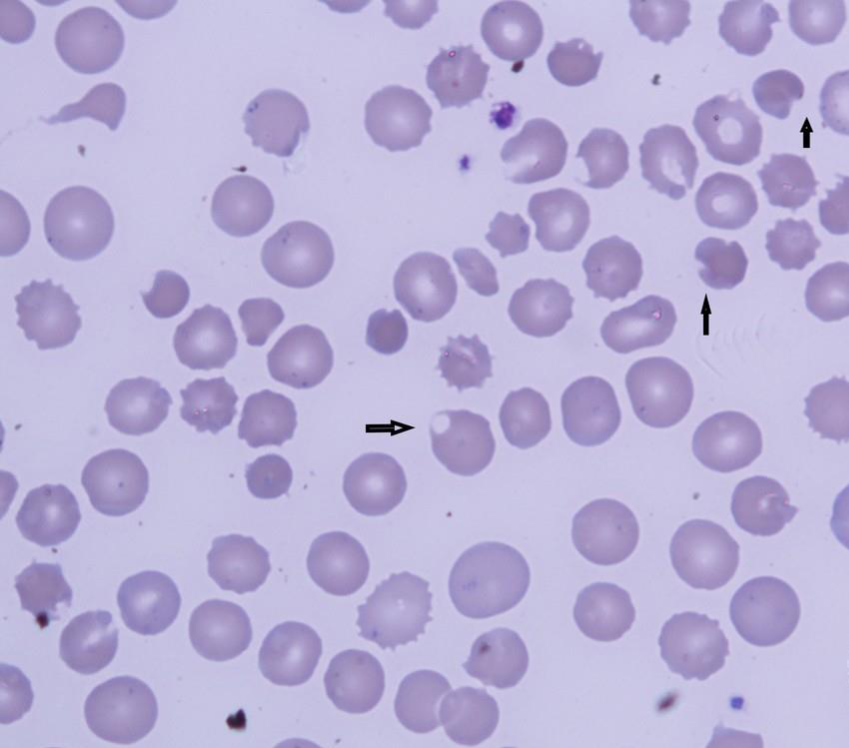}
\includegraphics[width=0.265\linewidth]{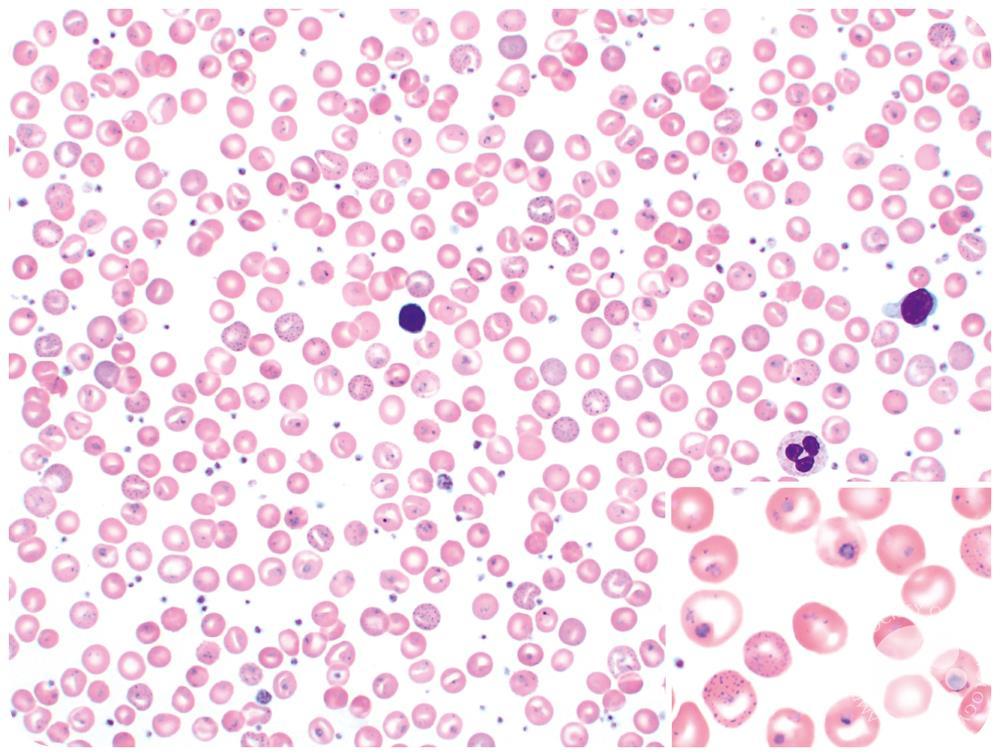}
\end{minipage}         
\vskip 0.1cm
\begin{minipage}{.48\textwidth}
\centering
\includegraphics[width=0.223\linewidth]{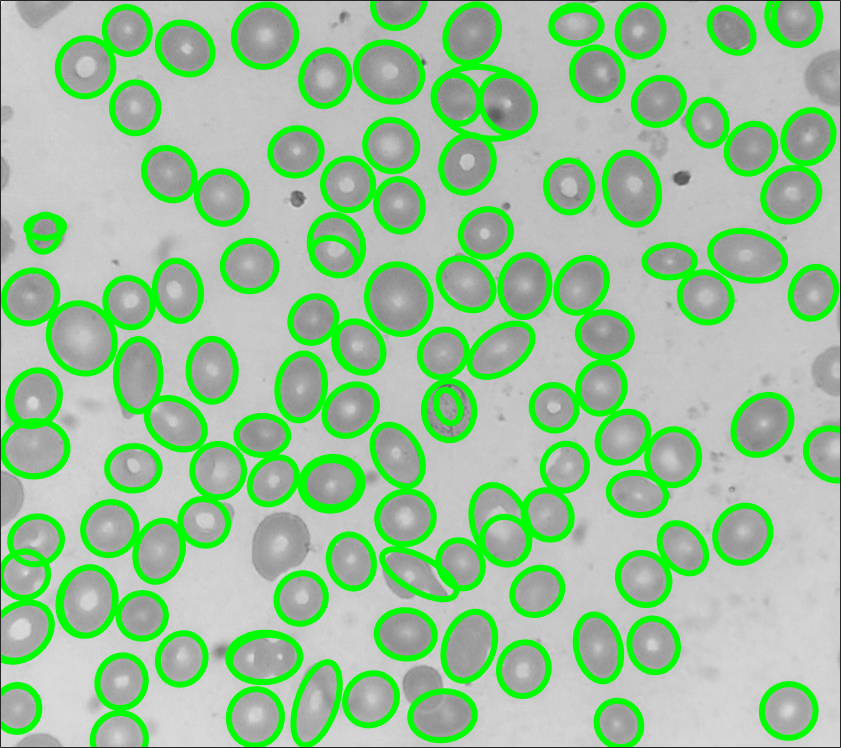}
\includegraphics[width=0.225\linewidth]{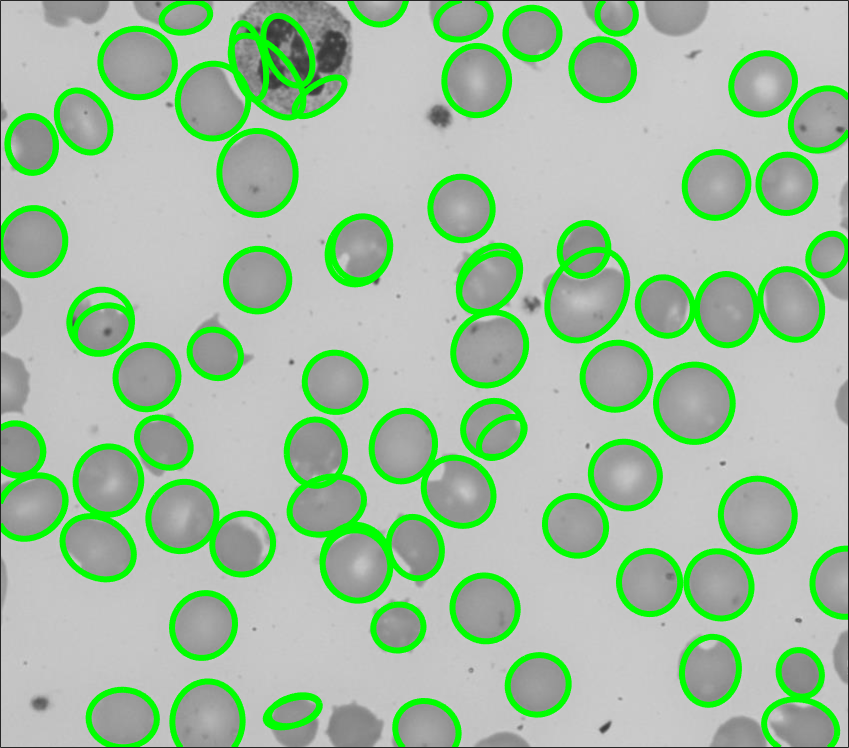}
\includegraphics[width=0.225\linewidth]{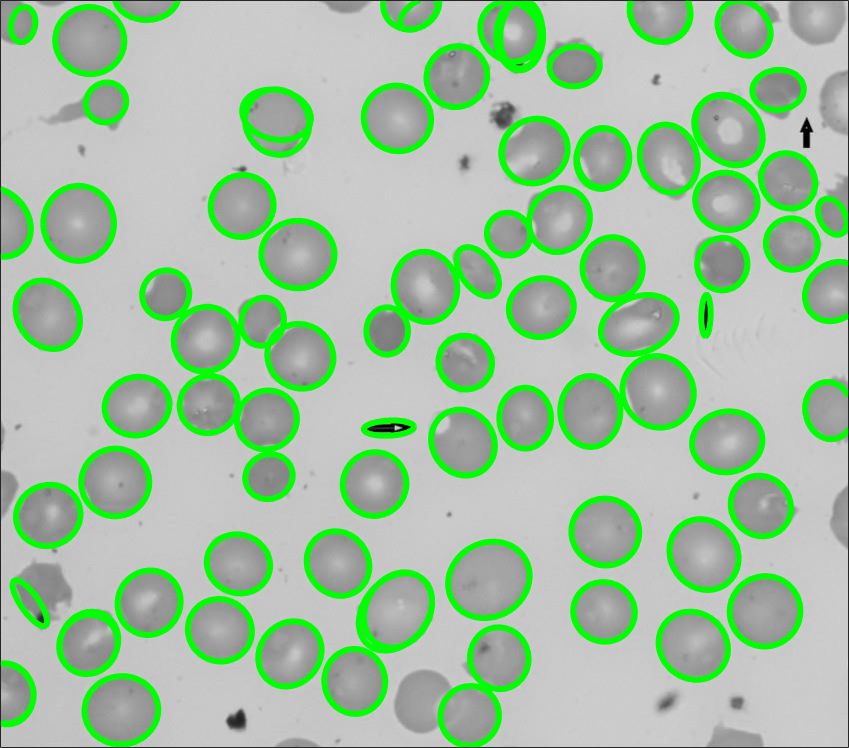}
\includegraphics[width=0.265\linewidth]{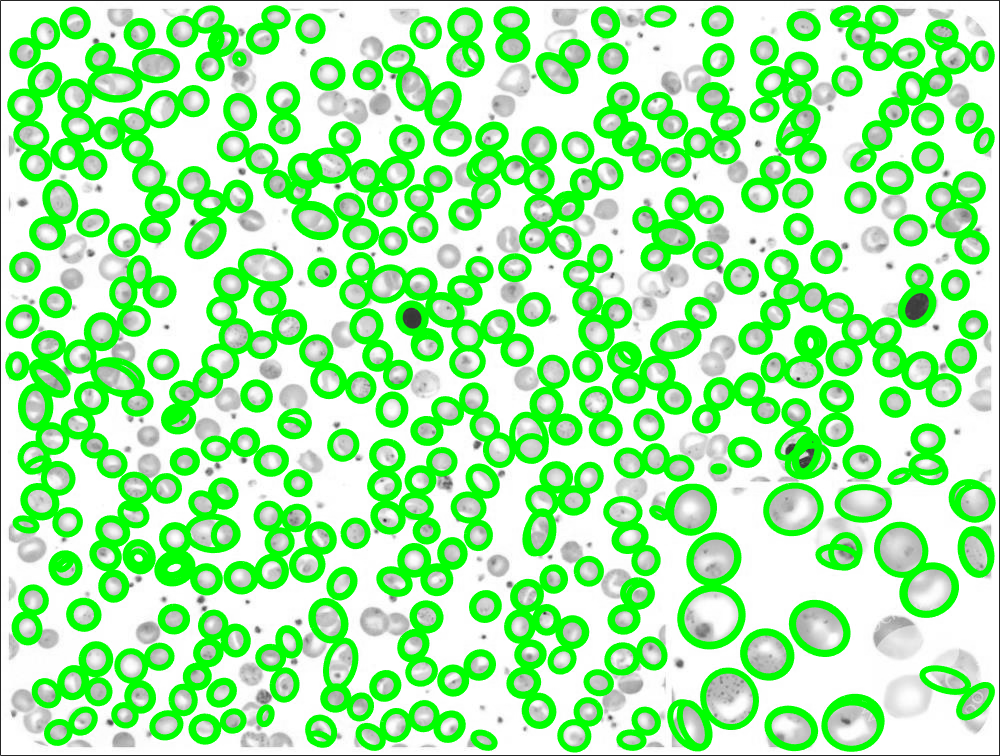}
\end{minipage}
\caption{\revise{Cell counting in microscopy blood cell images via BayFit.}}
\label{fig:blood}
	\vskip -0.3cm
\end{figure}

\revisemajor{\subsubsection{Application on Cell Counting}}
Finally, BayFit is deployed \revisemajor{for} cell counting in microscopy images, where multiple ellipse fittings are required. We conduct tests on both \emph{fluorescence} and \emph{blood} cell modalities, as depicted in Figs.~\ref{fig:fluorescence} and \ref{fig:blood}, respectively. The used fluorescence images are collected by~\cite{coelho2009nuclear}, while the blood cell images are sourced from the American Society of Hematology (ASH) image bank~\cite{ASH}. BayFit is applied to \revisemajor{the edge maps of these images} and returns ellipses one by one. In the bottom row of Fig.~\ref{fig:fluorescence}, BayFit not only delivers satisfactory approximations to various cell shapes but also provides highly accurate counts of the cell quantities. Even in challenging environments with a large number of irregular and small cells (Fig.~\ref{fig:blood}), BayFit still demonstrates its potential in automating the labor-intensive task of cell counting. It is noticeable that our ellipse fitting approach \revisemajor{in this context} is a general-purpose algorithm \revisemajor{that does not resort to} additional application-related information or post-verification, such as shape or appearance constraints, which may further facilitate the performance. \revisemajor{Other two applications including 3D reconstruction and elliptical object approximation are presented in Section 12.1 of the \emph{Supplementary Material}}.

\subsection{Assessments of 3D Ellipsoid Fitting}
In the subsequent sections, we evaluate the proposed method on the fitting of 3D ellipsoidal surfaces. Like~\cite{kesaniemi2017direct, lopez2017robust}, we quantitatively assess the fitting accuracy using two metrics: the offset bias $E_{\mathbf{c}}$ and the shape deviation $E_{\mathbf{a}}$. They are defined as
\begin{eqnarray}	
		E_{\mathbf{c}}=||\mathbf{c}_t-\hat{\mathbf{c}}||_2,~ E_{\mathbf{a}}=\frac{S_{\max}(\mathbf{\hat{A}^{-1}}\mathbf{A}_t)}{S_{\min}(\mathbf{\hat{A}^{-1}}\mathbf{A}_t)}-1, %
\end{eqnarray}where $\mathbf{c}_t$ and $\hat{\mathbf{c}}$, $\mathbf{A}_t$ and $\hat{\mathbf{A}}$ are the offset vectors and the affine matrices \revisemajor{representing} the ground truth and the fitted ellipsoids, respectively. $S_{\max}$, $S_{\min}$ \revisemajor{denote} the largest and the smallest singular values of the residual transformation $\hat{\mathbf{A}}^{-1}\mathbf{A}_t$. We conduct 100 independent trials for each test and report the average metrics in terms of $E_{\mathbf{c}}$ and $E_{\mathbf{a}}$.
\begin{table}[t]
\centering
\normalsize
\caption{Quantitative comparisons of 3D ellipsoidal surface fitting on outlier-contaminated measurements.}
\vskip -0.2cm
\scalebox{0.83}{
\begin{tabular}{|c|c|c|c|c|c|}
\hline
				{\textbf{Outlier} (\%)}&\textbf{Metric} &Tukey\cite{rousseeuw1991tutorial}&Huber\cite{huber2011robust}&RIX\cite{lopez2017robust}&BayFit\cr
				\hline
				\multirow{2}{*}{5} &$E_\mathbf{c}$&0.85&0.54&\best{0.30}&0.50\\ 
				&$E_\mathbf{a}$&0.06&\best{0.04}&0.07&0.05\\  \cline{1-6}
				\multirow{2}{*}{10}&$E_\mathbf{c}$ &0.81&0.55&\best{0.51}&0.55\\
				&$E_\mathbf{a}$&0.06&\best{0.04}&0.15&0.05\\ \cline{1-6}
				\multirow{2}{*}{20} &$E_\mathbf{c}$&1.05&0.84&1.61&\best{0.53}\\
				&$E_\mathbf{a}$&0.07&0.06&0.29&\best{0.05}\\ \cline{1-6}
				\multirow{2}{*}{30}&$E_\mathbf{c}$ &1.61&1.44&2.24&\best{0.54}\\
				&$E_\mathbf{a}$&0.11&0.11&0.34&\best{0.05}\\ \cline{1-6}
				\multirow{2}{*}{40}&$E_\mathbf{c}$ &2.06&2.00&6.39&\best{0.56}\\
				&$E_\mathbf{a}$&0.14&0.16&0.48&\best{0.05}\\ \cline{1-6}
				\multirow{2}{*}{50}&$E_\mathbf{c}$ &2.22&2.33&12.48&\best{0.54}\\
				&$E_\mathbf{a}$&0.17&0.22&0.55&\best{0.05}\\ \cline{1-6}
				\multirow{2}{*}{60}&$E_\mathbf{c}$ &2.31&3.74&13.39&\best{0.52}\\
				&$E_\mathbf{a}$&0.21&0.28&0.64&\best{0.05}\\ 
				\hline
			\end{tabular}
		}
		\label{tab:ellipsoid_outlier}
		\vskip -0.1cm
	\end{table}

\begin{table*}[!htbp]
	\vskip -0.2cm
		\centering
		\normalsize
		\caption{Quantitative comparisons on noisy measurements regarding 3D ellipsoidal surface fitting. The proposed method BayFit demonstrates the best overall fitting accuracy.
		}
		\vskip -0.2cm
				\scalebox{0.91}{
					\begin{tabular}{|c|c|c|c|c|c|c|c|c|c|}
						\hline
						\diagbox{\textbf{Noise} (\%)}{\textbf{Method}}&\textbf{Metric} &DLS\cite{li2004least}&HES\cite{kesaniemi2017direct}&MQF\cite{birdal2019generic}&Koop\cite{vajk2003identification}&Taubin\cite{taubin1991estimation}&GF\cite{bektas2015least}&RIX\cite{lopez2017robust}&BayFit (Ours)\cr
						\hline
						\multirow{2}{*}{5} &$E_\mathbf{c}$&3.43&3.42&3.47&4.06&3.89&1.31&\best{0.67}&1.03\\
						&$E_\mathbf{a}$&0.45&0.46&0.57&0.74&0.63&\best{0.14}&0.15&\best{0.14}\\  \cline{1-10}
						\multirow{2}{*}{10}&$E_\mathbf{c}$ &3.92&3.90&4.14&5.20&4.83&1.90&\best{1.17}&1.33\\
						&$E_\mathbf{a}$&0.47&0.48&0.65&0.88&0.71&0.21&0.23&\best{0.20}\\ \cline{1-10}
						\multirow{2}{*}{15} &$E_\mathbf{c}$&4.51&4.49&4.11&6.12&5.87&2.14&1.66&\best{1.58}\\
						&$E_\mathbf{a}$&0.48&0.49&0.76&1.11&0.86&0.29&0.29&\best{0.25}\\ \cline{1-10}
						\multirow{2}{*}{20}&$E_\mathbf{c}$ &4.61&4.60&3.62&7.83&6.86&3.23&2.20&\best{2.01}\\
						&$E_\mathbf{a}$&0.46&0.47&0.84&1.43&0.94&0.53&0.33&\best{0.32}\\ \cline{1-10}
						\multirow{2}{*}{25}&$E_\mathbf{c}$ &4.85&4.84&5.10&8.97&8.81&4.21&2.68&\best{2.16}\\
						&$E_\mathbf{a}$&0.44&0.46&1.09&1.65&1.18&0.94&\best{0.38}&0.40\\  \cline{1-10}
						\hline
					\end{tabular}
				}
				\label{tab:ellipsoid_noise}
				\vskip -0.3cm
			\end{table*}
			
\begin{figure}[t]
	\centering
	\subfigure[500 data points]{
		\includegraphics[width=0.11\textwidth]{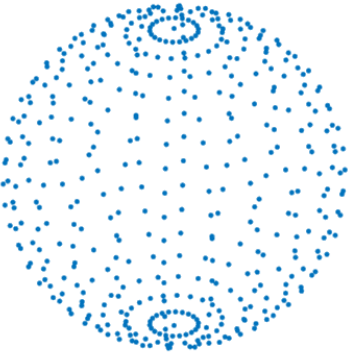}
		\includegraphics[width=0.11\textwidth]{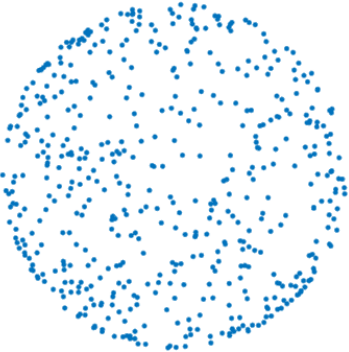}
	}
	\subfigure[1,000 data points]{
		\includegraphics[width=0.11\textwidth]{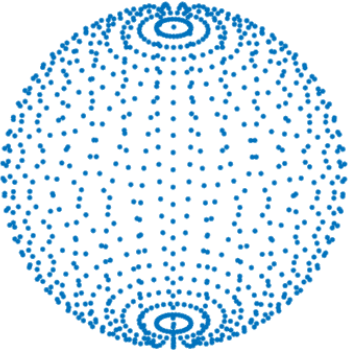}
		\includegraphics[width=0.11\textwidth]{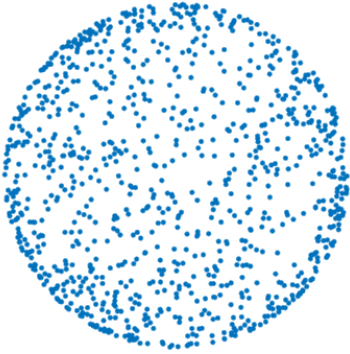}
	}
	\vskip -0.3cm
	\caption{\revise{Sampling with parametric (first) and random (second) schemes under different point sizes, where the first sampling manner generates more evenly coverage.}}
	\label{fig:sampling}
	\vskip -0.3cm
\end{figure}				
\begin{figure}[!htbp]
	\centering
\begin{minipage}{0.116\textwidth}
\includegraphics[width=0.85\textwidth]{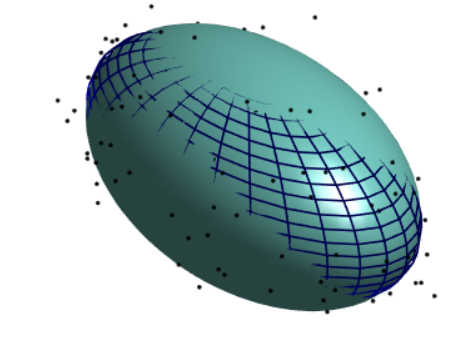}
\end{minipage}
\begin{minipage}{0.116\textwidth}
\includegraphics[width=0.85\textwidth]{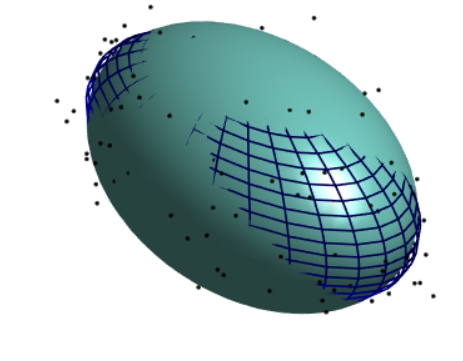}
\end{minipage}
\begin{minipage}{0.116\textwidth}
\includegraphics[width=0.85\textwidth]{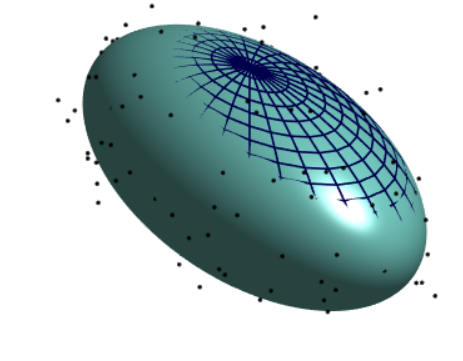}
\end{minipage}
\begin{minipage}{0.116\textwidth}
\includegraphics[width=0.9\textwidth]{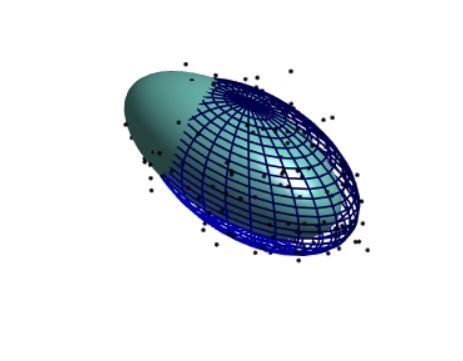}
\end{minipage}

\begin{minipage}{0.116\textwidth}
\centering
{\scriptsize{(a) DLS~\cite{li2004least}}}
\end{minipage}
\begin{minipage}{0.116\textwidth}
	\centering
{\scriptsize{(b) HES~\cite{kesaniemi2017direct}}}
\end{minipage}
\begin{minipage}{0.116\textwidth}
	\centering
{\scriptsize{(c) MQF~\cite{birdal2019generic}}}
\end{minipage}
\begin{minipage}{0.116\textwidth}
	\centering
{\scriptsize{(d) Koop~\cite{vajk2003identification}}}
\end{minipage}

\begin{minipage}{0.116\textwidth}
\includegraphics[width=0.9\textwidth]{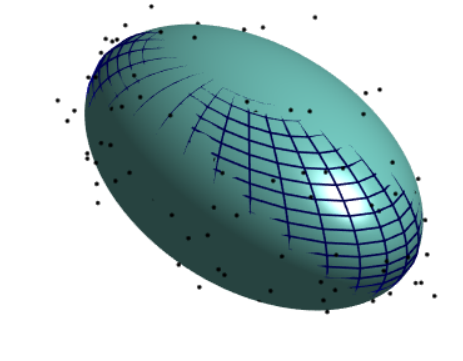}
\end{minipage}
\begin{minipage}{0.116\textwidth}
\includegraphics[width=0.85\textwidth]{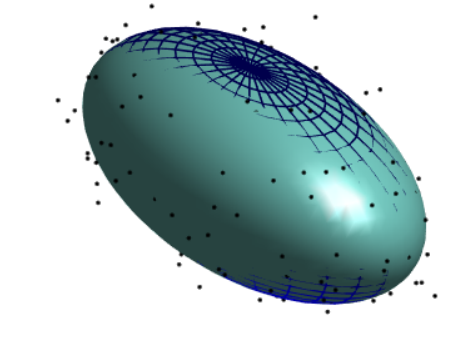}
\end{minipage}
\begin{minipage}{0.116\textwidth}
\includegraphics[width=0.85\textwidth]{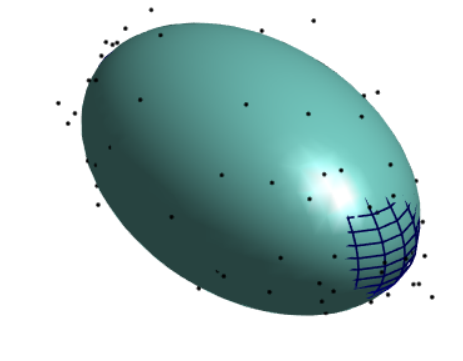}
\end{minipage}
\begin{minipage}{0.116\textwidth}
\includegraphics[width=0.85\textwidth]{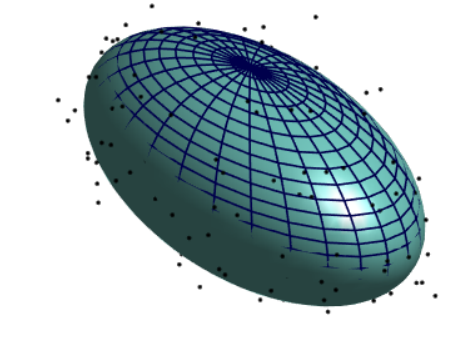}
\end{minipage}

\begin{minipage}[b]{0.116\textwidth}
	\centering
	{\scriptsize{(e) Taubin~\cite{taubin1991estimation}}}
\end{minipage}
\begin{minipage}[b]{0.116\textwidth}
	\centering
	{\scriptsize{(f) GF~\cite{bektas2015least}}}
\end{minipage}
\begin{minipage}[b]{0.116\textwidth}
	\centering
	{\scriptsize{(g) RIX~\cite{lopez2017robust}}}
\end{minipage}
\begin{minipage}[b]{0.116\textwidth}
	\centering
	{\scriptsize{(h) BayFit}}
\end{minipage}
\vskip -0.2cm
\caption{Qualitative fitting results by different methods under 10\% Gaussian noise and 1\% outliers, where the net ellipsoid represents the ground truth surface. As observed, BayFit outperforms competitors with higher approximation accuracy.}\label{fig:ellipsoid_noise_exam}
\vskip -0.3cm
\end{figure}

\begin{figure*}[t]
	\centering
	\begin{minipage}{.01\textwidth} 
		\centering
		{\rotatebox{90}{$\eta=60\%$}} 
	\end{minipage}
	\begin{minipage}{.97\textwidth}
		\centering
		\includegraphics[width=0.192\textwidth]{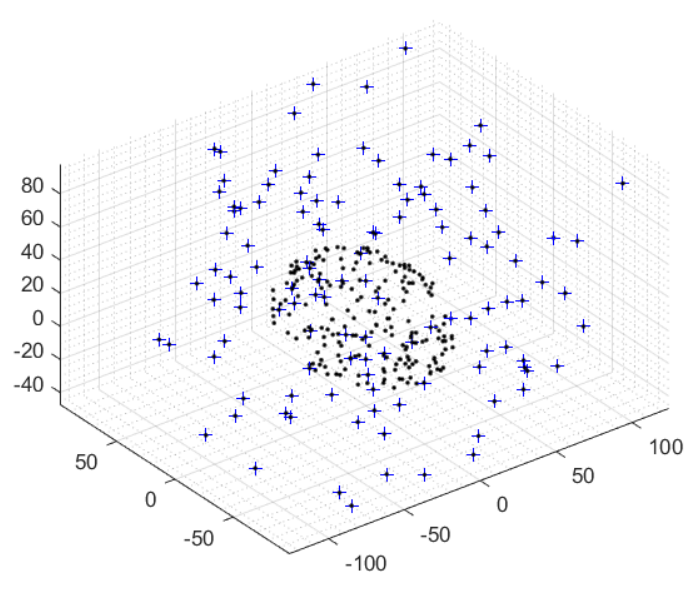}
		\includegraphics[width=0.192\textwidth]{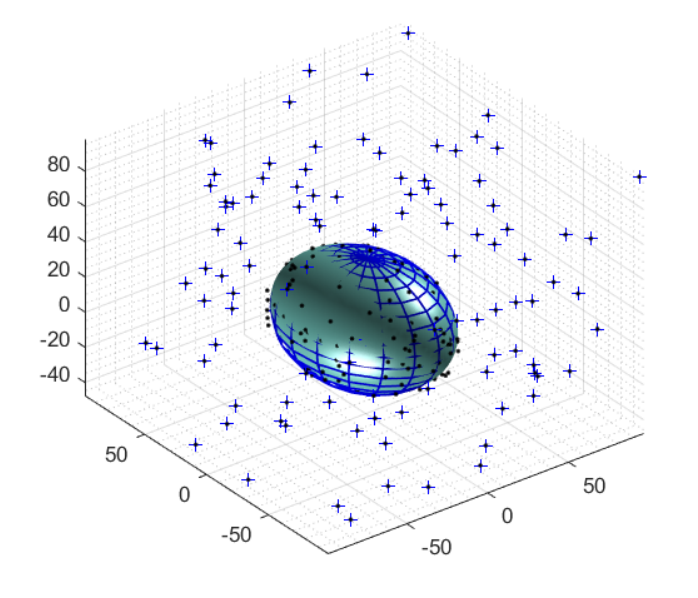}
		\includegraphics[width=0.192\textwidth]{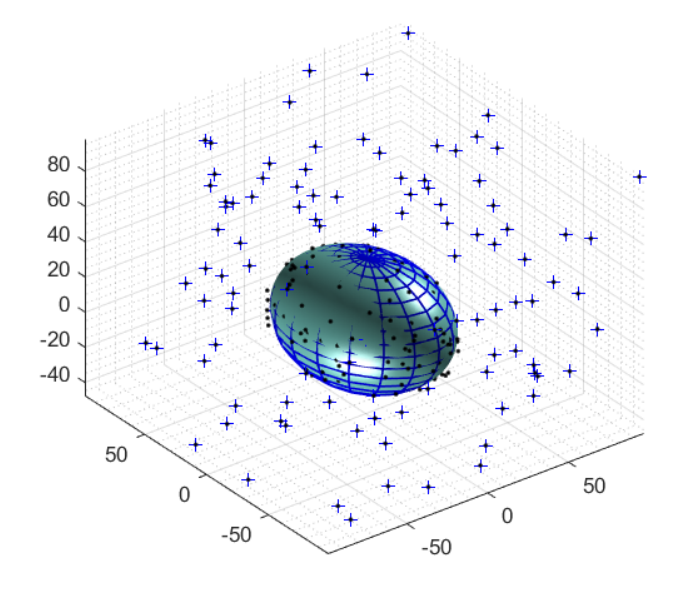}
		\includegraphics[width=0.192\textwidth]{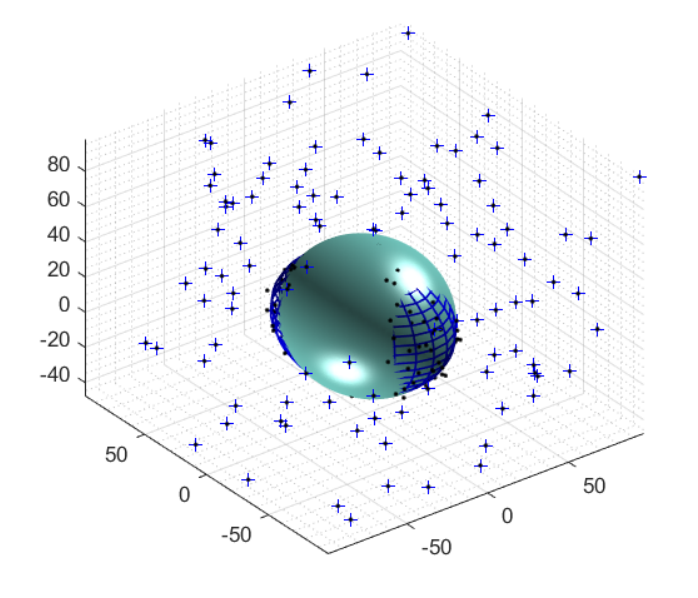}
		\includegraphics[width=0.192\textwidth]{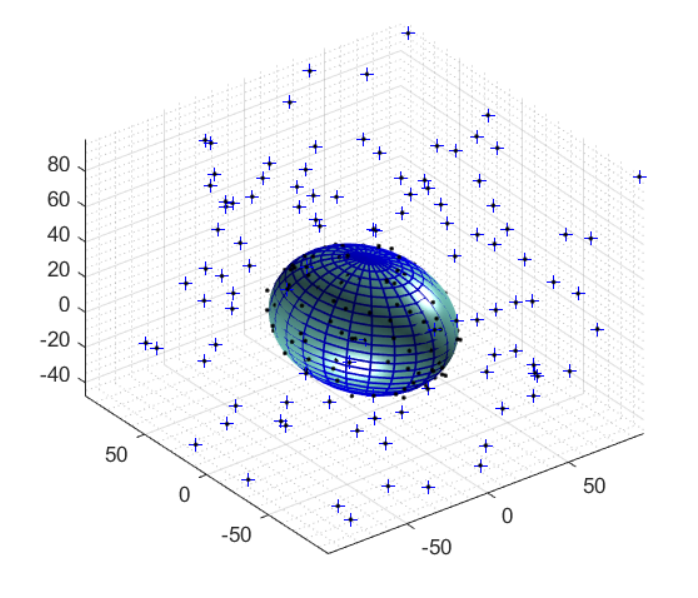}	
	\end{minipage}
	\vskip -0.11cm
	\begin{minipage}{.01\textwidth} 
		\centering
		{\rotatebox{90}{$\eta=80\%$}} 
	\end{minipage}
	\begin{minipage}{.97\textwidth}
		\centering
		\includegraphics[width=0.192\textwidth]{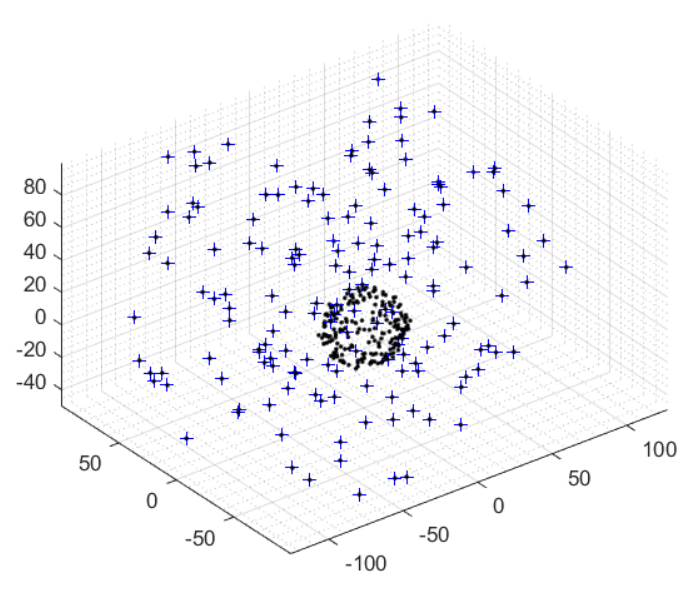}
		\includegraphics[width=0.192\textwidth]{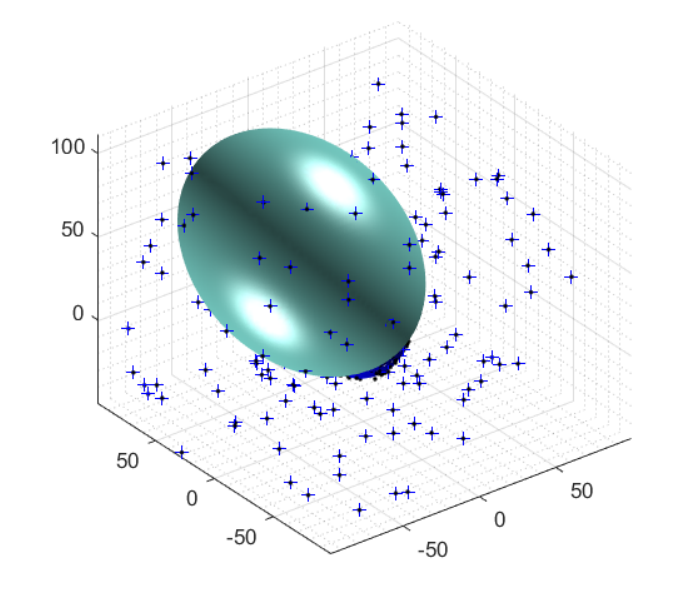}
		\includegraphics[width=0.192\textwidth]{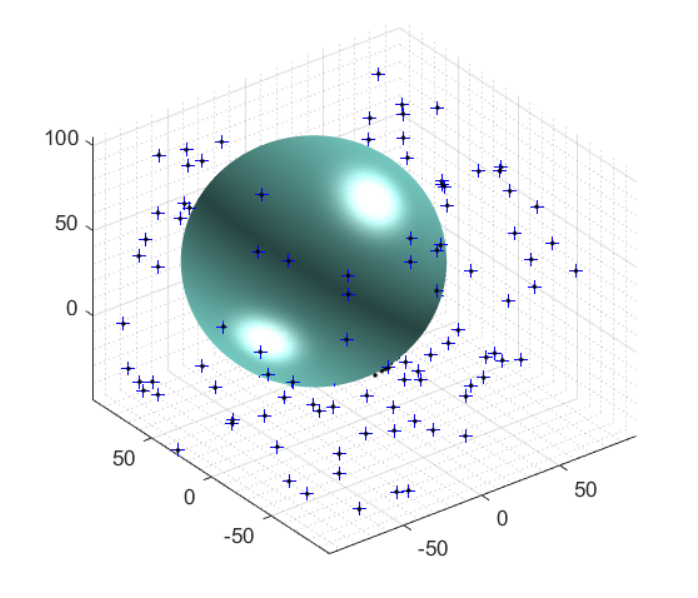}
		\includegraphics[width=0.192\textwidth]{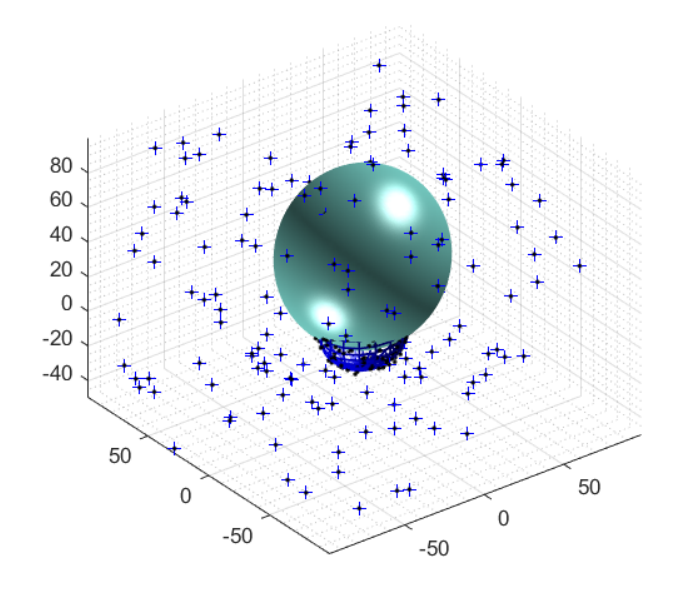}
		\includegraphics[width=0.192\textwidth]{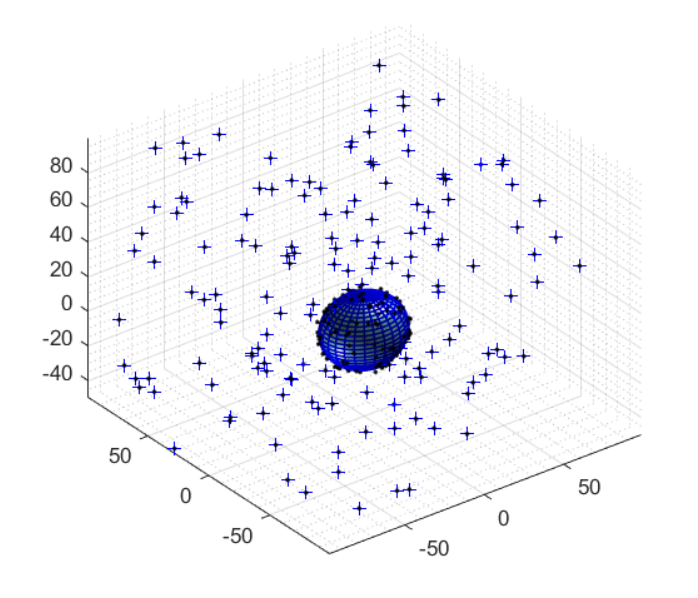}	
	\end{minipage}
	\vskip -0.11cm
	\begin{minipage}{.01\textwidth} 
		\centering
		{\rotatebox{90}{$\eta=100\%$}} 
	\end{minipage}
	\begin{minipage}{.97\textwidth}
		\centering
		\subfigure[Input]{\includegraphics[width=0.192\textwidth]{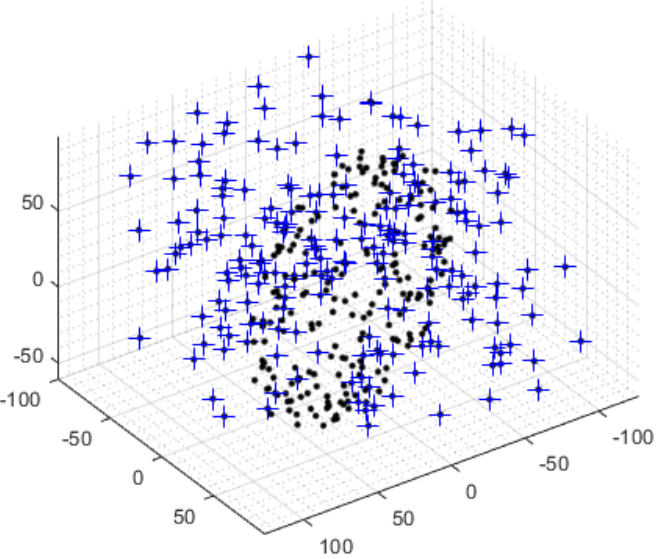}}
		\subfigure[{Tukey~\cite{rousseeuw1991tutorial}}]{\includegraphics[width=0.192\textwidth]{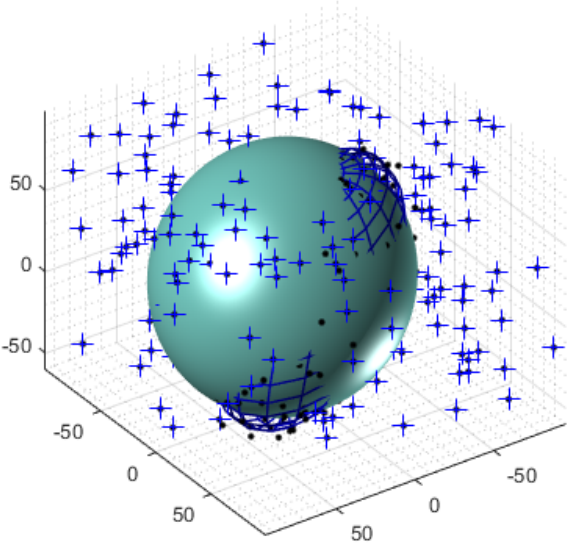}}
		\subfigure[{Huber~\cite{huber2011robust}}]{\includegraphics[width=0.192\textwidth]{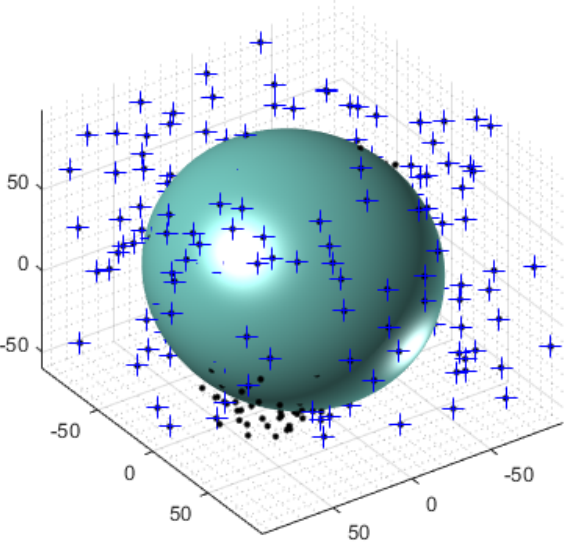}}
		\subfigure[RIX~\cite{lopez2017robust}]{\includegraphics[width=0.192\textwidth]{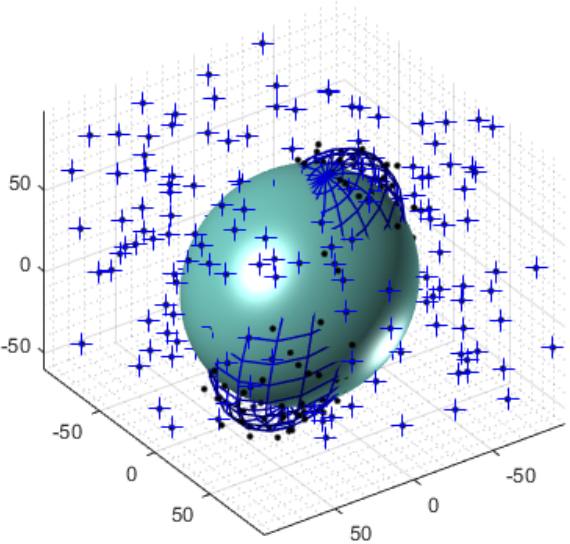}}
		\subfigure[BayFit]{\includegraphics[width=0.192\textwidth]{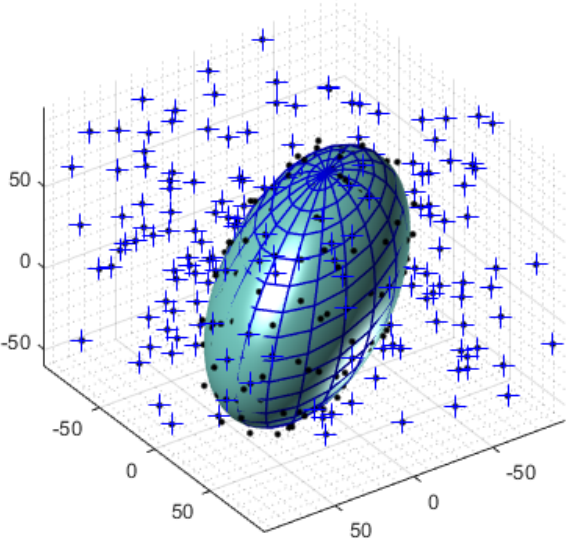}}	
	\end{minipage}
\vskip -0.2cm
\caption{{Qualitative comparisons of 3D ellipsoid fitting on outlier-contaminated scenes ('\revise{+}' represents the outlier). When the outlier percentage $\eta$ increases from $60\%$ to $100\%$, competitors lead to significant deviations, whereas BayFit still keeps quite stable fittings.}}\label{fig:outlier_exam}
\label{fig:axis}
\vskip -0.3cm
\end{figure*}				
\begin{table}[t]
	\centering
	\caption{\revise{Gradually increased volume occlusion $O(t)$ $(\%)$.}}
	\vskip -0.3cm
	\begin{tabular}{lcccccc}
		\toprule
		Cutting plane~~$t$&4 & 3 & 2 & 1 & 0 & -1 \\
		\midrule
		Occlusion part~~$O(t)$&45.2   &41.6   &40.4   &42.8  & 50.0   &63.2    \\
		\bottomrule
	\end{tabular}
	\label{table:occlusion}
	\vskip -0.3cm
\end{table}	
\begin{figure}[t]
	\centering
	\includegraphics[width=\linewidth]{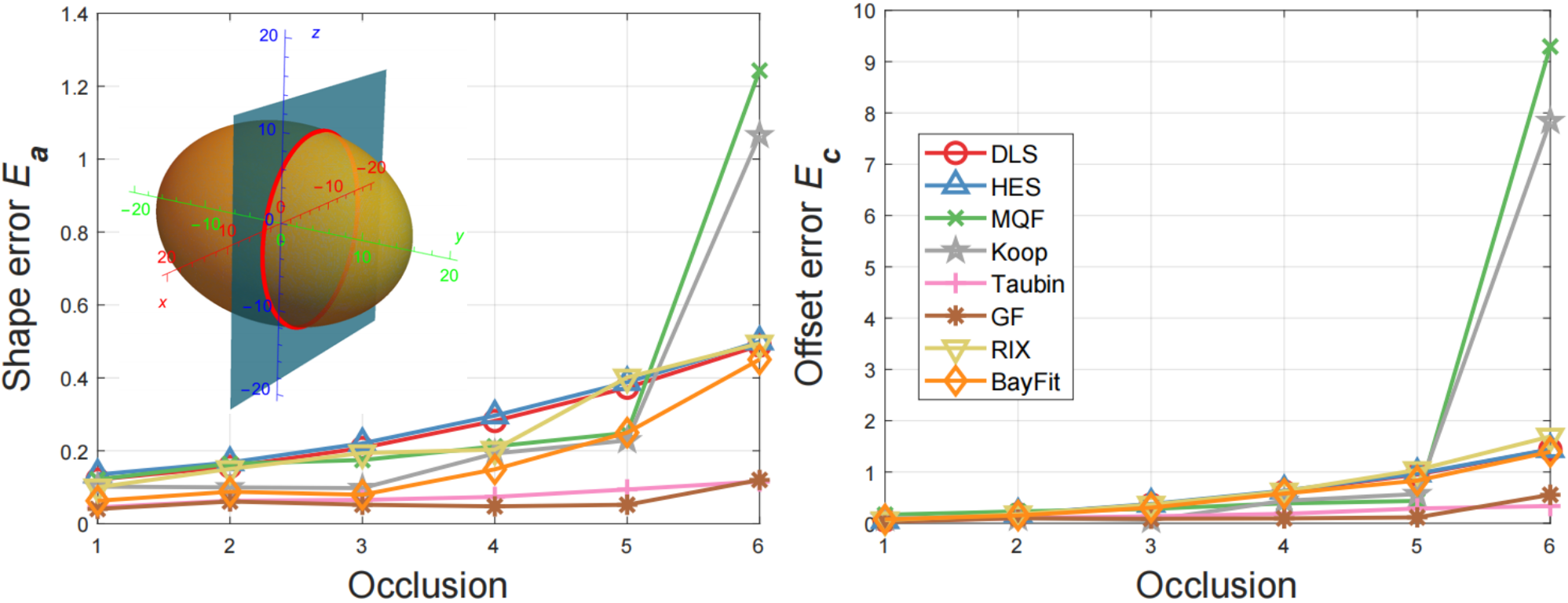}
	\vskip -0.3cm
	\caption{\revise{Occlusion tests on 3D ellipsoid fitting with the cutting plane moving from $y=1$ to $y=-1$. The values in $x$-axis denote the occlusion length of the ellipsoidal axis along the $y$ direction. Inset: Schematic description of the occlusion ellipsoidal surface, where the red ellipse represents the intersection line between the ellipsoidal surface and the orthogonal cutting plane $y=t$.}}
	\label{fig:ell_plane}
	\vskip -0.5cm
\end{figure}

\begin{figure*}[t]
	\vskip -0.25cm
	\centering
	\subfigure[{DLS~\cite{li2004least}}]{
		\begin{minipage}[b]{0.11\textwidth}
			\includegraphics[width=1\textwidth]{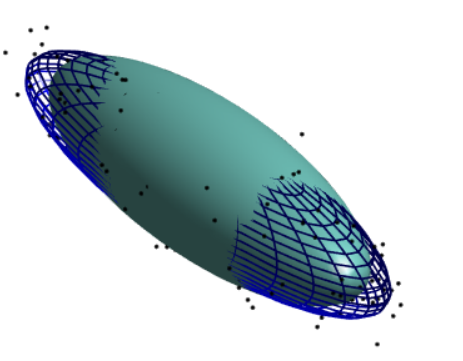}\\
			\includegraphics[width=1\textwidth]{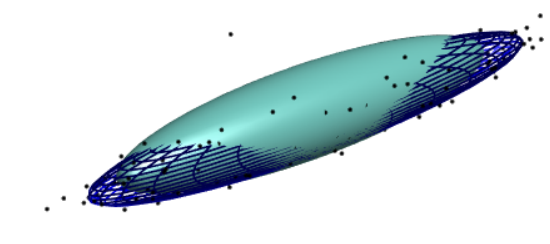}\\
			\includegraphics[width=1\textwidth]{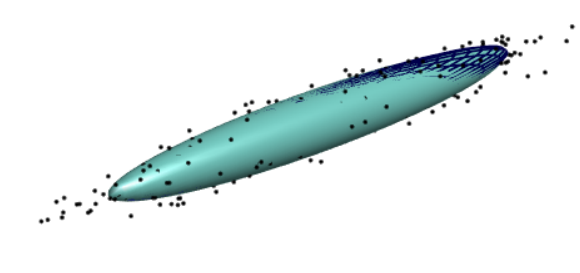}
		\end{minipage}
	}
	\subfigure[{HES~\cite{kesaniemi2017direct}}]{
		\begin{minipage}[b]{0.11\textwidth}
			\includegraphics[width=1\textwidth]{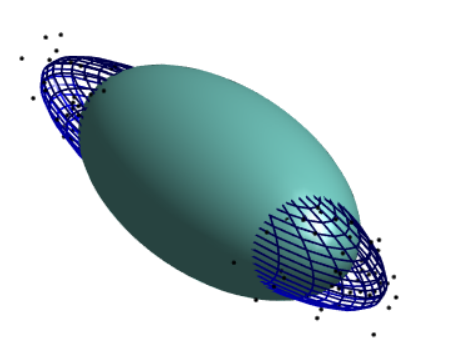}\\
			\includegraphics[width=1\textwidth]{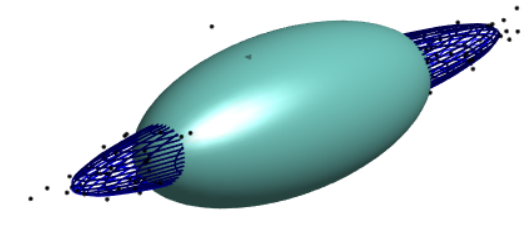}\\
			\includegraphics[width=1\textwidth]{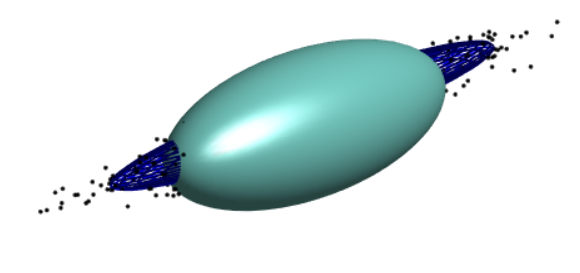}
		\end{minipage}
	}
	\subfigure[{MQF~\cite{birdal2019generic}}]{
		\begin{minipage}[b]{0.11\textwidth}
			\includegraphics[width=1\textwidth]{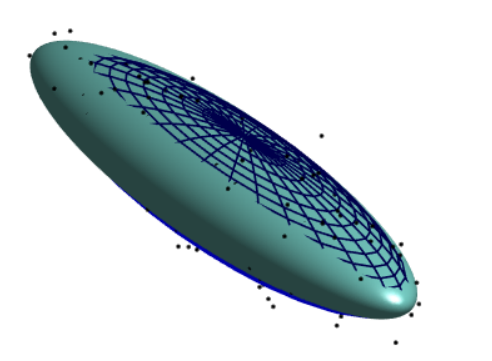}\\
			\includegraphics[width=1\textwidth]{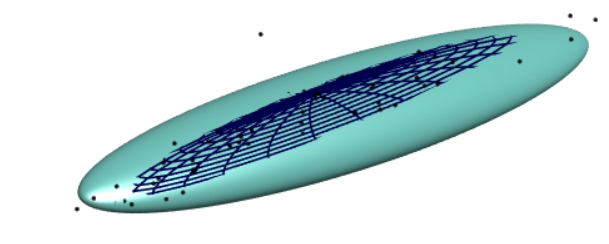}\\
			\includegraphics[width=1\textwidth]{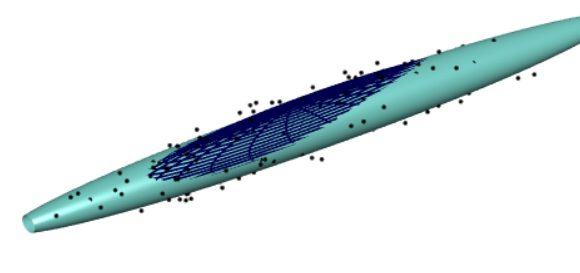}
		\end{minipage}
	}
	\subfigure[{Koop~\cite{vajk2003identification}}]{
		\begin{minipage}[b]{0.11\textwidth}
			\includegraphics[width=1\textwidth]{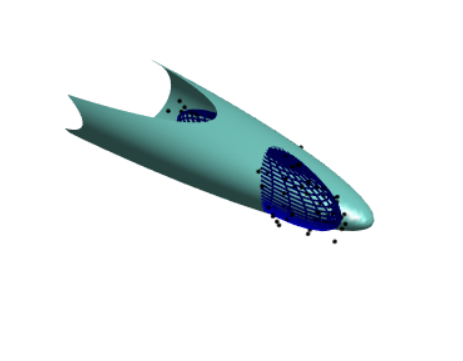}\\
			\includegraphics[width=1\textwidth]{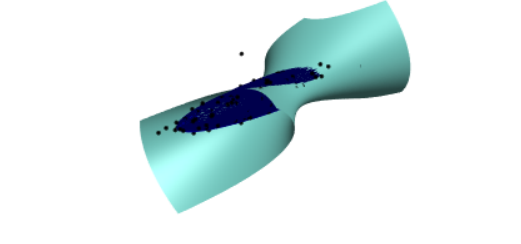}\\
			\includegraphics[width=1\textwidth]{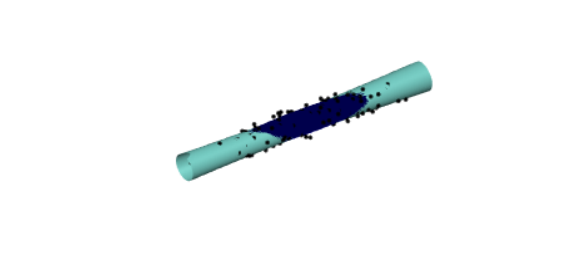}
		\end{minipage}
	}
	\subfigure[{Taubin~\cite{taubin1991estimation}}]{
		\begin{minipage}[b]{0.11\textwidth}
			\includegraphics[width=1\textwidth]{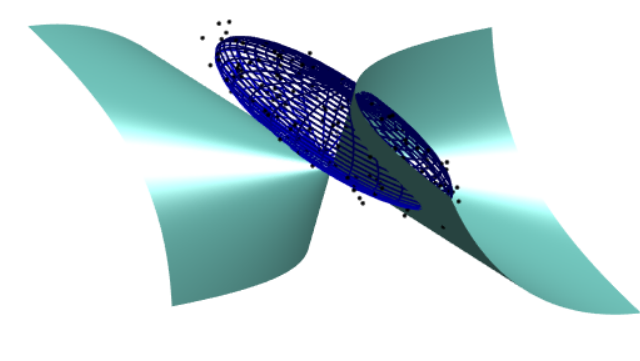}\\
			\includegraphics[width=1\textwidth]{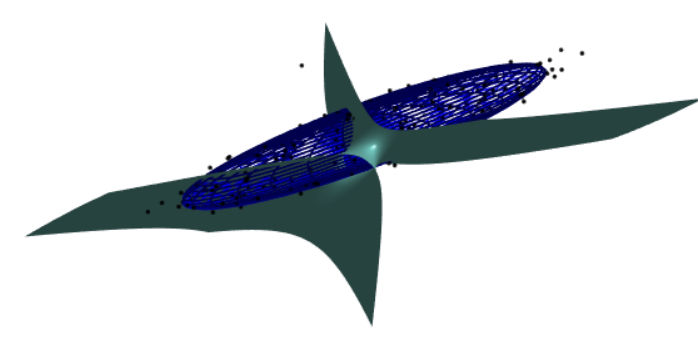}\\
			\includegraphics[width=1\textwidth]{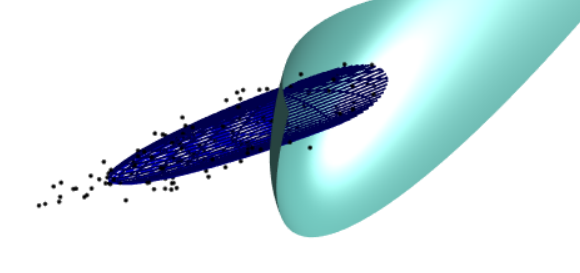}
		\end{minipage}
	}
	\subfigure[{GF~\cite{bektas2015least}}]{
		\begin{minipage}[b]{0.11\textwidth}
			\includegraphics[width=1\textwidth]{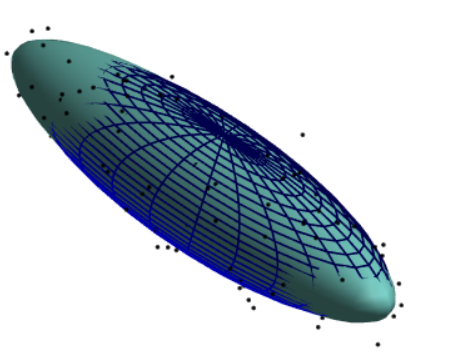}\\
			\includegraphics[width=1\textwidth]{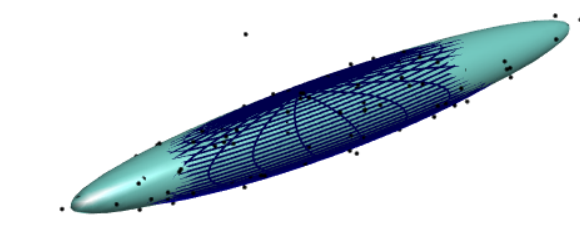}\\
			\includegraphics[width=1\textwidth]{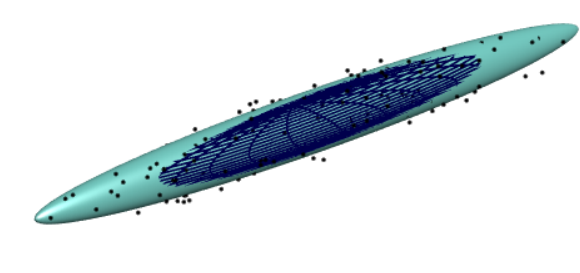}
		\end{minipage}
	}
	\subfigure[{RIX~\cite{lopez2017robust}}]{
		\begin{minipage}[b]{0.11\textwidth}
			\includegraphics[width=1\textwidth]{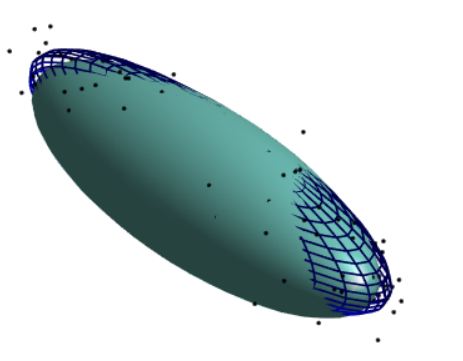}\\
			\includegraphics[width=1\textwidth]{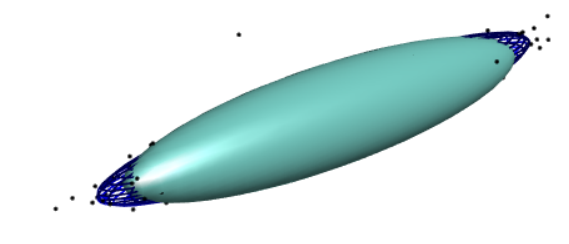}\\
			\includegraphics[width=1\textwidth]{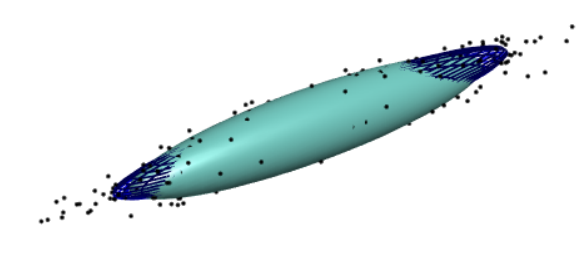}
		\end{minipage}
	}
	\subfigure[{Ours}]{
		\begin{minipage}[b]{0.11\textwidth}
			\includegraphics[width=1\textwidth]{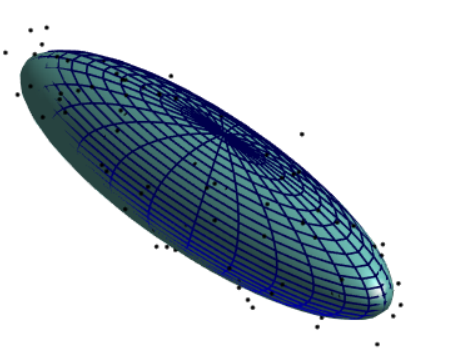}\\
			\includegraphics[width=1\textwidth]{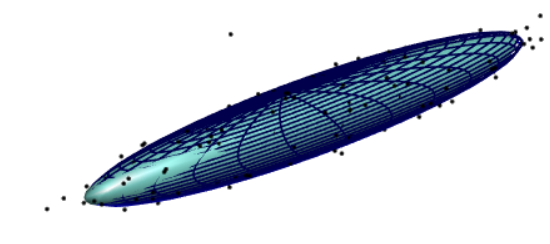}\\
			\includegraphics[width=1\textwidth]{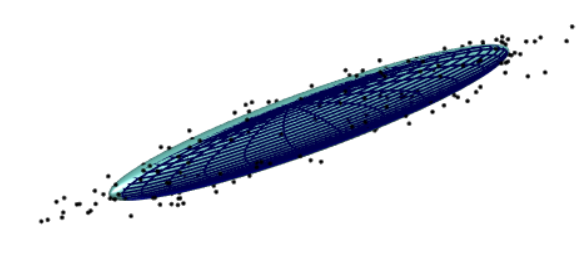}
		\end{minipage}
	}
	\vskip -0.2cm
	\caption{Qualitative comparisons on the fitting of challenging ellipsoids with large axis ratio $r_{ax}$. From the first to last rows, $r_{ax}$ is equivalent to $5, 8, 10$ separately. Several methods tend to generate non-ellipsoidal quadrics, indicating their limitations in representing thin or elongated ellipsoids, whereas BayFit ensures consistent ellipsoid-specific results and acceptable fitting accuracy simultaneously.
	}
	\label{fig:axis}
	\vskip -0.3cm
\end{figure*}			
\begin{figure}[t]
\centering
\includegraphics[width=0.24\textwidth]{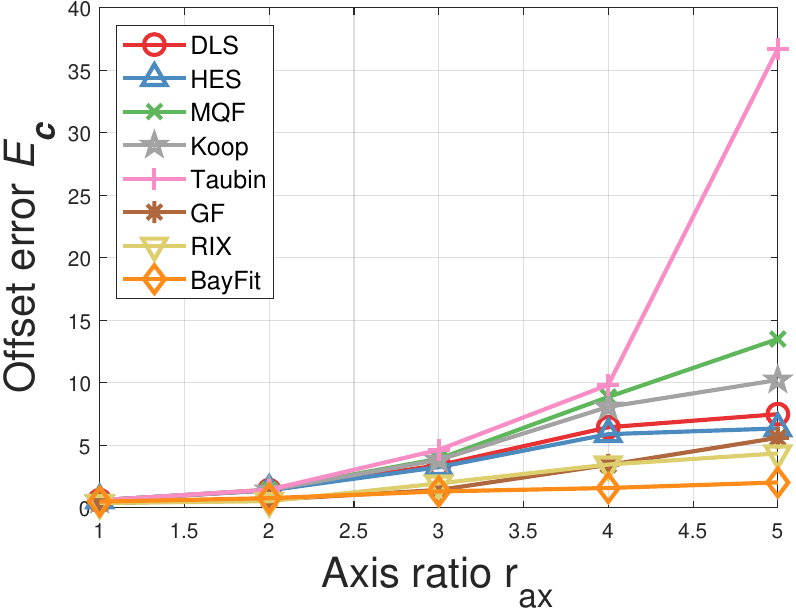}
\includegraphics[width=0.242\textwidth]{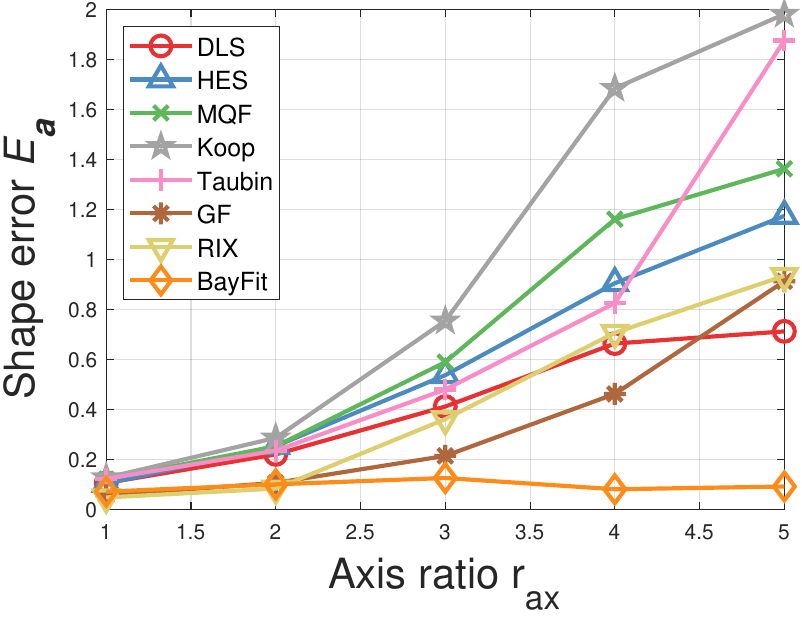}
\vskip -0.3cm
\caption{Test on challenging scenarios with various axis ratios. BayFit exhibits enhanced robustness and stability against large axis ratios. 	
}\label{fig:axis_ratio}
\vskip -0.3cm
\end{figure}	

\begin{figure}[t]
	\centering
	\includegraphics[width=0.20\textwidth]{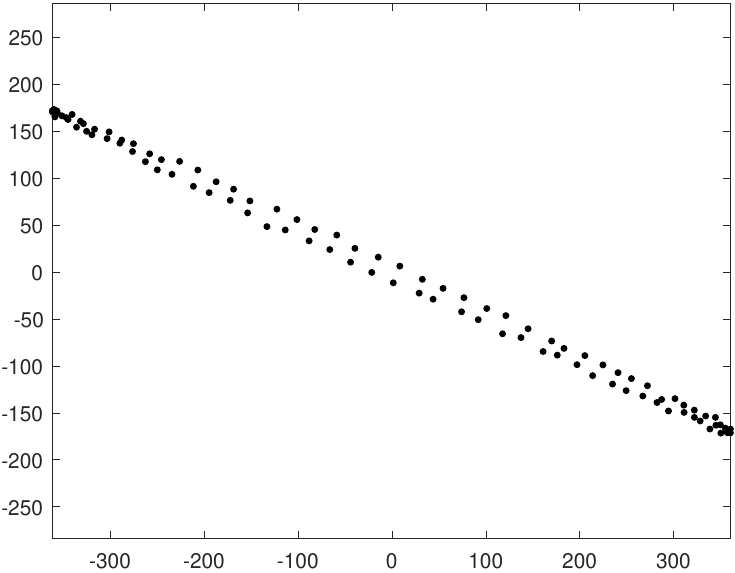}
	\includegraphics[width=0.28\textwidth]{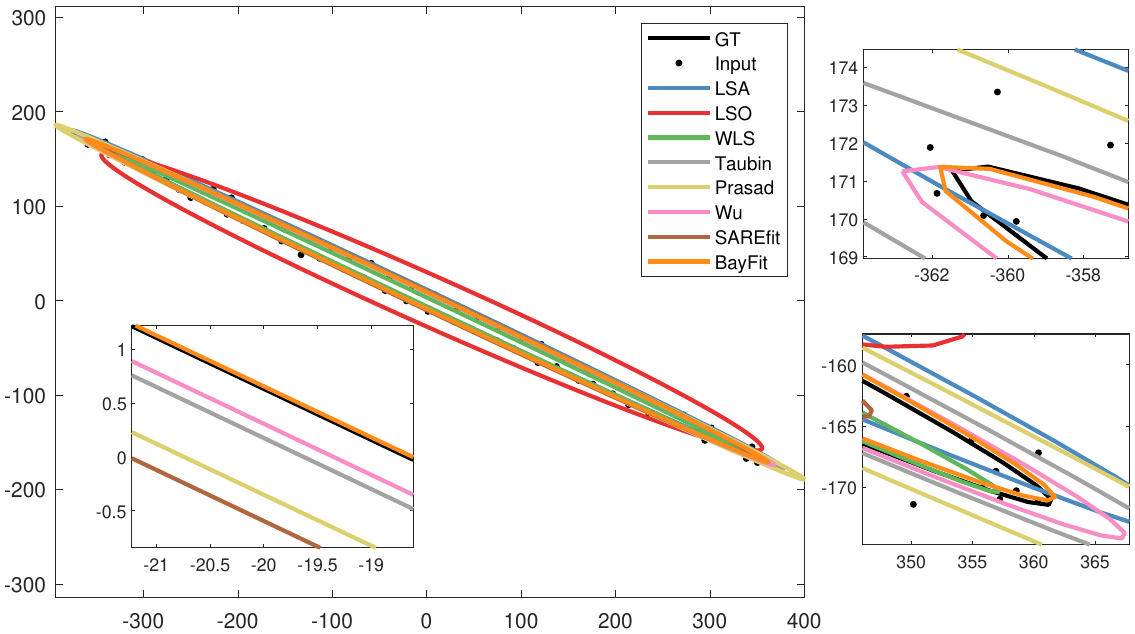}
	\vskip -0.1cm
	\caption{\revise{Qualitative comparisons on ellipse fitting with a very high axis ratio ($r_{ax}=50$). BayFit outperforms competitors with higher accuracy.}}
	\label{fig:axis_ratio_2D}
	\vskip -0.3cm
\end{figure}
\begin{figure}[t]
	\centering
	\includegraphics[width=0.49\textwidth]{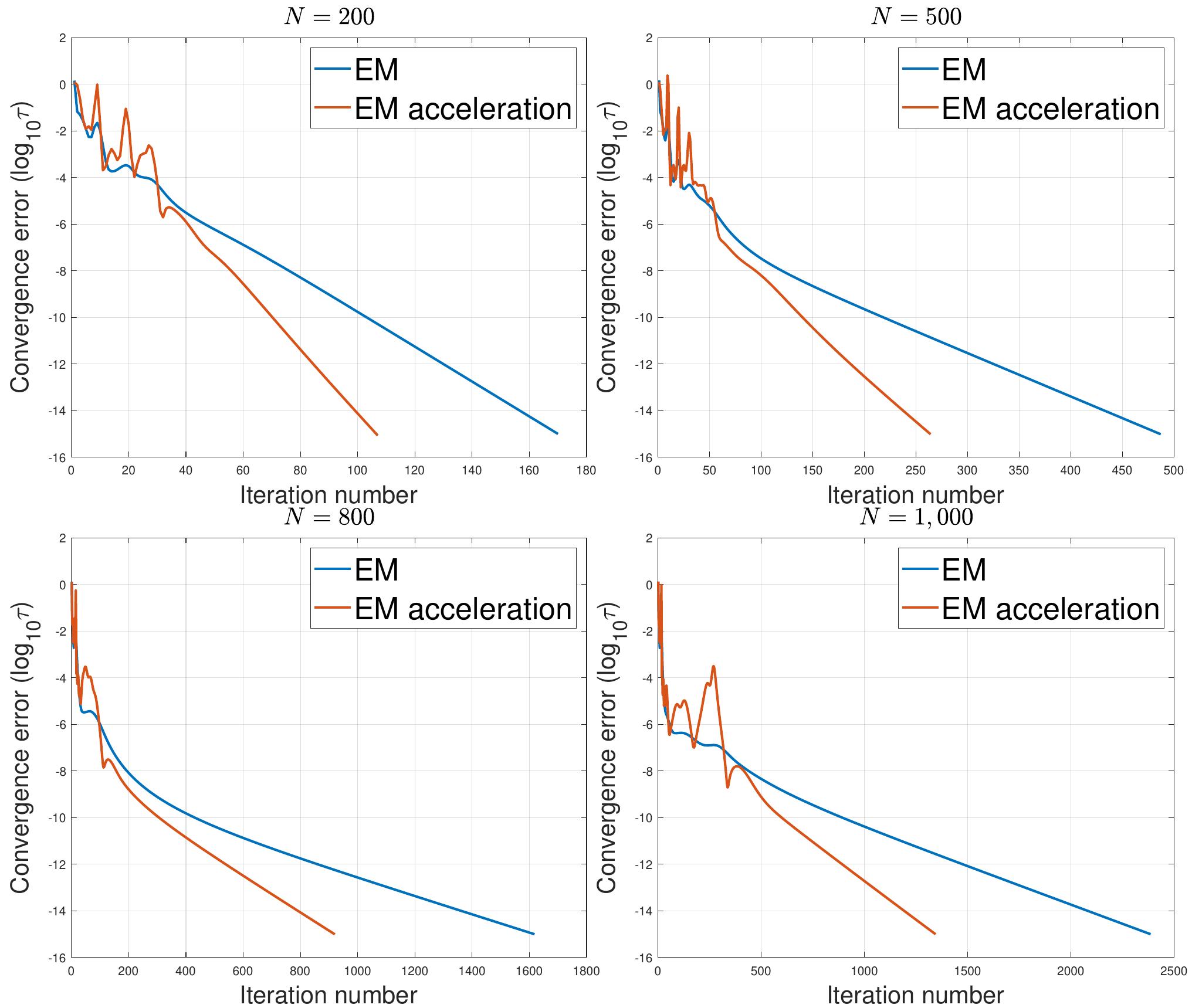}
	\vskip -0.3cm
	\caption{A comparison of convergence speeds between the $\varepsilon$-accelerated EM (orange line) and the primary one (blue line). We vary the point size $N$ during fitting tests. It can be concluded that the acceleration rate becomes much more significant as either the sample size or the desired convergence precision increases. 
	}\label{fig:em_acc}
	\vskip -0.3cm
\end{figure}
\revise{\subsubsection{Sampling from Ellipsoidal Surfaces}\label{sec:sampling}
In synthetic experiments, we employ the following methods to generate sample points from ellipsoidal surfaces for the evaluation of different approaches. The data $\mathbf{Y}$ defined in Eq.~\ref{eq:sum} are} {sampled according to the parameterization form of Eq.~\ref{eq:parameterization}. Our sampling approach bears similarity to recursive sampling as proposed in~\cite{vaskevicius2017revisiting}, which aims to provide a complete coverage of the target surface.} {The input test point clouds are first sampled from a unit sphere created via a Gaussian distribution~\cite{lopez2017robust}, using \revisemajor{the following  
concise} three-line Matlab implementation:
\begin{equation*}
	\begin{small}
		\begin{aligned}
			&\texttt{Samples=randn(Dimension,NumSamples)}\\
			&\texttt{SampleNorms=sqrt(sum(Samples.$*$Samples,1))}\\
			&\texttt{Samples\!=\!Samples.\!/\!repmat\!(\!SampleNorms,\![\!Dimension 1]\!)\!} \\
		\end{aligned}
		\label{im:sampling}
	\end{small}
\end{equation*}Then, we deploy a spatial transformation defined by the desirable ellipsoidal parameters to generate the final ellipsoidal point clouds. We present more detailed descriptions regarding the sampling in Section 5 of the \emph{Supplementary Material}. Fig.~\ref{fig:sampling} illustrates two sets of ellipsoidal point clouds with various data sizes obtained through parametric and random samplings, where the samples cover the complete range of ellipsoidal surfaces. Since our evaluation metrics primarily rely on the shape parameters rather than the \revisemajor{individual shortest} distance from points to surfaces, the generated point samples are sufficient for our purpose. \revisemajor{However, for more advanced sampling methods, particularly those involving near equal-distance schemes applied to superquadrics, readers are encouraged to explore the excellent work by~\cite{pilu1995equal,vaskevicius2017revisiting,liu2022robust} for further study.}
}

\subsubsection{Noisy Evaluations}
We add zero-mean Gaussian noise with $\sigma\in[5\%, 25\%]$ to 200 data points that are sampled from a randomly synthetic ellipsoidal surface. We compare BayFit with state-of-the-art algebraic methods, including DLS~\cite{li2004least}, HES~\cite{kesaniemi2017direct}, MQF~\cite{8578470,birdal2019generic}, and Koop~\cite{vajk2003identification}, as well as geometric methods such as GF~\cite{bektas2015least} and Taubin~\cite{taubin1991estimation}. Additionally, we include the robust method RIX~\cite{lopez2017robust}, which is specifically designed to handle outliers. We re-implement MQF by Matlab due to the unavailable code. Besides, we manually tune the hyper-parameters in RIX to make it work convincingly. For example, \revisemajor{we vary} the maximal step size of RIX from 50 to 100, and the scale factor is tuned from 1.5 to 6, as we found that the default settings of RIX often lead to noticeable deviations. Table~\ref{tab:ellipsoid_noise} indicates that the proposed method achieves the overall best fitting and \revisemajor{remains} highly stable even in the presence of heavy noise. Although GF and RIX obtain small \revisemajor{biases} than other least-squares-based competitors, they also produce significant deviations when $\sigma>20\%$. The fitting samples of compared approaches are presented in Fig.~\ref{fig:ellipsoid_noise_exam}.

\subsubsection{Robustness Against Outliers}
Next, we contaminate the ground truth data points by a set of outliers with the percentage $\eta$ increasing from $5\%$ to $60\%$ and the noise level equivalent to $\sigma=5\%$. Two popular M-estimators \emph{Tukey}~\cite{rousseeuw1991tutorial} and \emph{Huber}~\cite{huber2011robust} are used for comparison. They use \emph{influence functions} to depress outlier disturbances and \revisemajor{transform} robust fitting \revisemajor{into} a \emph{iteratively re-weighted least-squares} problem. Table~\ref{tab:ellipsoid_outlier} suggests that all methods \revisemajor{yield} similar results when $\eta<20\%$, however, RIX starts producing significant deviations when $\eta>30\%$. Tukey and Huber share similar performance and they are indeed more stable than RIX, which benefits from their weighting principle for distinct residuals. Nevertheless, when $\eta>50\%$, they also \revisemajor{produce} substantial deviations. \revisemajor{In contrast}, BayFit consistently performs well across different outlier percentages. Its evaluation metrics remain fairly small, especially for the shape estimate $E_a$. Even when $\eta$ up to $60\%$, it is still capable of guaranteeing accurate fittings. We present several comparison results in Fig.~\ref{fig:outlier_exam}.

\revise{\subsubsection{Occlusion Evaluations}}
{
We also assess the effectiveness of our designed method in ellipsoid fitting involving occlusion. For simplicity, we adopt a canonical ellipsoid surface with semi-axis lengths equal to $3$, $5$, and $4$ in the $x$, $y$, and $z$ directions, respectively. To \revisemajor{create} occlusion data, we \revisemajor{introduce} a cutting plane orthogonal to the $y$ axis, \ie, $y=t$, to intercept the ellipsoidal surface, as illustrated in the inset of Fig.~\ref{fig:ell_plane}.  We take the ellipsoidal surface with $y>t$ as the occlusion part.

By moving $t=4$ to $t=-1$, we increase the volume occlusion percentage $O(t)$, as reported in Table~\ref{table:occlusion} (\revisemajor{the} derivation of $O(t)$ is presented in Section 6 of the \emph{Supplementary Material}). The quantitative statistical results are detailed in Fig~\ref{fig:ell_plane}. It can be observed that as the occlusion percentage increases, all approaches tend to exhibit larger deviations. MQF and Koop produce significant estimation deviations in terms of both shape and offset metrics when the occlusion percentage exceeds $50\%$. \revisemajor{In comparison}, our method demonstrates performance comparable to other algebraic fitting approaches under these occlusion conditions. The geometric fitting methods, including Taubin and GF, produce higher quality fittings. This highlights the advantages of incorporating geometric information in the fitting. For further insights, we present two sets of qualitative comparisons in Section 6 of the \emph{Supplementary Material}.  }

\subsubsection{Influence of the Axis Ratio}
Subsequently, we evaluate the performance of BayFit under the challenging settings where the axis ratio $r_{ax}$ varies, given that many current ellipsoid-specific fitting algorithms require \emph{a priori} knowledge of or are susceptible to $r_{ax}$. We randomly generate a series of 3D ellipsoidal surfaces with $r_{ax}\in[1, 5]$ and report the corresponding statistical results in   Fig.~\ref{fig:axis_ratio}. Note that we have excluded non-ellipsoidal quadrics \revisemajor{from the statistics} as failed fittings will make the error metrics unreliable. As observed, BayFit achieves substantially stable estimates among different $r_{ax}$ and the corresponding fitting biases are significantly lower than competitors. Instead, Taubin and Koop are more sensitive to large $r_{ax}$, while RIX produces more shape biases than center biases as $r_{ax}$ increases. We present three randomly generated ellipsoidal surfaces with $r_{ax}=5, 8, 10$ together with the corresponding fitting results in Fig.~\ref{fig:axis}.

{Furthermore, we present a validation of our method's ability to handle very high axis ratio fitting in 2D. This test aims to demonstrate that our approach can effectively enhance current ellipse fitting methods, although their tolerated $r_{ax}$ in $\mathbb{R}^2$ is up to $+\infty$ (as shown in Table~\ref{tab:axis_ratio}). In the left of Fig.~\ref{fig:axis_ratio_2D}, a highly elongated ellipse with $r_{ax}=50$ is generated. The fitting results displayed on the right side of Fig.~\ref{fig:axis_ratio_2D} along with zoom-in subfigures regarding the ellipse ends, confirm that our method consistently achieves excellent fittings, outperforming competitors in terms of accuracy, especially on parts with \emph{high curvatures}. Additional 2D fitting results under high axis ratio conditions are reported in Section 13 of the \emph{Supplementary Material}.}

\subsubsection{Investigation of the Vector $\varepsilon$ Technique}
Furthermore, we investigate the effect of the $\varepsilon$-accelerated EM algorithm on ellipsoid fitting. We randomly sample $N=200$, 500, 800, and 1,000 data points from an ellipsoidal surface and compare the convergence behavior of the $\varepsilon$-accelerated EM algorithm with the primary one. The comparison curves in Fig.~\ref{fig:em_acc} \revisemajor{reveal} that (1) for \revisemajor{a fixed number of data points} $N$, $\varepsilon$-accelerated EM converges more quickly than the primary EM as the required convergence accuracy $\tau$ increases {(for scenarios involving lower convergence accuracy such as $\tau=10^{-4}$, we provide additional insights in Section 9 of the \emph{Supplementary Material})}; (2) conversely, for a fixed value of $\tau$, as $N$ increases, the difference in the number of iterations between the $\varepsilon$-accelerated EM and EM becomes increasingly larger. This indicates a faster convergence process achieved by $\varepsilon$-accelerated EM, particularly for large-scale observation samples.

\begin{figure}[t]
	\centering
	\includegraphics[width=0.239\textwidth]{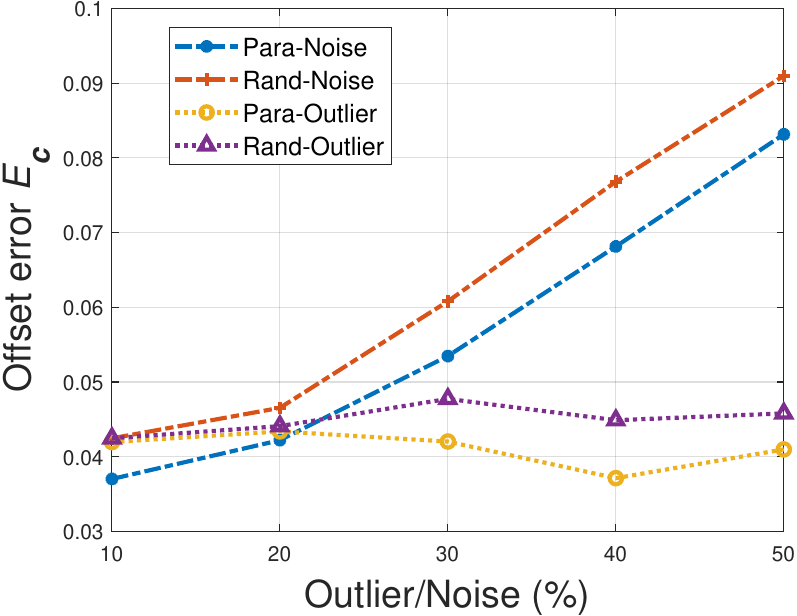}
	\includegraphics[width=0.24\textwidth]{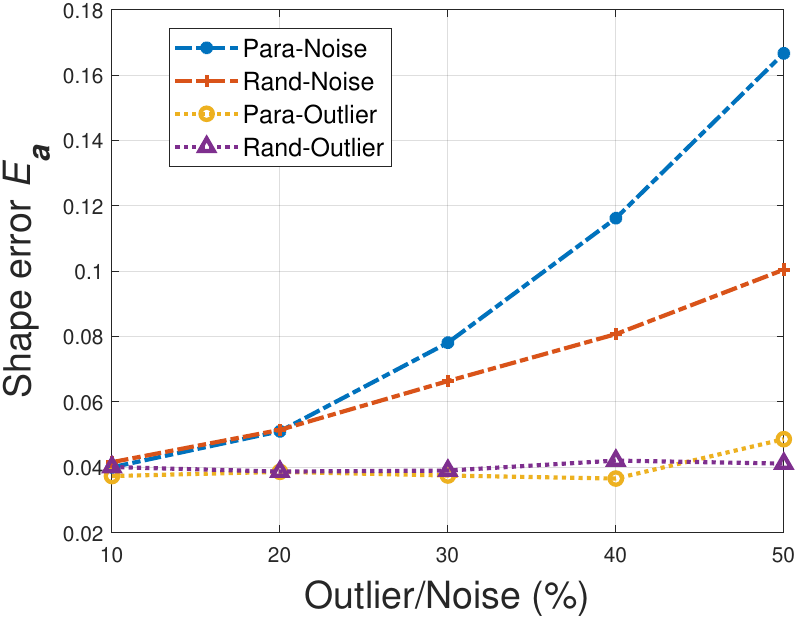}
	\vskip -0.3cm
	\caption{\revise{Comparisons between parametric and random sampling strategies. They both yield similar results under different noise and outlier settings.}}
	\label{fig:ablation_sampling}
	\vskip -0.3cm
\end{figure}

\revise{\subsubsection{Ablation Study of the Sampling Schemes}
We adopt the parametric form defined in Eq.~\ref{eq:parameterization} to generate $\mathbf{Y}$. To compare the differences between this parametric manner and the random sampling method introduced in Section~\ref{sec:sampling},} {we also adopt the random strategy to generate $\mathbf{Y}$. Fig.~\ref{fig:ablation_sampling} summarizes statistical results using these two types of sampling strategies under outlier and noise settings, denoted as Para-* and Rand-*. It can be observed that both parametric and random samplings yield similar fitting accuracy under outlier disturbances. In noisy settings, parametric and random samplings achieve slightly higher accuracy in offset and shape estimates separately, but these differences are relatively small.}

\begin{table}[t]
		\centering
		\vskip -0.3cm
		\caption{{Geometry of the ellipsoidal axes tested in high-dimensional spaces.}}
		\vskip -0.2cm
		\renewcommand{\arraystretch}{1.5}   
		\begin{tabular}{c|c|cc}
			\hline
			\textbf{Space}&\textbf{Square of the semi-axis length}&\textbf{Axis ratio}\cr
			\hline
			{$\mathbb{R}^4$}&[3, 4, 36, 6] &2$\sqrt{3}$\\ \hline
			{$\mathbb{R}^{12}$}
			&{[1, 5, 16, 25, 4, 12, 4, 10, 8, 12, 4, 1]}&5\\
			\hline
		\end{tabular}
		\label{tab:high}
		\vskip -0.5cm
	\end{table}

\begin{figure}[b]
\centering
\includegraphics[width=0.235\textwidth]{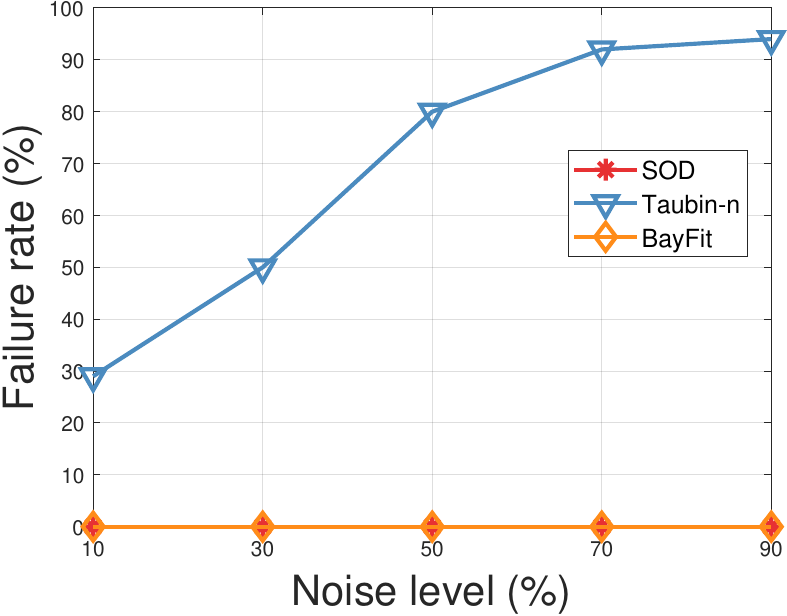}
\includegraphics[width=0.235\textwidth]{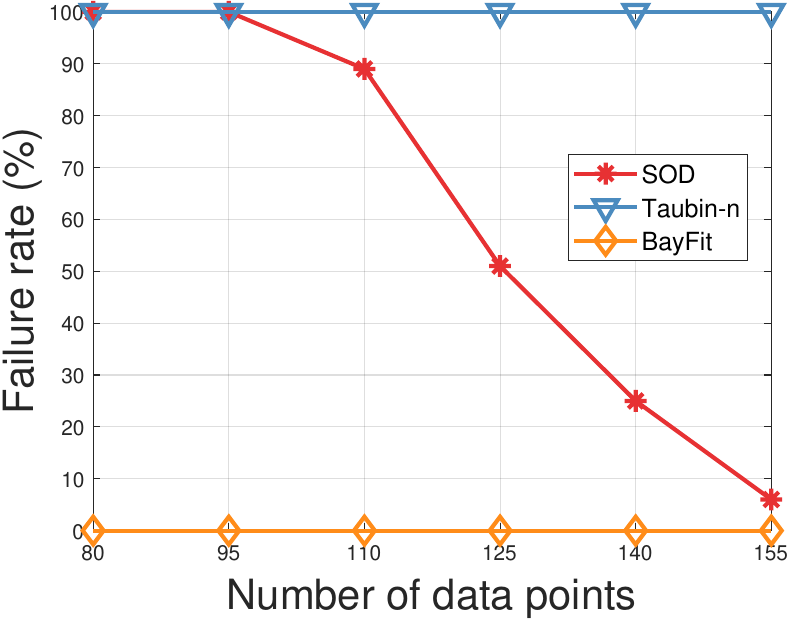}
			
\vskip -0.1cm
\caption{\revise{Experiments in high-dimensional spaces including $\mathbb{R}^4$ (left) and $\mathbb{R}^{12}$ (right).  Our method maintains a zero failure rate consistently regardless of the noise intensity and the number of points, and is therefore able to generate ellipsoid-specific fittings in multidimensional spaces.} 
}\label{fig:high}
\end{figure}

\begin{figure}[t]
	\centering
	\includegraphics[width=0.235\textwidth]{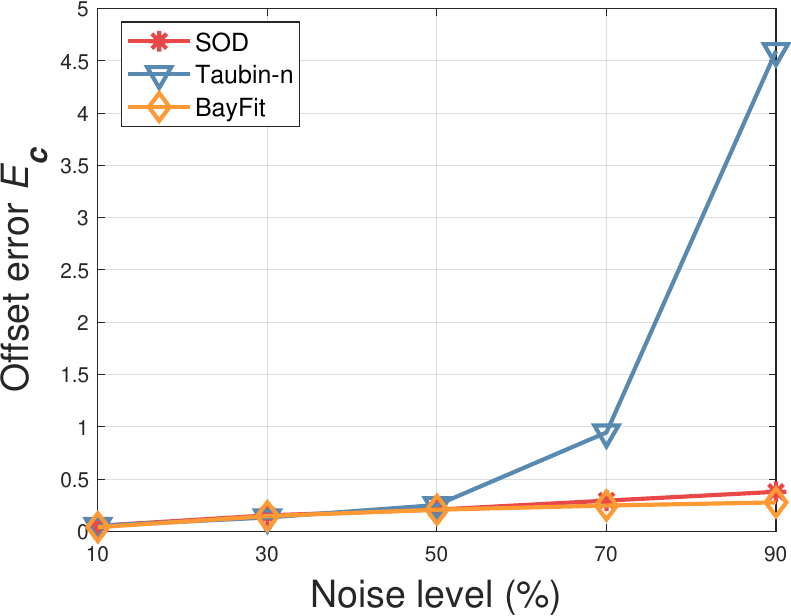}
	\includegraphics[width=0.235\textwidth]{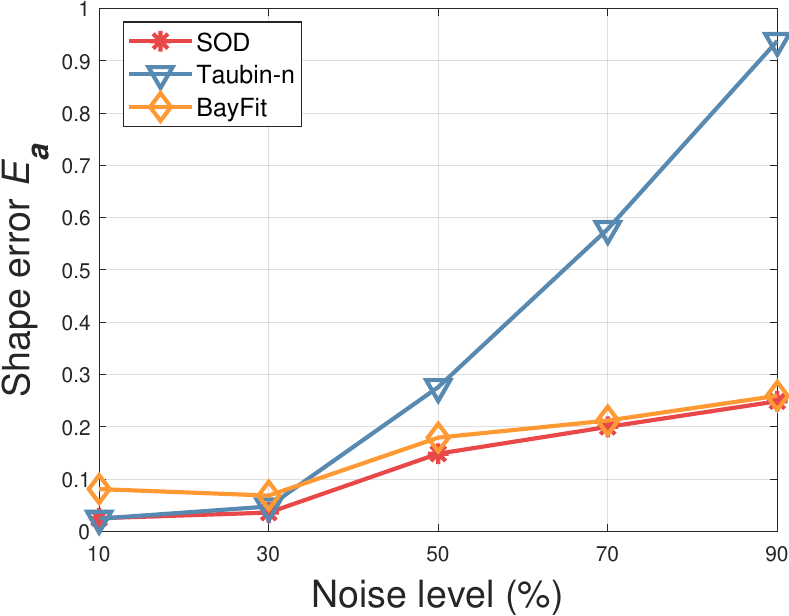}

	\includegraphics[width=0.235\textwidth]{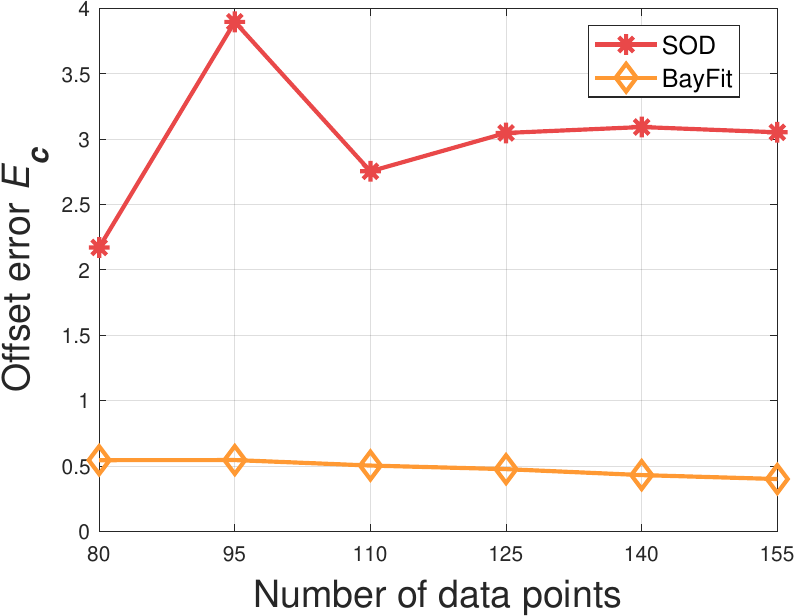}
	\includegraphics[width=0.235\textwidth]{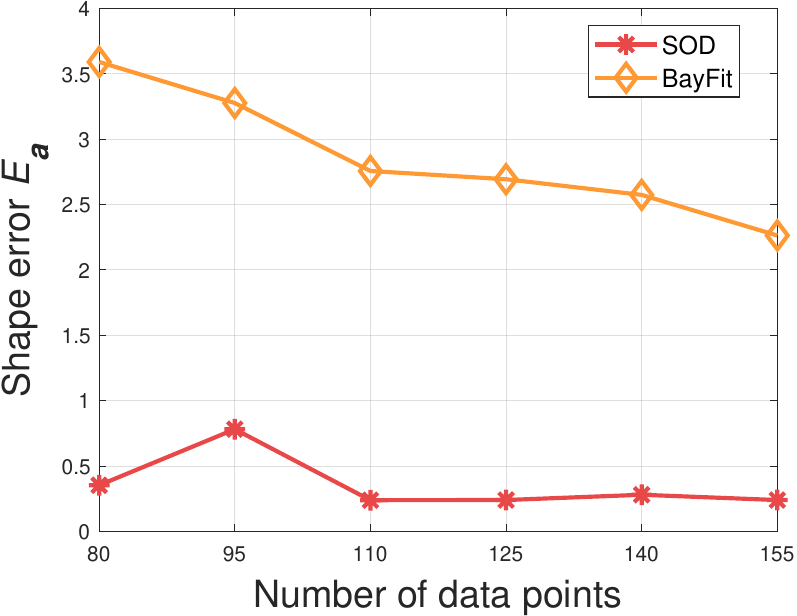}

	\vskip -0.3cm
	\caption{\revise{Qualitative comparisons in high-dimensional spaces regarding noise and point quantity.  Top: The noisy test result in $\mathbb{R}^{4}$. Bottom: The point size test result in $\mathbb{R}^{12}$.}  
	}\label{fig:high_accuracy}
	\vskip -0.3cm
\end{figure}
\begin{figure}[!htbp]
	\centering
		\includegraphics[width=0.235\textwidth]{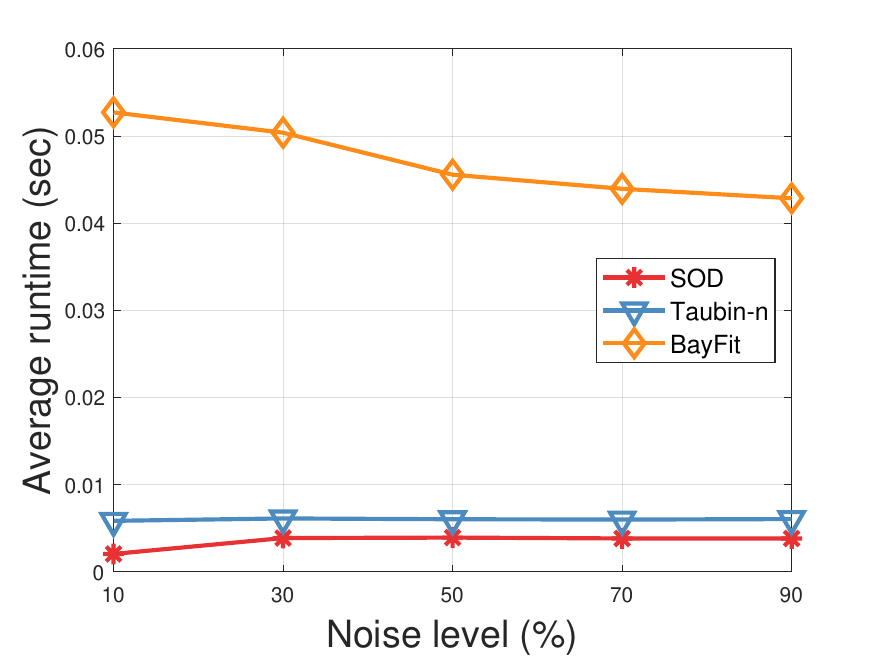}
		\includegraphics[width=0.235\textwidth]{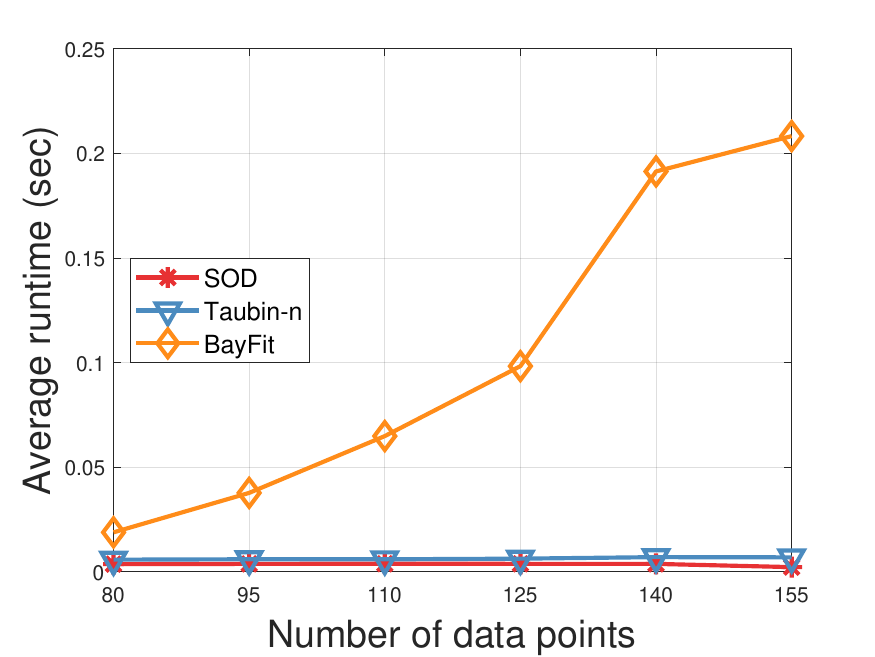}
	\vskip -0.3cm
	\caption{\revisemajor{The average runtime in the corresponding high-dimensional spaces of $\mathbb{R}^4$ (left) and $\mathbb{R}^{12}$ (right). }
	}\label{fig:high_time}
	\vskip -0.5cm
\end{figure}

\subsection{Generalization to High-Dimensional Spaces} 
Finally, we demonstrate the generalization capability of the designed algorithm on the high-dimensional space $\mathbb{R}^n$. We consider ellipsoidal surfaces embedded in $\mathbb{R}^4$ and $\mathbb{R}^{12}$ for test. The square of the semi-axis length and the corresponding axis ratios are reported in Table~\ref{tab:high}. 
BayFit is compared with two baseline methods, \revisemajor{namely} SOD and Taubin-n~\cite{kesaniemi2017direct}, which are specifically customized for high-dimensional ellipsoid fittings. However, it is noticeable that despite the elaboration of ellipsoid-specific constraints from 3D to higher-dimensional spaces, SOD and Taubin-n still cannot guarantee ellipsoid-specific surfaces, in other words, they may result in non-ellipsoidal quadrics under \revisemajor{certain} fitting settings (\eg, Proposition~\ref{prop:1}). \revisemajor{Consequently}, we report the failure rate with respect to non-ellipsoidal surfaces in 100 tests.

{First, we sample 100 data points from a ellipsoidal surface in $\mathbb{R}^4$. The noisy test result in the left panel of Fig.~\ref{fig:high} indicates that the noise intensity influences the fitting results, and heavy noise has deviated Taubin-n from ellipsoidal surfaces to other quadrics significantly, whereas SOD and BayFit \revisemajor{consistently maintain} all ellipsoid-specific fittings in 100 runs. The corresponding accuracy of these fittings is detailed in the top panel of Fig.~\ref{fig:high_accuracy}. Specifically, when the noise level surpasses a certain threshold, Taubin-n exhibits considerable deviations in both the offset and shape metrics. In contrast, the SOD and BayFit methods maintain a relatively more stable and accurate fitting performance.}

\revise{We then extend our ellipsoid fitting analysis to a significantly higher-dimensional space of $\mathbb{R}^{12}$ and investigate the impact of data quantity on ellipsoid-specific fitting. The fitting outcomes with a noise intensity of $50\%$ are summarized in the right panel of Fig.~\ref{fig:high}. We observe that when the number of data points is less than $100$, both SOD and Taubin-n exhibit noticeably higher failure rates in achieving ellipsoid-specific fits. \revisemajor{In contrast}, our algorithm consistently delivers successful ellipsoidal surface fits across a range of data sizes. The fitting accuracy, as presented in the bottom of Fig.~\ref{fig:high_accuracy}, highlights that BayFit outperforms SOD in terms of offset estimation accuracy, while SOD excels in achieving more accurate shape estimation. However, the accuracy of BayFit tends to improve as the dataset size grows.}

\revisemajor{We also report the average runtime for the tested methods in both high-dimensional spaces. As shown in the left panel of Fig.~\ref{fig:high_time}, all methods exhibit a relatively consistent fitting speed as the noise level increases. While BayFit is slightly slower compared to the other approaches, it still maintains real-time fitting with the} {runtime below 0.06s. In the higher-dimensional space of $\mathbb{R}^{12}$, as the number of data points increases, BayFit experiences a longer runtime. However, it still achieves fast fitting within the time less than 0.25s and, notably, demonstrates a significant improvement in fitting accuracy, thereby further enhancing its practical applicability.}

\revise{\subsection{Limitations}
In this section, we summarize the limitations of BayFit.}
{\subsubsection{Heavy Occlusion}
While our method demonstrates superiority over predecessor} {approaches in maintaining ellipsoid-specificity regardless of spatial dimensions and input data, effectively handling large axis ratios, robustness against noise and outliers, as well as generalization to high-dimensional spaces, it may return less accurate fittings when data points are heavily occluded. Indeed this challenge of occlusion is a common issue in the field of ellipse and ellipsoid fitting~\cite{kanatani2016ellipse}. To mitigate this issue, it is desirable to incorporate geometric information, such as the Sampson distance, into our robust fitting framework.}
\revise{\subsubsection{Determination of the Optimal Parameter $k$}		
During the initialization process of $M$ and $w$ \revisemajor{in our method}, calculating $\RDOS$ necessitates the specification of $k$. However, there is currently a lack of an automated technique to efficiently determine the optimal value of $k$, despite the possibility of exploring a range of $k$ values to find the best fit. It is also worth noting that the differences in fitting results among most $k$ settings are generally minor.
}
\revise{\subsubsection{Uniform Sampling}}	
{Another limitation of the proposed method is related to the current sampling schemes, which rely on parametric and random approaches. While these schemes are simple and efficient, enabling complete coverage of the ellipsoidal surfaces across various spatial dimensions, they do not guarantee strictly uniform coverage. Additionally, since the closed-form solution of Eq.~\ref{eq:integral_expec} is typically intractable, employing uniform sampling following an affine transformation still results in a biased Monte Carlo estimation of the predictive distribution. This bias can potentially lead to decreased fitting accuracy, particularly in cases with larger axis ratios and higher-dimensional spaces. To mitigate this limitation, we suggest exploring more advanced yet potentially more complex near equal-distance strategies from superquadrics~\cite{pilu1995equal,vaskevicius2017revisiting,liu2022robust} as an alternative approach.}

\section{Conclusions and Future Directions}
\revise{We presented a comprehensive solution termed BayFit for fitting ellipsoidal surfaces in multidimensional spaces, effectively addressing various challenges encountered by previous approaches. Our method offers substantial advantages including the ability to produce ellipsoid-specific results, dimension independence, robustness to axis ratios and outliers, and computational efficiency.} {We employ a Bayesian optimization framework by considering each model point as a possible source for generating each measurement data. This is established based on the insight of the predictive distribution within the Bayesian context.} To ensure ellipsoid-specific results, we incorporate a uniform prior and effectively solve the geometric parameters through maximum a posterior probability or a degenerate maximum likelihood optimization. Moreover, \revisemajor{to} enhance the robustness of BayFit, \revisemajor{we incorporate} an additional uniform distribution into the predictive distribution, which significantly improves the method's ability to handle outliers. Besides, we introduce an $\varepsilon$-accelerated EM algorithm for fast and stable fitting.

Our key contributions in this work constitute the following aspects: As opposed to previous ellipsoid-specific constraints that are typically subject to spatial dimensions (Proposition~\ref{prop:1}), we introduce a Bayesian prior distribution to constrain the search for primitive parameters within an ellipsoid domain. This dimension-independent approach makes the employment of BayFit in multidimensional scenarios straightforward and flexible. From a Bayesian information criterion perspective, this also equates to selecting the ellipsoidal quadric with the largest posterior probability. Our method does not suffer from limitations related to axis ratios, a challenge for many previous methods. It can accurately estimate parameters for ellipsoids with large axis ratios, providing a significant advantage \revisemajor{in handling ellipsoids with extreme shapes.} As for robustness, especially for outliers that are overlooked by many baseline algorithms but often exist in captured point clouds, we incorporate a uniform distribution into the predictive distribution, resulting in a highly robust fitting mechanism. The necessity of the uniform component is also theoretically proven through the Kullback-Leibler Information Criterion. In terms of the computational efficiency, we introduce the vector $\varepsilon$ technique in EM to accelerate BayFit converge much more quickly. This is particularly advantageous for large-scale point clouds or scenarios requiring high fitting precision.

\revisemajor{Extensive experimental evaluations have been conducted to validate the performance of BayFit}, demonstrating its effectiveness in fitting popular 2D ellipses and 3D ellipsoids, as well as higher-dimensional cases, particularly for settings with no \emph{a priori} knowledge or under various disturbances. Moreover, BayFit has showcased its versatility and effectiveness in a wide range of practical applications, making it of great interest and utility within the computer vision, robotics, and graphics communities.

There are several promising avenues for future work that can extend the applicability and enhance the performance of BayFit.  
One feasible direction is generalizing BayFit for robustly \revisemajor{estimating} common conics or quadrics. The Bayesian prior distribution and the optimization process presented in this work can be readily adapted to other geometric primitives, \revisemajor{allowing for a broader range of shape fitting applications}. \revise{Additionally, exploring the composition of ellipsoids for 3D shape reconstruction or abstraction, as seen in the pioneering work~\cite{wu2022primitive},} {holds potential for future research.} Another interesting direction is  integrating the Bayesian optimization framework with the Bayesian Information Criterion (BIC) to fit type-agnostic geometric primitives. Such an algorithm is expected to autonomously select the best model or provide an acceptable fit in a fully automatic manner, guided by the maximum a posterior probability.

\section*{Acknowledgements} 
This work is partially funded by the Strategic Priority Research Program of the Chinese Academy of Sciences (XDB0640000), National Science and Technology Major Project (2022ZD0116305), National Natural Science Foundation of China (62172415,12325110), the innoHK project, and the CAS Project for Young Scientists in Basic Research (YSBR-034).

	\ifCLASSOPTIONcaptionsoff
	\newpage

\fi

	
	
	%
	%
	%
	
	\bibliographystyle{ieeetr}
	\bibliography{references}

\end{document}